
\input epsf
\input harvmac
\overfullrule=0pt
\def\Title#1#2{\rightline{#1}\ifx\answ\bigans\nopagenumbers\pageno0\vskip1in
\else\pageno1\vskip.8in\fi \centerline{\titlefont #2}\vskip .5in}

%
%

\baselineskip=18pt plus 2pt minus 2pt
\def\cI{{\cal I}}

\def\Om{{\Omega}}
\def\sq{{\vbox {\hrule height 0.6pt\hbox{\vrule width 0.6pt\hskip 3pt
   \vbox{\vskip 6pt}\hskip 3pt \vrule width 0.6pt}\hrule height 0.6pt}}}
\def\CD{{\cal D}}
\font\ticp=cmcsc10
\def\sq{{\vbox {\hrule height 0.6pt\hbox{\vrule width 0.6pt\hskip 3pt
   \vbox{\vskip 6pt}\hskip 3pt \vrule width 0.6pt}\hrule height 0.6pt}}}
\def\ajou#1&#2(#3){\ \sl#1\bf#2\rm(19#3)}
\def\frac#1#2{{#1 \over #2}}

\def\pmi{{\partial_-}}
\def\p+{{\partial_+}}
\def\la{{\lambda}}
\def\air{{\rightarrow}}
\def\ci{{\cal I}}
\def\half{{1 \over 2}}
\def\overleftrightarrow#1{\buildrel\leftrightarrow\over#1}
\ifx\epsfbox\UnDeFiNeD\message{(NO epsf.tex, FIGURES WILL BE IGNORED)}
\def\figin#1{\vskip2in}
\else\message{(FIGURES WILL BE INCLUDED)}\def\figin#1{#1}\fi
\def\ifig#1#2#3{\xdef#1{fig.~\the\figno}
\goodbreak\midinsert\figin{\centerline{#3}}%
\smallskip\centerline{\vbox{\baselineskip12pt
\advance\hsize by -1truein\noindent\footnotefont{\bf Fig.~\the\figno:} #2}}
\bigskip\endinsert\global\advance\figno by1}
%
%
\gdef\journal#1, #2, #3, 1#4#5#6{{\sl #1} {\bf #2}
(1#4#5#6) #3}
\lref\hhay{S.W. Hawking and J.D. Hayward, \ajou  Phys. Rev.
&D49 (94) 5252.}
\lref\hast{S. W. Hawking and J.M. Stewart, hep-th/9207105,\ajou  Nucl. Phys.
&B400 (93) 393.}
\lref\tdu{T.Tada and S. Uehara, {\sl Consequence of Hawking Radiation from
2d Dilaton Black Holes}, hep-th/9409039.}
\lref\thrd{ K.~Kuchar, \journal J.~Math.Phys., 22, 2640, 1981\semi
N.~Cadderni and M.~Martellini, \journal
Int.~J.~Theor.~Phys., 23, 23, 1984\semi  A. Jevicki, in {\sl Frontiers in
Particle Physics '83},
D. Sijacki et. al., eds., World Scientific (1984)\semi
I. Moss, in {\sl Field Theory, Quantum Gravity and Strings, II}
eds. H. deVega and N. Sanchez,
Springer, Berlin (1987)\semi T.~Banks, \journal Nucl.~Phys.,
B309, 643, 1988\semi V. Rubakov, \journal Phys. Lett., B214, 503, 1988\semi
S.B.~Giddings and A.~Strominger, \journal Nucl.~Phys.,
B321, 481, 1989\semi W. Fischler, I. Klebanov, J. Polchinski and L. Susskind
\journal Nucl.~Phys.,
B327, 157, 1989\semi A.~Strominger, \journal Phil.~Trans.~R.~Soc.~Lond.,
A329, 395, 1989.}
\lref\bhexp{See {\it e.g.} the reviews R. P. van der Marel {\sl
Black Holes in Galactic Nuclei: the Dynamical Evidence}, astro-ph/9410012\semi
L. Stella, G.L.
 Israel, S. Mereghetti and D. Ricci, {\sl The Search for Black Holes in
X-Ray Binaries: an Update}, astro-ph/9410073.}
\lref\hawk{S.W.~Hawking, \journal Comm.~Math.~Phys., 43, 199, 1975.}
\lref\hawktwo{S.W.~Hawking, \journal Phys. Rev., D14, 2460, 1976.}
\lref\CGHS{C.G. Callan, S.B. Giddings, J.A. Harvey, and A. Strominger,
\ajou Phys. Rev. &D45 (92) R1005.}
\lref\RST{J.G.~Russo, L.~Susskind, and L.~Thorlacius,
\ajou Phys.~Lett. &B292 (92) 13.}
\lref\lpst{D.A.~Lowe, J.~Polchinski, A.~Strominger and L.~Thorlacius,
unpublished (1994).}
\lref\suss{L.~Susskind, {\sl Comment on a Proposal by Strominger},
Preprint SU-ITP-94-14, hep-th/9405103.}
\lref\suent{L.~Susskind and J. Uglum,\ajou Phys. Rev. &D50 (94) 2700.}
\lref\asmg{A.~Strominger, {\sl Unitary Rules for Black Hole Evaporation},
hep-th/9410187.}
\lref\polst{J.~Polchinski and A.~Strominger,
hep-th/9407008, \ajou Phys. Rev. &D50 (94) 7403.}
\lref\bek{J. D. Bekenstein,  \ajou Phys. Rev. &D7
 (73) 2333 \semi  \ajou Phys. Rev. &D9 (74) 3292.}
\lref\sbgtr{S.B.~Giddings,  {\sl Quantum Mechanics
of Black Holes}, hep-th/9412138, to appear in Proceedings of the Spring School
on String Theory, Gauge Theory and Quantum Gravity,
Trieste, April 11-19, 1994, R. Dijkgraaf
{\it et al.} eds.}
\lref\pnrs{See e.g. S. W. Hawking and G. F. R. Ellis, {\sl The Large
Scale
Structure of Space-time}, Cambridge University Press (1973).}
\lref\wald{R.M.~Wald, {\sl General Relativity},
The University of Chicago Press,
Chicago (1984).}
\lref\dxbh{S.B.~Giddings and A.~Strominger, \journal Phys. Rev., D46,
627, 1992.}
\lref\WittTwod{E. Witten, \ajou Phys. Rev&
D44 (91) 314.}
\lref\Mandal{G.~Mandal, A.~Sengupta, and S.~Wadia,
\ajou Mod. Phys. Lett. &A6 (91) 1685.}
\lref\Mann{R.B.~Mann,  in Proc. of 4th Canadian
Conf. on General Relativity and Relativistic Astrophysics, Winnipeg,
Canada,
and references therein.}
\lref\rom{R. Jackiw,  in {\sl Quantum Theory of Gravity},
S.~Christensen, ed.
(Hilger, Bristol U.K. 1984)\semi D.~Cangemi and R.~Jackiw, \ajou
Phys.~Rev.~Lett. &69 (92) 233.}
\lref\tei{C.~Teitelboim  in {\sl Quantum Theory of Gravity},
S.~Christensen, ed.
(Hilger, Bristol U.K. 1984)}
\lref\cham{A.~Chamseddine, \ajou  Nucl. Phys. &B368 (92) 98.}
\lref\CrFu{S. M. Christensen and S. A. Fulling,
\ajou Phys. Rev. &D15 (77) 2088.}
\lref\bcmn{B. K. Berger, D. M. Chitre, V. E. Moncrief,
and Y. Nutku. \ajou Phys. Rev.  &D5 (72 2467.}
\lref\poly{A. M. Polyakov,
\ajou Phys. Lett. &103B (81) 207.}
\lref\BDDO{T. Banks, A. Dabholkar, M.R. Douglas, and M O'Loughlin,
\ajou Phys. Rev.
&D45 (92) 3607.}
\lref\nummer{T.~Piran and A.~Strominger, \ajou
Phys. Rev. &D48 (93) 4729.}
\lref\past{Y.~Park and A.~Strominger, \ajou Phys. Rev. &D47
 (93) 1569.}
\lref\jptb{S.R.~Das, S.~Naik and S.R.~Wadia,
\ajou Mod. Phys. Lett. &A4 (89) 1033 \semi
J. Polchinski, \ajou Nucl. Phys. &B324 (89) 123\semi
T. Banks and J. Lykken, \ajou Nucl. Phys. &B331 (90) 173.}
\lref\deAli{S.P.~de Alwis, \ajou Phys.~Lett. &B289 (92) 282\semi
\ajou Phys.~Rev. &D46
(92) 5429.}
\lref\BiCa{A. Bilal and C. Callan,  \ajou Nucl.~Phys. &B394 (93) 73.}
\lref\verl{E.~Verlinde and H.~Verlinde, \ajou Nucl. Phys. &B406 (93) 43 \semi
K. Schoutens, E.~Verlinde and H.~Verlinde, \ajou Phys. Rev. &D48 (93) 2690
and in these proceedings.}
\lref\bc{T.D.~Chung and H.~Verlinde, \journal Nucl.~Phys., B418, 305,
1994,
hep-th/931107\semi S. Das and S. Mukherji, \journal
Phys.~Rev., D50, 930, 1994\semi A.~Strominger and L.~Thorlacius, \journal
Phys.~Rev., D50, 5177, 1994,
hep-th/9405084.}
\lref\asst{A.~Strominger and S.P.~Trivedi, unpublished.}
\lref\rst{J.G.~Russo, L.~Susskind, and L.~Thorlacius, \ajou
Phys.~Rev. &D46 (93) 3444\semi \ajou Phys.~Rev. &D47 (93) 533.}
\lref\AS{to be supplied}
\lref\psf{to be supplied}
\lref\ande{A.~Anderson and B.~DeWitt, \ajou Found. Phys. &16 (86) 91.}
\lref\abdi{J.~Preskill, hep-th/9204058 and in {\sl Black Holes, Membranes,
Wormholes, and Superstrings}, ed.~S.~Kalara and D.V.~Nanopoulos (World
Scientific, Singapore) 1993.}
\lref\fpst{Thomas M.~Fiola, J.~Preskill, A.~Strominger, and Sandip
P.~Trivedi, \ajou Phys. Rev. &D50 (94) 3987.}
\lref\zur{W.H.~Zurek, \ajou Phys. Rev. Lett. &49 (82) 1683.}
\lref\infloss{G.'t Hooft, \ajou Nucl. Phys. &B335 (90) 138
\semi \ajou Phys. Scr. &T36 (91) 247 and references therein.}
\lref\theo{T.~Jacobson, \ajou Phys.~Rev. &D44, (91) 173\semi \ajou ibid.,
&D48 (93) 728.}
\lref\blshift{L.~Susskind, \ajou Phys. Rev. Lett. &71, (93) 2367\semi
L.~Susskind and L.~Thorlacius, \ajou Phys. Rev. &D49 (94) 966\semi
L.~Susskind, \ajou ibid. &49 (94) 6606.}
\lref\prkr{L.~Parker, \ajou Phys.~Rev.~Lett. &21 (82) 562.}
\lref\brdv{N.~D.~Birrell and P.C.W.~Davies, {\sl Quantum fields in
curved space}, section 3.5,
Cambridge University Press, Cambridge 1982.}
\lref\caru{V.A.~Rubakov, \ajou Nucl.~Phys. &B203 (82) 311\semi
C.G.~Callan,  \ajou Phys.~Rev. &D25 (82) 2141\semi \ajou ibid. &D26,
(82) 2058\semi \ajou Nucl.~Phys. &B212 (83) 391.}
\lref\jpia{ J.~Polchinski,
 \ajou Nucl.~Phys. &B242  (84) 345\semi I.~Affleck and J.~Sagi,
\ajou Nucl.~Phys.  &B417 (94) 374.}
\lref\bos{T.~Banks, M.O'Loughlin, and A.~Strominger, \ajou
Phys.~Rev. &D47 (93)  4476\semi see also T.~Banks and M.O'Loughlin,
\ajou
Phys.~Rev. &D47  (93) 540.}
\lref\stas{A.~Strominger and S.P.~Trivedi, \ajou Phys.~Rev. &D48
 (93) 5778.}
\lref\carwil{R. D. Carlitz and R. S. Willey, \ajou Phys.~Rev. &D36
 (87) 2336.}
\lref\sbg{S.B.~Giddings, \ajou Phys.~Rev. &D49 (94) 4078\semi
\ajou Phys.~Rev. &D49 (94) 947.}
\lref\garst{D.~Garfinkle and A.~Strominger, \ajou Phys.~Lett. &B256
(91) 146.}
\lref\remn{F.~Dowker, J.P.~Gauntlett,
D. Kastor and J. Traschen,\ajou Phys.~Rev. &D49 (94) 2909\semi
F.~Dowker, J.P.~Gauntlett, S.B.~Giddings, and G.~Horowitz,
\ajou Phys.~Rev. &D50 (94) 2662,
hep-th/9312172\semi P. Yi, {\sl Toward One-Loop Tunneling Rates of
Near Extremal Magnetic Black Holes}, hep-th/9407173 and {\sl Magnetic Black
Hole Pair Production} gr-qc/9410035 \semi S. W. Hawking,
Gary T. Horowitz, and
 Simon F. Ross, {\sl Entropy, Area, and Black Hole Pairs },
hep-th/9410045\semi  David
 Brown, {\sl Pair Creation of Electrically Charged Black Holes },
gr-qc/9412018.}
\lref\banrev{T.~Banks,
{\sl Lectures on Black Holes and Information Loss}, hep-th/9412131,
to appear in Proceedings of the Spring School
on String Theory, Gauge Theory and Quantum Gravity,
Trieste, April 11-19, 1994, R. Dijkgraaf
{\it et al.} eds.}
\lref\cash{Y. Aharonov, A. Casher and S. Nussinov, \ajou Phys. Lett.
&191B
(87) 51.}
\lref\bps{T.~Banks, M.~Peskin, and L.~Susskind, \ajou
Nucl.~Phys. &B244 (84) 125.}
\lref\sidcc{ S.~Coleman, \ajou Nucl.~Phys. &B310 (88) 643.}
\lref\worm{ S.~Coleman, \ajou Nucl.~Phys. &B307 (88) 864 \semi S.B.~Giddings
and A.~Strominger, \ajou Nucl.~Phys. &B307 (88) 854.}
\lref\wrloss{A.~Strominger, \journal Phys.~Rev.~Lett., 52, 1733, 1984\semi
S.W.~Hawking, \journal Phys. ~Lett., B195, 337, 1987.}
\lref\qabh{J.~Harvey and A.~Strominger, in TASI '92, {\sl From Black Holes and
Strings to
Particles}, (World Scientific, Singapore) 1993.}
\lref\strb{A.~Strominger, in TASI '88, {\sl Particles,
Strings and Supernovae}, (World Scientific, Singapore) 1989.}
\lref\cpt{A.~Strominger, \ajou Phys.~Rev. &D48 (93) 5784.}
\lref\ggs{D.~Garfinkle, S.B.~Giddings, and A.~Strominger, \ajou
Phys.~Rev. &D49  (94) 958.}
\lref\brek{J.D.~Bekenstein, {\sl Do We Understand Black Hole Entropy?},
gr-qc/9409015.}
\lref\gibint{G.W.~Gibbons, in {\sl Fields and Geometry}, ed.~A.~Jadczyk,
(World Scientific, Singapore) 1989.}
\lref\qtdg{S. Giddings and A. Strominger, hep-th/9207034, \ajou
Phys.~Rev. &D47  (93) 2454.}
\lref\GiNe{S.B. Giddings and W.M. Nelson, \ajou Phys.~Rev. &D46 (92) 2486.}
\lref\kzm{Y. Kazama, Y. Satoh, and A. Tsuchiya, {\sl A Unified Approach
to Solvable Models of Dilaton Gravity in Two Dimensions Based on Symmetry},
preprint UT-Komaba 94-16.}
\lref\kzr{Y. Kazama, {\sl On Quantum Black Holes}, hep-th/941224.}
\lref\lowe{D. A. Lowe, \ajou Phys. Rev.  &D47 (93) 2446.}
\lref\tsvpa{T. Piran and R. Parentani, \ajou Phys. Rev. Lett. &73 (94) 2805.}
\lref\tomb{E. Tomboulis, \journal Phys.~Lett., B70,
361, 1977.}
\lref\asst{A. Strominger and S. P. Trivedi, unpublished.}
\lref\sbgl{S. B. Giddings, {\sl Why Aren't Black Holes
Infinitely Pair-Produced?}, hep-th/9412159. }
\lref\snow{P.C. Argyres, A.O.
 Barvinski, V. Frolov, S.B. Giddings, D.A. Lowe, A. Strominger, and L.
 Thorlacius {\sl Quantum Aspects of Gravity}, astro-ph/9412046, to
appear in
Proceedings of the APS Summer Study on Particle and Nuclear
Astrophysics and Cosmology in the Next Millenium, Snowmass,
Colorado, June 29 - July 14, 1994, E.~Kolb {\it et al.} eds.}
\lref\laref{L. Thorlacius,
{\sl Black Hole Evolution}, hep-th/9411020 to appear
in Proceedings of the Spring School
on String Theory, Gauge Theory and Quantum Gravity,
Trieste, April 11-19, 1994, R. Dijkgraaf
{\it et al.} eds.}
\lref\dpage{D.N. Page, {\sl Black Hole Information}, hep-th/9305040
\semi  \ajou Int. J. Mod. Phys. &D3 (94) 93.}

\Title{\vbox{\baselineskip12pt
\hbox{
}
}}
{\vbox{
\centerline{}
\centerline{}\centerline {Les Houches Lectures on Black Holes}}
}

\centerline{{\ticp Andrew Strominger}
}

\vskip.1in
\centerline{\sl Department of Physics}
\centerline{\sl University of California}
\centerline{\sl Santa Barbara, CA 93106-9530}


\Date{}

\centerline{\bf Table of Contents}
\line{{\bf 1. Introduction}\dotfill 2}
\line{{\bf 2. Causal Structure and Penrose Diagrams}\dotfill 4}
\line{\indent {\it 2.1 Minkowski Space}\dotfill 4}
\line{\indent {\it 2.2 1+1 Dimensional Minkowski Space}\dotfill 8}
\line{\indent {\it 2.3 Schwarzchild Black Holes}\dotfill 9}
\line{\indent {\it 2.4 Gravitational Collapse and the Vaidya
Spacetimes}\dotfill 11}
\line{\indent {\it 2.5 Event Horizons, Apparent Horizons, and Trapped
Surfaces}\dotfill 13}
\line{{\bf 3. Black Holes in Two Dimensions}\dotfill 15}
\line{\indent{\it 3.1 General Relativity in the $S$-Wave Sector}\dotfill 15}
\line{\indent {\it 3.2 Classical Dilaton Gravity}\dotfill 16}
\line{\indent {\it 3.3 Eternal Black Holes} \dotfill 17}
\line{\indent {\it 3.4 Coupling to Conformal Matter}\dotfill 19}
\line{\indent {\it 3.5 Hawking Radiation and the Trace Anomaly}\dotfill 21}
\line{\indent {\it 3.6 The Quantum State}\dotfill 23}
\line{\indent {\it 3.7 Including the Back-Reaction}\dotfill 25}
\line{\indent {\it 3.8 The Large $N$ Approximation}\dotfill 26}
\line{\indent {\it 3.9 Conformal Invariance and Generalizations of
Dilaton Gravity}\dotfill 30}
\line{\indent {\it 3.10 The Soluble $RST$ Model}\dotfill 32}
\line{{\bf 4.  The Information Puzzle in Four Dimensions}\dotfill 37}
\line{\indent {\it 4.1 Can the Information Come Out Before the
Endpoint?}\dotfill 39}
\line{\indent {\it 4.2 Low-Energy Effective Descriptions of
the Planckian Endpoint}\dotfill 45}
\line{\indent {\it 4.3  Remnants?}\dotfill 46}
\line{\indent {\it 4.4 Information Destruction?}\dotfill 50}
\line{\indent {\it 4.5  The Superposition Principle}\dotfill 53}
\line{\indent {\it  4.6 Energy Conservation}\dotfill 55}
\line{\indent {\it 4.7  The New Rules}\dotfill 59}
\line{\indent {\it 4.8 Superselection Sectors, $\alpha$-parameters, and the
Restoration of Unitarity}\dotfill 61}
\line{{\bf 5. Conclusions and Outlook}\dotfill 64}
\line{{\bf Acknowledgements}\dotfill 66}
\line{{\bf References}\dotfill 67}
\vfill\eject
\newsec{Introduction}
The experimental evidence in favor of quantum mechanics
is fantastically compelling. Evidence in favor of
black holes is incomplete but mounting \bhexp. When this evidence is
combined with
indirect theoretical arguments, it is hard to deny
the existence of black holes. Yet Hawking has
argued \refs{\hawk,\hawktwo} that the two cannot
coexist in the same universe: black holes swallow information
and then disappear without releasing it. This is
inconsistent with quantum mechanical determinism, and as such
the very
foundations of physics are in jeopardy.
Hawking's arguments (reviewed herein) appear
to be very simple and general, and in
particular insensitive to unknown details of the short-distance
laws of physics.

It is the author's belief -- shared by many -- that Hawking has
raised a deep and important puzzle. This puzzle involves the
laws of physics that we believe we already
know and understand. We should therefore either
be able to solve it,
or to understand why it is necessary to go
beyond the known laws of physics.

In the decade following Hawking's seminal work,
a variety of objections\foot{Examples
of such objections are that
the backreaction was ignored, the semiclassical expansion was
inconsistent, gravity was not quantized,
energy was not conserved or that
the derivation secretly depended on short-distance physics.}
to his calculation were
raised,
and the self-consistency of his proposed non-deterministic
laws of physics was questioned.
Attempts to settle the debates were bogged down both by the
nonrenormalizability of quantum gravity and the innate difficulty of
trying to keep track of the information carried by the
many degrees of freedom involved in
the formation/evaporation of a macroscopic black hole. A possible way around
this
impasse was recently found with the
discovery \CGHS\ of two-dimensional models for
black holes (reviewed herein). These models were derived
as the $S$-wave sector of four-dimensional black hole
dynamics, and accordingly contain
black hole formation/evaporation. The information puzzle
thus arises in a simple form, disentangled
from the many technical difficulties encountered in
four dimensions: In two dimensions quantum gravity
is renormalizable and the number of degrees of freedom involved is far less.
The objections raised to Hawking's four-dimensional arguments can
also be raised in two dimensions. Many debates can in this simplified context
be settled by concrete calculation.
The
sharpened understanding gained from two dimensions may then be
applied back to the four-dimensional problem.

A primary goal in the subject of two-dimensional black holes is to construct
a fully self-consistent, quantum mechanical model in which black holes
form and evaporate. For some time it appeared as if the $RST$ model \rst\
(a soluble two-dimensional model reviewed herein) was the starting point in an
expansion of such
a consistent model, and one
in which information is indeed destroyed as argued by Hawking. Influenced
by this, many people -- including the author -- began to believe
that such theories could be fully self-consistent and that
information may indeed be destroyed in the real world.
However, rather recently it was realized \refs{\lpst,\asmg} that the $RST$
model is not self-consistent even at leading order.
The inconsistencies of the $RST$ model arise from a general,
model independent conflict with energy conservation and the superposition
principle.
This conflict was uncovered in thinking about two
dimensions, but it turns out that Hawking's
original prescription for information destruction in four dimensions
suffers from precisely the same inconsistencies \asmg.
Repairing the damage is possible, but surprisingly
leads to a radically different picture \polst, in which
the information is not destroyed, but is
slowly released as the black hole decays back to the vacuum.
This picture of black hole formation/evaporation is reviewed in the last
several subsections.

The outline of these lectures is as follows.  Section 2 contains a review
of classical four-dimensional black holes and their causal structure. Section 3
begins with a discussion of the connection between the $S$-wave sector
of general relativity and two-dimensional black hole models.
In $3.2$-$3.4$  gravitational collapse in classical dilaton gravity  is
reviewed.
In $3.5$-$3.8$ quantum effects are systematically included into the model.
Generalized models and the connection with conformal field theory
are described in $3.9$. The $RST$ model is
described and solved in $3.10$. The section ends with a discussion of
the inconsistency of the $RST$ model. Having introduced the
basic ingredients in the simplified two-dimensional setting, in
Section 4 we turn to four dimensions and a general discussion of
the information puzzle.
In $4.1$ we review the argument that the information cannot come out
before the black hole becomes planckian (in a version which emerged during
dinner conversations at Les Houches). Sections $4.2$-$4.4$ review
remnants (including a new discussion of absorption of pair-production
infinities
by renormalization of Newton's constant)
and Hawking's proposal for information destruction.
Sections $4.5$-$4.6$ review constraints introduced from the superposition
principle and energy conservation. In $4.7$-$4.8$ we review a possible
resolution of the information puzzle which is compatible with these
constraints, and with the insight gained from the two-dimensional models.
We end with conclusions and outlook in Section 5.

This is not meant to be an exhaustive review of all recent developments
in quantum black hole physics. The content basically follows
lectures/discussions at Les Houches, although the lectures on
quantum field theory in curved space have been omitted (see the excellent
text \brdv), and sections 3.6, 3.9, 3.10 and 4.3 have
been added for completeness.
Perhaps the most serious omission is a
discussion of the fascinating and mysterious generalized second law \bek.
A recent discussion  of the two-dimensional view on this can be found in
\fpst. Other recent general reviews -- representing a rich variety of
viewpoints -- include
\refs{\dpage, \banrev,\laref,\sbgtr,\snow,\kzr \brek}. Parts of these lectures
were adapted -- with varying amounts of editing and updating -- from
my previous writings\refs{\qabh,\cpt,\asmg}. I am particularly
grateful to Jeff Harvey for permission to adapt sections of \qabh.

\newsec{Causal Structure and Penrose Diagrams}

The most basic question one can ask about two spacetime points $x$ and
$x^\prime$ concerns their causal relation.  Is $x^\prime$ in, on or
outside of the past or future light cone of $x$? {\it Causal structure}
becomes particularly important and subtle in the context of black holes.
{\it Penrose diagrams} are an indispensable aid in understanding the
causal structure of a spacetime.  We illustrate them here with several
examples of increasing complexity.  More details can be found in
\refs{\pnrs,\wald}.

\subsec{Minkowski Space}

The line element for Minkoswki space
in spherical coordinates $(t, r, \theta , \phi)$
is given by
\eqn\oneone{
ds^2 = (-dt^2 + dr^2) + r^2 (d \theta^2 + \sin^2 \theta d \phi^2 ) \equiv
(-dt^2 + dr^2 ) + r^2 d \Omega_{II}^2 .}
At each point $(r,t)$ with $ - \infty < t < \infty , ~ 0 < r < \infty$
there is an $S^2$ of area $4 \pi r^2$.
In what follows we focus on the $(r,t)$ plane and suppress the presence of
the two\--spheres.
It is often useful to introduce light\--cone coordinates
\eqn\onetwo{
\eqalign{
u &= t - r, \cr
v &= t + r, \cr}}
so that $-dt^2 + dr^2 = - du dv$.

\ifig\fone{Relation between $(r+t)$ coordinates and light-cone
coordinates $(u, v)$ and various asymptotic regions of Minkowski space.}
{\epsfysize=4.5in \epsfbox{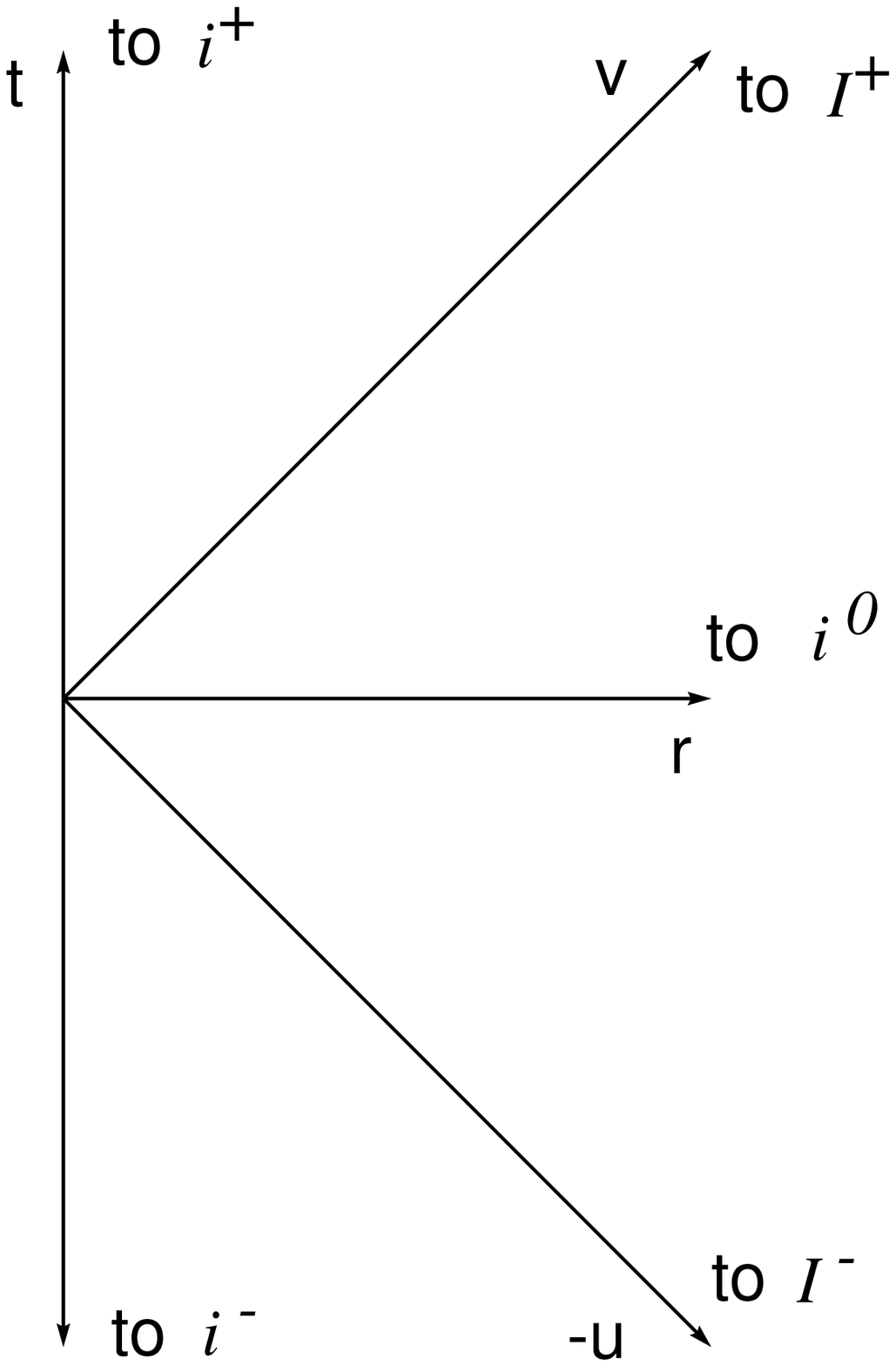}}

The relation between $(r, t)$ and $(u, v)$ and
various asymptotic regions which will play a role in the
following discussion are indicated in \fone.
These are:

\bigskip

$\qquad i^+ = \{ t \air + \infty$ at fixed $r \} =$ future timelike infinity,

\medskip

$\qquad i^- = \{ t \air - \infty$ at fixed $r \} = $ past timelike infinity,

\medskip
$\qquad i^0 = \{ r \air \infty$ at fixed $t \} =$ spacelike infinity,

\medskip
$\qquad \ci^+ = \{ v \air \infty$ at fixed $ u \} =$ future null infinity,

\medskip

$\qquad \ci^- = \{ u \air - \infty$ at fixed $v \} =$ past null infinity.

Future and past null infinity are useful concepts when dealing with radiation.
For example, to measure the mass of an object one needs to know the
deviation of the metric from flat space at large distances.
If the object emits a pulse of radiation at time $t$ and we want to know the
resulting change of mass then, at radius $r$, we must wait a time $t \geq
r$ until the radiation is past to measure the new metric.
As $r \air \infty$, we end up making the measurement at $\ci^+$.

\ifig\ftwo{Penrose diagram for Minkowski space (shaded region). Each point
represents a two-sphere at fixed radius and time. The origin
corresponds to the vertical boundary on the left.}
{\epsfysize=4.5in\epsfbox{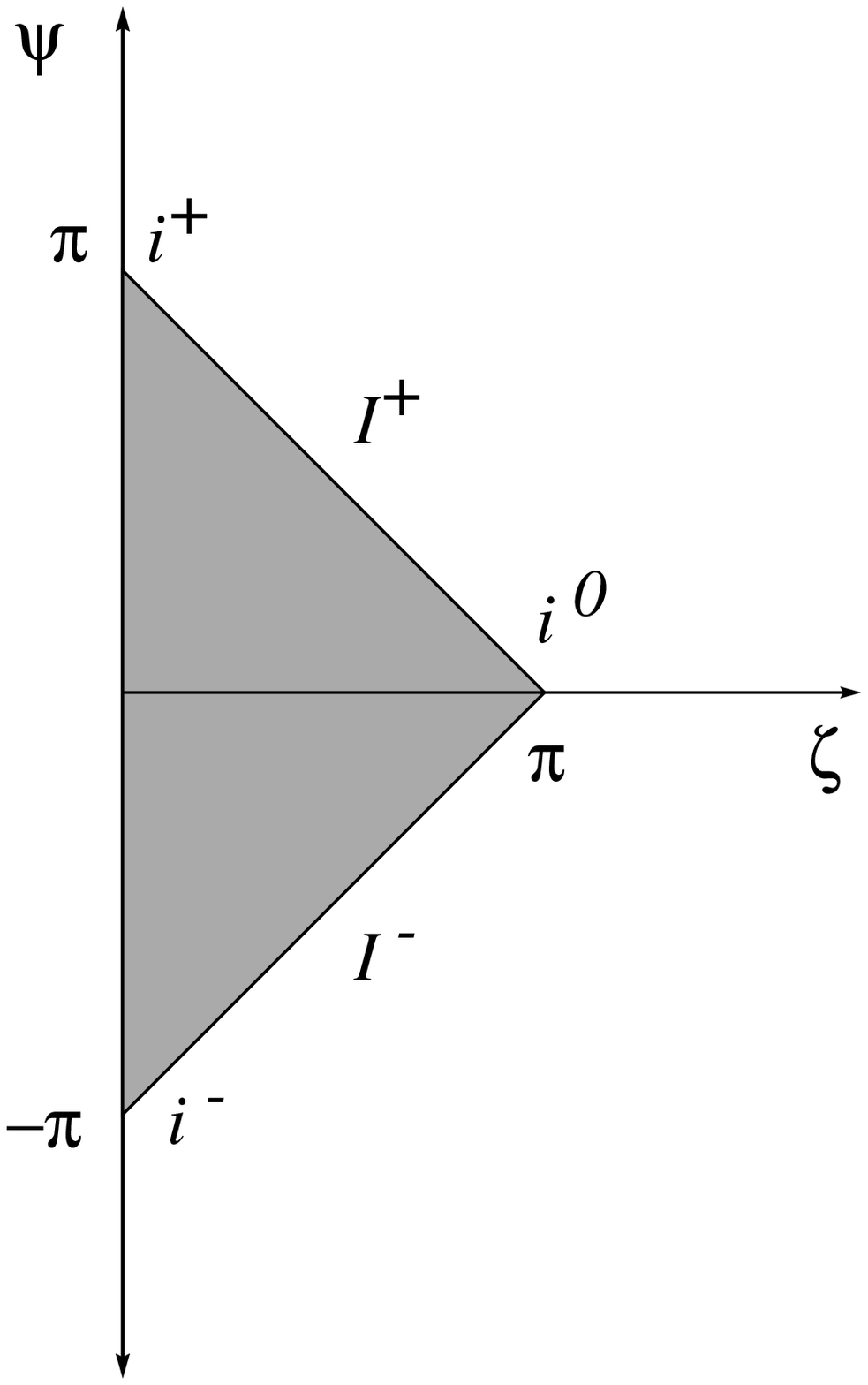}}

However, it is awkward to study ${\cal I}^+$ in $(u,v)$ coordinates because it
is at an infinite value of $v$.
We therefore introduce coordinates $( \psi , \zeta )$ with
\eqn\onethree{\eqalign{
v &= t+r = \tan \frac{1}{2} ( \psi + \zeta ), \cr
u &= t-r = \tan \frac{1}{2} ( \psi - \zeta ), \cr}}
so that
\eqn\onefour{
ds^2 = \Omega^2 ( \psi, \zeta ) ( -d \psi^2 + d \zeta^2 ) + r^2 ( \psi, \zeta )
d \Omega_{II}^2,
}
with
\eqn\onefive{
\Omega^{-2} ( \psi, \zeta ) = 4 \cos^2 \frac{1}{2} ( \psi + \zeta )
\cos^2 \frac{1}{2} ( \psi - \zeta ) .
}
The new coordinates
$(\psi,\zeta)$ range over the half-diamond $\zeta \pm \psi < \pi,~~\zeta >0$.
We then introduce an unphysical metric $\bar{g}_{\mu \nu}$ which is
conformal to the actual metric $g_{\mu \nu}$
\eqn\onesix{
\bar{g}_{\mu \nu} = \Omega^{-2} g_{\mu \nu} .
}
Although distances measured with the $\bar g$ metric differ
(by a possibly infinite factor) from those measured with the
$g$ metric, {\it the causal relation of any two points is the
same in both metrics}. Thus the causal structure of the $g$-spacetime is
equivalent to that of the $\bar g$-spacetime.
The unphysical metric $\bar{g}$ is well behaved at the values
of $( \psi, \zeta )$ which correspond to the asymptotic regions of $g$ as
shown in \ftwo .

The Penrose diagram of \ftwo\ brings the previous asymptotic regions
into finite points.
Furthermore, even though $\bar{g}$ is not the physical metric, statements
about the asymptotic behavior of fields in the spacetime with metric $g$
can be translated into simple statements about the behavior of fields at
the finite points corresponding to $i^\pm , i^0 , \ci^\pm$ in the
spacetime with metric $\bar{g}$.
This type of discussion can also be applied to solutions such as the
Schwarzschild metric which have an appropriate notion of asymptotic
flatness.
See \wald\ for further details.

The basic feature of a Penrose diagram is that null geodesics are always
represented by $45^o$ lines. Thus it is easy to discern if two
points are in causal contact, which makes the diagrams very useful.
For example a glance at \ftwo\ reveals that all of Minkowski space is in the
causal past of an observer at $i^+$. The price one pays for this is
that distances are not accurately portrayed: two points finitely separated on a
Penrose diagram may or may not be an infinite geodesic distance apart.

\ifig\fthree{Penrose diagram for $1+1$ dimensional Minkowski space (shaded
region).}{\epsfysize=4.5in\epsfbox{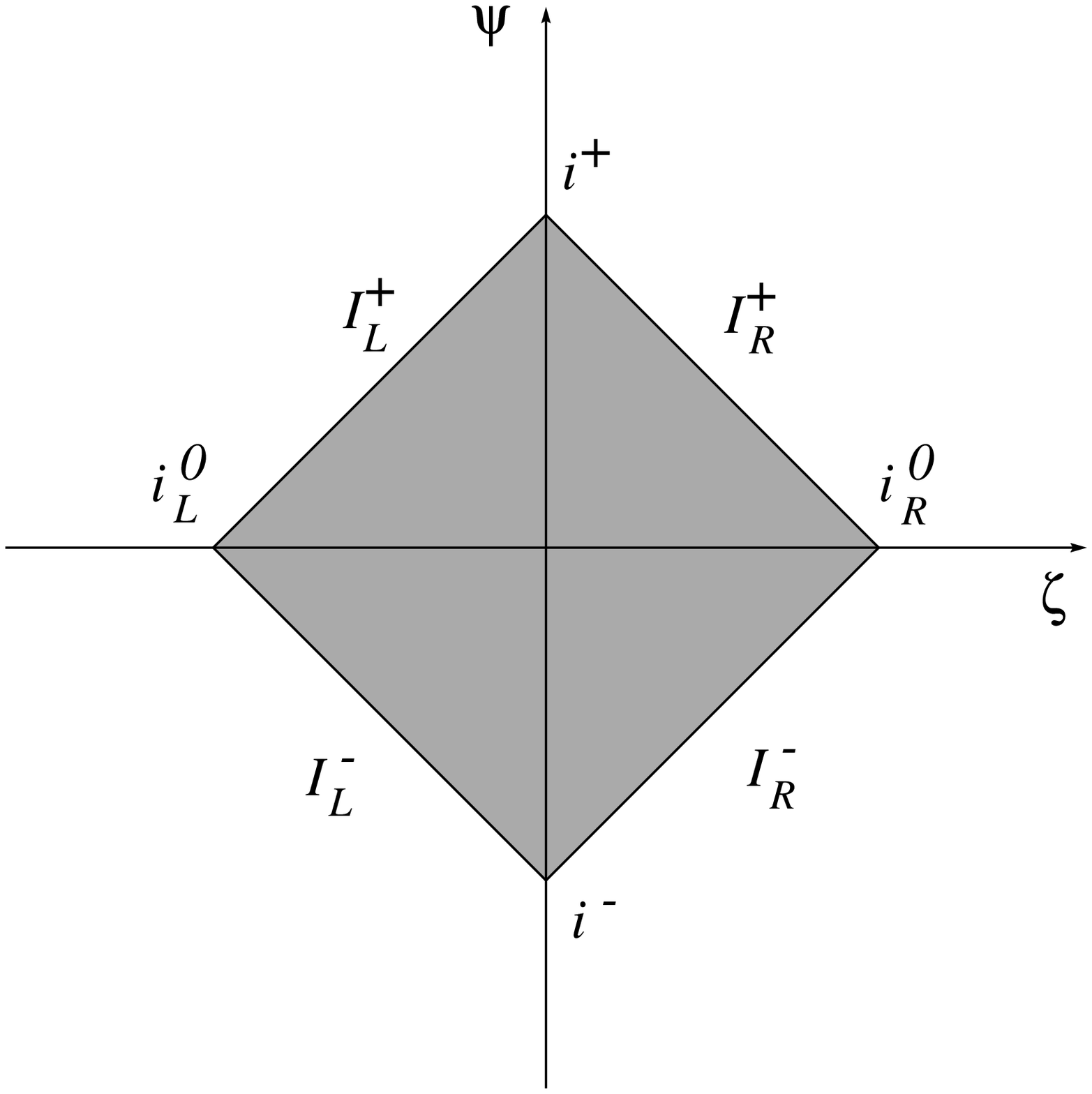}}

\subsec{$1+1$ Dimensional Minkowski Space}

We have the line element
\eqn\oneseven{
ds^2 = - dt^2 + dx^2 = - dx^+ dx^-,
}
with $x^\pm = t \pm x$.
Letting
\eqn\oneeight{
x^\pm = \tan \frac{1}{2} ( \psi \pm \zeta ),
}
where now, since  $- \infty < x <
\infty$, $(\zeta, \psi)$ range over the full diamond $|\zeta \pm \psi|<\pi$.
It follows as in the previous discussion that
the Penrose diagram consists of two copies of \ftwo\ as
shown in \fthree .
There are now two spacelike infinities, $i_{R,L}^0$, corresponding to $x
\air \pm \infty$, and two past and two future null infinities, $\ci_R^\pm
, \ci_L^\pm$ with for example $\ci_R^+$ being where right\--moving light
rays go and $\ci_L^+$ where left\--moving light rays go.

\subsec{Schwarzschild Black Holes}

The Schwarzschild black hole with line element
\eqn\onenine{
ds^2 = - (1 - \frac{2M}{r}) dt^2 + \frac{dr^2}{(1 - \frac{2M}{r})} + r^2
d \Omega_{II}^2
}
is probably the most familiar non-trivial solution to the vacuum Einstein
equations
$R_{\mu \nu} = 0$.
As is well known, at the origin $r=0$ there is a curvature singularity as
may be verified by calculation of the invariant $R_{\mu \nu \lambda \psi}
R^{\mu \nu \lambda \psi}$.
The singularity in the metric at $r = 2M$ is not a curvature singularity
but instead represents a breakdown of this particular coordinate system.

\ifig\ffour{Maximal analytic extension of the Schwarzschild black hole in
null Kruskal coordinates.}{\hskip.5in\epsfysize=3.6in\epsfbox{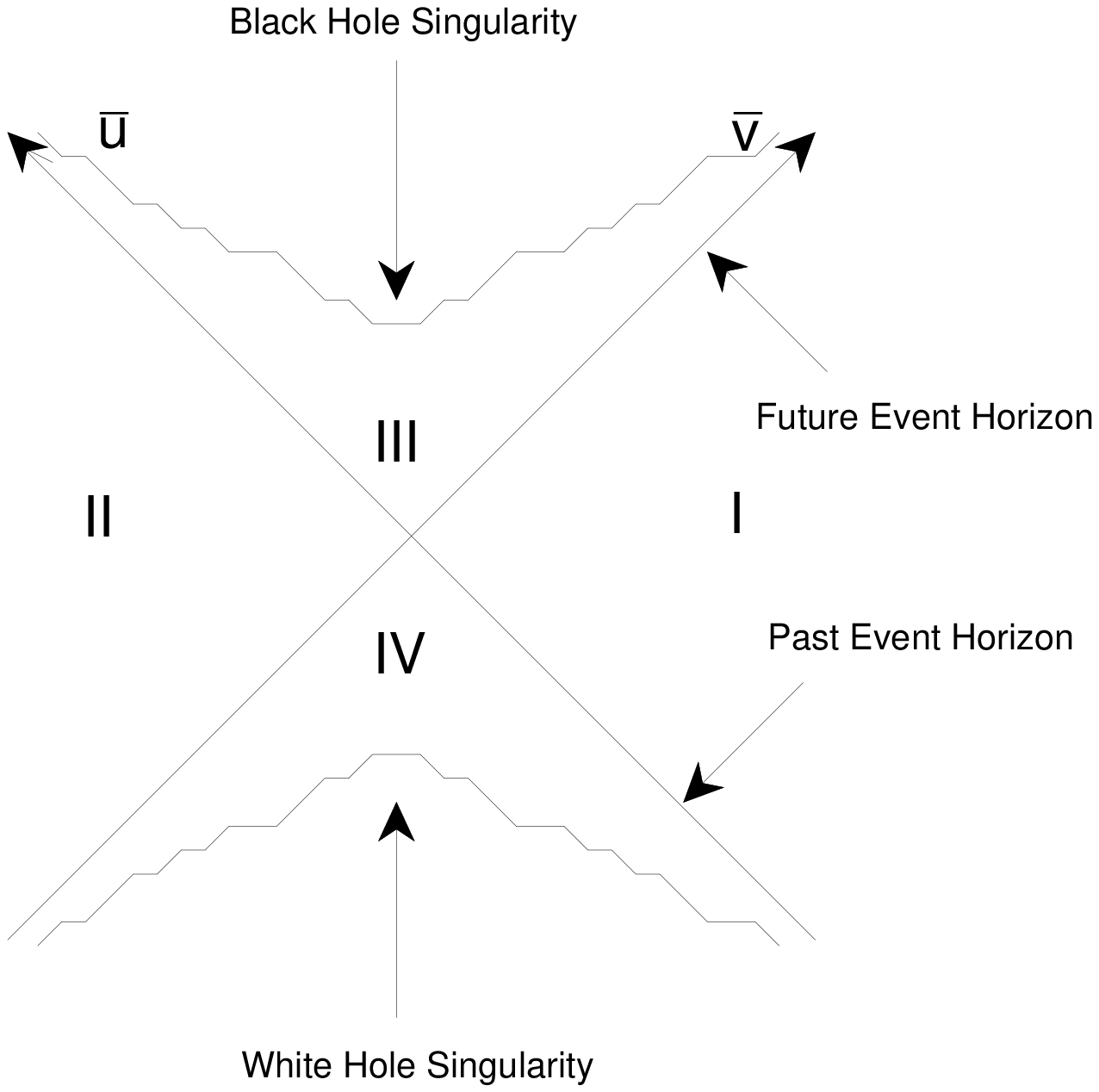}}

The most convenient method to study the behavior near $r = 2M$ is to
introduce coordinates along ingoing and outgoing radial null geodesics.
We thus introduce the tortoise coordinate
\eqn\oneten{
r^* = r + 2M \ln (\frac{r}{2M} - 1 )
,}
with $dr = (1 - 2M/r)dr^*$ and
\eqn\oneeleven{
ds^2 = (1 - \frac{2M}{r}) ( -dt^2 + d{r^*}^2) + r^2 (r^*) d \Omega_{II}^2 .
}
It is clear from \oneeleven\ that null geodesics correspond to $t = \pm r^*$.
Also note that $r = 2M$ is at $r^* = - \infty$.
The appropriate null coordinates are
\eqn\onetwelve{\eqalign{
u &= t - r^*, \cr
v &= t + r^* . \cr
}}
The next step is to introduce the null Kruskal coordinates
\eqn\onethirteen{\eqalign{
\bar{u} &= - 4Me^{-u/4M}, \cr
\bar{v} &= 4Me^{v/4M} . \cr
}}
The region $r \geq 2M$ or $- \infty < r^* < \infty$ maps onto the region
$- \infty < \bar{u} < 0$, $0 < \bar{v} < \infty$.
But now inspection of the metric shows that
\eqn\onefourteen{
ds^2 = -\frac{2M}{r} e^{-r/2M} d \bar{u} d \bar{v} + r^2 d \Omega_{II}^2
,}
where $ r (\bar{u}, \bar{v})$ is defined implicitly by \oneten\ --
\onetwelve\ and
it is clear that the metric components are non\--singular at $r = 2M$.
We can thus analytically continue the solution to the whole region $-
\infty < \bar{u} , \bar{v} < \infty$.
The resulting Kruskal diagram of the extension of the Schwarzschild black
hole is shown in \ffour.
\ifig\ffive{Penrose diagram of the analytic extension of
the Schwarzschild black hole.}
{\hskip.25in\epsfysize=2.5in \epsfbox{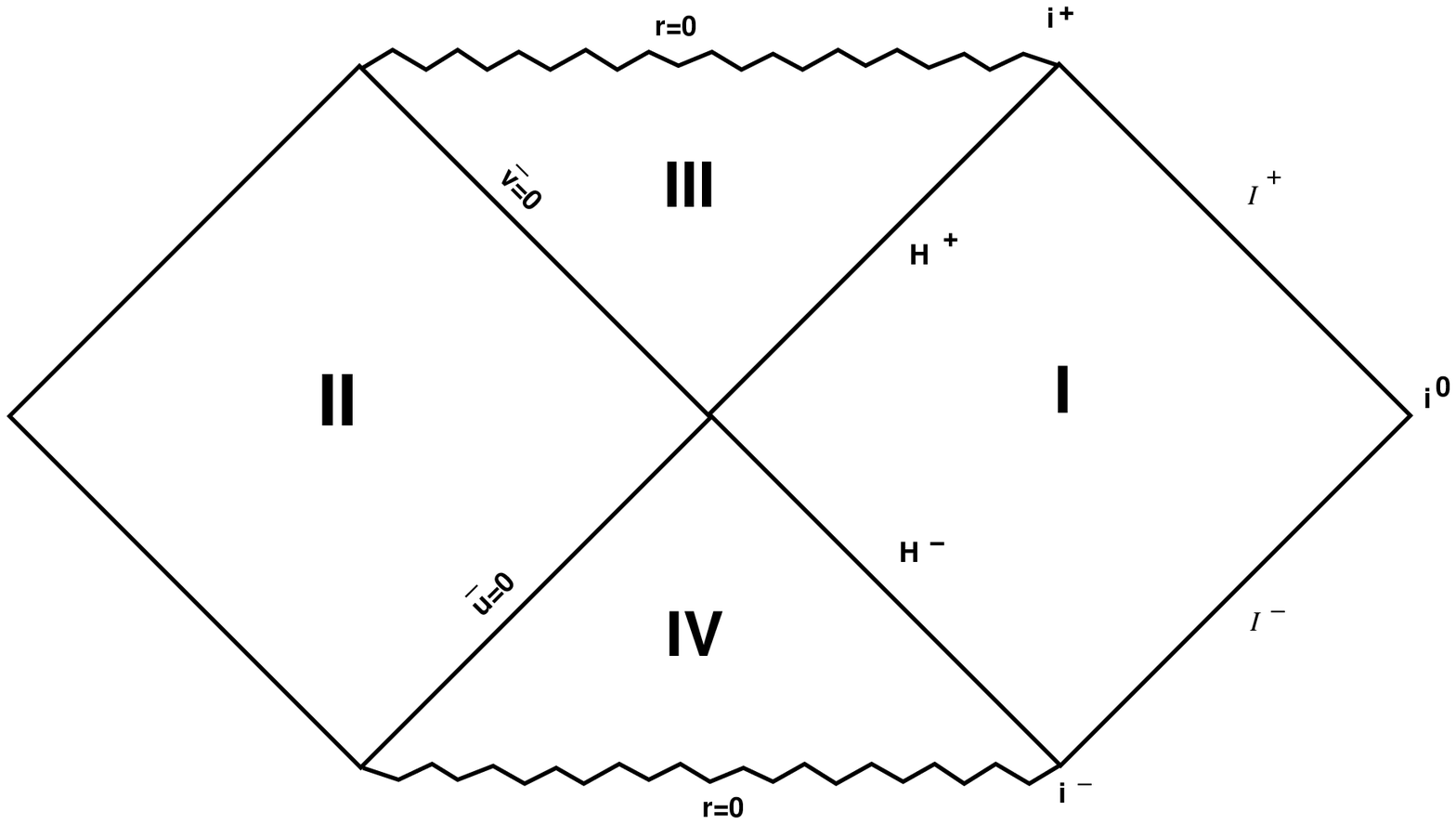}}
A procedure similar to that described earlier for Minkowski space allows
one to bring the asymptotic regions of \ffour\ into finite points in terms
of an unphysical metric $\bar{g}$.
The resulting Penrose diagram for the Schwarzschild black hole is shown
in \ffive .
In this extension of the Schwarzschild metric there are two
asymptotically flat regions denoted I, II in \ffour\ and \ffive .
Also, in addition to the black\--hole singularity (where $r(\bar u, \bar v)$
vanishes)
which reaches $i^+$ in
the infinite future, there is a white\--hole singularity which emerges
from $i^-$ in the infinite past.

\subsec{Gravitational Collapse and the Vaidya Spacetimes}
\ifig\fstar{Penrose diagram for a black hole
formed by spherically symmetric collapse of a null shock wave.
The solid line is the apparent horizon, which bounds the shaded
region of trapped
surfaces or apparent black hole. The dashed line is the event horizon,
which coincides with the apparent horizon after the collapse is completed.}
{\hskip .5in \epsfysize= 4.8in \epsfbox{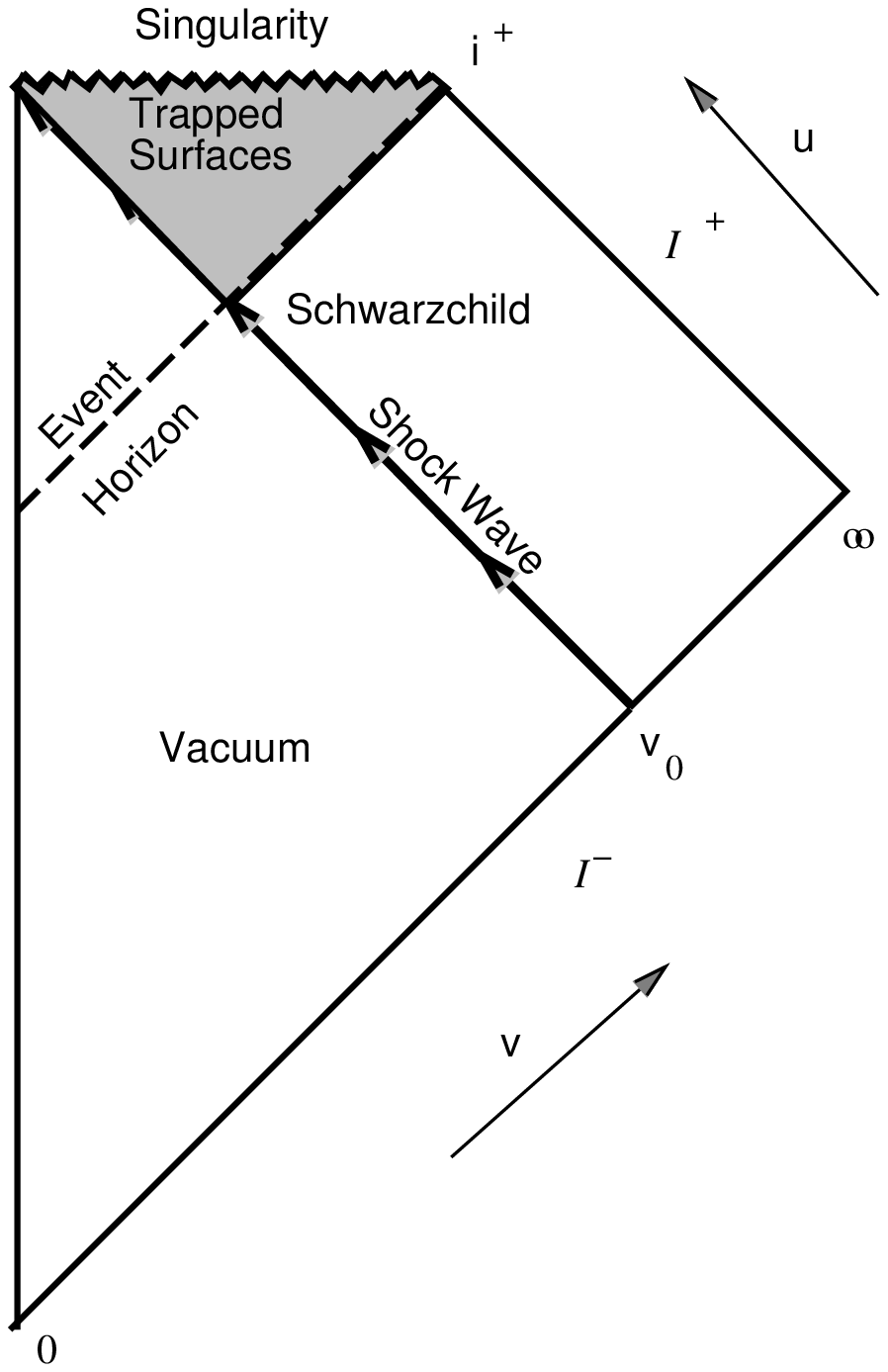}}

It is reasonable to ask how much of this structure is relevant for
classical black holes formed by the collapse of infalling matter.
Only region I and part of region III will exist
for such a physical black hole. This can be seen analytically in the
Vaidya spacetimes. These represent a black hole formed by collapse of
null matter whose stress tensor takes the form
\eqn\vstr{T_{vv} = \frac{{\cal E}(v)}{4\pi r^2},}
with all other components equal to zero.  The metric is simplest in
infalling $(r,v)$ coordinates:
\eqn\umet{ds^2 = -\left(1-\frac{2m(v)}{r}\right) dv^2 + 2dr dv + r^2
d \Omega_{II}^2 }
where
\eqn\oneseventeen{m(v) = \int\limits^v_{-\infty} dv^\prime {\cal
E}(v^\prime)}
is the total mass inside $v$. Consider the special case that the matter
is a shock wave, for which
$T_{vv}$
is nonzero only along $v_0$:
\eqn\onse{T_{vv} = \frac{M\delta(v-v_0)}{4\pi r^2}.}
In this case
\eqn\oneeighteen{\eqalign{m & =0 \quad v<v_0,\cr
m &= M \quad v>v_0\ .}}
Then the region below $v=v_0$ is just flat space, and the corresponding
portion of the Penrose diagram is the region below a null line in
the Minkowski diagram of \ftwo. Similarly, the region above $v=v_0$ is
identically Schwarzschild, and is represented by a region above an
ingoing null line $v=v_0$ in \ffive.  This region
does {\it not} include regions
II or IV. The complete Penrose diagram obtained by patching together the
two regions is then as illustrated in \fstar. This geometry is perhaps
the simplest explicit example of gravitational collapse. A two-dimensional
version will be discussed at length in section 3.

\subsec{Event Horizons, Apparent Horizons and Trapped Surfaces}

In this subsection we will describe the notions of event horizons,
apparent
horizons, and trapped surfaces. We will not give precise definitions for
general surfaces or general spacetimes,
as there
are many subtleties involved. Rather we will attempt to give a flavor of the
ideas in the highly simplified context of spherically symmetric spacetime
geometries and symmetric surfaces.  The statements made in this
section refer only to such surfaces and geometries, although many of them can
be generalized. The reader interested in
precise
statements instead of the general flavor should refer to \wald\ and
\pnrs.

A {\it future event horizon} is the null surface from behind which it is
impossible to escape to ${\cal I}^+$ without exceeding the speed of light. A
{\it past
event horizon} is the time reverse of this: a surface which it is
impossible to get behind starting from ${\cal I}^-$. Schwarzchild contains both
a past and future event horizon as indicated in \ffour\ and \ffive,
while the spacetime representing a black hole formed by
gravitational collapse contains only a future event horizon as indicated
in \fstar.

The interior of a black hole generally contains a region of trapped
surfaces. To illustrate this notion, consider a two-sphere in flat
Minkowski space. There are two families of null geodesics which
emanate from the two-sphere, those that go out and those that go in.
The former diverge, while the latter converge. A {\it trapped surface}
is one for which both families of null geodesics are everywhere
converging, due to gravitational forces. It is easy to check that two-spheres
of constant radius behind
the future horizon in Schwarzchild are trapped. Outgoing null geodesics from
the two-sphere exactly at ${r} = {2M}$ of course generate the
horizon itself, whose area is constant for Schwarzchild. This two-sphere
is therefore marginally trapped.

An {\it apparent horizon} is the outer boundary of a region of trapped
surfaces. We will also find it convenient to refer to a region of
trapped surfaces as an {\it apparent black hole}.

The notions of an apparent horizon and an event horizon are quite different,
although the two are sometimes confused as they happen to coincide for
Schwarzchild. An event horizon is a global concept, and the entire
spacetime must be known before its existence or location can be
determined. The location of an apparent horizon, in contrast, can be
determined from the initial data on a spacelike slice.

To illustrate this, consider a black hole geometry with an apparent
horizon at time ${t_0}$. Throwing matter into the black hole at
a time ${t>}{t_0}$ (relative to any smooth time slicing which goes through the
black
hole) will have no effect on the area or location of the apparent horizon
at time ${t_0}$ (although it will increase its area for later times). However,
the infalling matter does cause the event horizon at the earlier time
${t_0}$ to move out to larger radius. The apparent and event horizons for
a black hole formed by collapsing radiation are illustrated in \fstar .

In classical general relativity, the apparent horizon is typically a null
or spacelike surface which lies inside or coincides with
the event horizon (assuming cosmic censorship) \pnrs.
This is not true when the effects of Hawking radiation are taken into account,
in which case -- as will be illustrated in section three -- the apparent
horizon  can shrink, become timelike and
move outside the event horizon.

It is important to stress that there is no evidence for the existence of
black hole event horizons (as opposed to apparent horizons) in the real
world. In order to answer this question one must follow the apparent
black hole all the way to the endpoint of Hawking evaporation.

\newsec{Black Holes in Two Dimensions}
\subsec{General Relativity in the ${\bf S}$-Wave Sector}

The time-dependent dynamics of classical --- let alone quantum --- black
holes are extremely complex.  Great simplifications can be achieved by
restricting the metric and matter fields to have spherical symmetry.  We
shall see that implementing
 this restriction does not throw out the baby with the
bath water --- virtually all of the interesting and puzzling features of
black holes are present in the $S$-wave sector.

The most general spherically symmetric metric can be expressed in the
form
\eqn\ssmt{ds^2 = g_{\mu\nu} dx^\mu dx^\nu + \frac{1}{\lambda^2}
\, e^{-2\phi} d^2
\Omega}
where
$\mu, \nu=0,1$,\  $(x^0,x^1) \sim (t,r)$,
$\phi$ and $g$ are functions of $x$ and the dimensionful constant
$\lambda$ is introduced so that the field $\phi$ is dimensionless.
The vacuum Einstein
equations become
\eqn\twotwo{\eqalign{G_{\mu\nu} & = 2\nabla_\mu \nabla_\nu \phi -
2\nabla_\mu \phi \nabla_\nu \phi + 3 g_{\mu\nu} (\nabla\phi)^2\cr
& - 2 g_{\mu\nu}\ \sq \phi - \lambda^2 g_{\mu\nu} e^{2\phi}}}
\eqn\nstn{\eqalign{^{(4)}G_{\theta\theta} & = \sin^2\theta
\ ^{(4)}G_{\phi\phi}\cr
& = \frac{1}{\lambda^2}\ e^{-2\phi} \left[(\nabla \phi )^2 - \sq \phi -
\half R\right]}}
where all curvatures and connections are constructed from the
two-dimensional metric $g_{\mu\nu}$ unless marked with a superscript (4).
(We
apologize for using $\phi$ to denote both a field and an angle -- the
meaning should be clear from the context.) These equations follow from
the effective
two-dimensional action
\eqn\zactn{
{S} = {{1}\over{2\pi}} {\int} {d^2}{x}{\sqrt{-g}} e^{-2 \phi}
\left[{R} + {2} ({\nabla}
{\phi})^2 + {2}{\lambda^2}e^{2 \phi}\right]\ ,}
\noindent where the cosmological constant $2{\lambda^2}$
is a relic of the
components of the scalar curvature
tangent to the two-sphere. \zactn\ may also be directly derived by
substitution of the ansatze \ssmt\ in to the Einstein-Hilbert action
(in units with Newton's constant $G_N=\pi / 2 \lambda^{2}$).

Before leaving four dimensions there
are several useful entries in the dictionary relating four- and
two-dimensional quantities we would like to explain. In a spherically symmetric
four-dimensional
spacetime of the form \ssmt,
the area of the two-spheres is given by the function $\frac{4\pi}{\lambda^2}
e^{-2\phi}({\sigma^+}, {\sigma^-})$ where ${\sigma^+} = t\pm r$ are null
coordinates. The two-sphere at ${\sigma^+}, {\sigma^-}$, will be
trapped if is decreasing in both null directions,
i.e. ${\partial_{\pm}} e^{-2\phi}<{0}$.
Therefore  a {\it trapped
point} in the two-dimensional theory is a point at which
\eqn\tpnt{{\partial_{\pm}}{\phi}>{0}.}
\noindent An {\it apparent horizon} is then the outer boundary of such
a region at which ${\partial _+}{\phi} = {0}$ \RST\ (since asymptotically
${\partial_+}{\phi}<{0}$ while ${\partial_-}{\phi}>{0}$.) We will
also use the phrase {\it apparent black hole} to refer to
a region of trapped points. This is distinct from a real black hole,
which is a region from which it is impossible to escape to
${\cal I}_R^+$.

\subsec{Classical  Dilaton Gravity}
In the following we will be discussing a $1+1$ dimensional theory of gravity
coupled to a dilaton field $\phi$ with action
\eqn\twseven{S_D= {1 \over 2 \pi} \int d^2 x \sqrt{-g} e^{-2 \phi} \left[
  R + 4 (\nabla \phi)^2 + 4 \lambda^2 \right] . }
\twseven\ differs from \zactn\  in the numerical coefficient of the
dilaton kinetic energy term and the $\phi$-dependence of the
potential. These differences do not qualitatively change the
physics.  There are still black holes and, as shall be seen shortly,
Hawking evaporation.  However, the theory described by \twseven\ is
dramatically simpler to study: the classical solutions  can be presented
in explicit closed form. This is our main reason for studying \twseven
\ rather
than \zactn.

The action \twseven\ arises in a low-energy effective description of
certain dilatonic black holes in string theory.  This connection was our
original motivation for studying \twseven\ \refs{\CGHS,\dxbh} and is
described at length in the review \refs{\qabh}.
The model also arises in the related context of
 two-dimensional non-critical string theory and as such
its black hole solutions were first discovered in \WittTwod\ and \Mandal.
Previous work on two-dimensional black holes which is closely related can be
found
in \Mann, and on models of two-dimensional gravity with scalars in
\refs{\bcmn, \rom,\tei, \cham}.

The classical equations of motion which follow from \twseven\ are
\eqn\tweight{
2 e^{-2 \phi} \left[ \nabla_\mu \nabla_\nu \phi + g_{\mu \nu}
( (\nabla \phi)^2 - \nabla^2 \phi - \lambda^2 ) \right] =0,}
\eqn\twnine{ e^{-2 \phi} \left[
R + 4 \lambda^2 + 4 \nabla^2 \phi - 4 (\nabla \phi)^2  \right] = 0,}
where the first equation results from variation of the metric
and the second is the dilaton equation of motion.  We first
note that there is a solution (often called the linear dilaton
vacuum) characterized by
\eqn\thirty{
R = \nabla^2 \phi = 0 , \qquad (\nabla \phi)^2 = \lambda^2 .}
We shall refer to this simply as the vacuum.
We can introduce coordinates $(\sigma , \tau)$ so that
\eqn\thone{
g_{\mu \nu} = \eta_{\mu \nu} , \qquad \phi = -\lambda \sigma,}
in the vacuum. Note that the vacuum is not translationally invariant.
The ``origin'', where $e^{-2\phi}=0$, is at $\sigma=-\infty$, while the
asymptotic region with infinite-area two spheres, is at
$\sigma=+\infty$.
 The natural
coupling constant in this theory is $g_s = e^{\phi}$
which depends on $\sigma$ and is inversely proportional to
the square root of the area, $e^{-2\phi}$. Thus
the vacuum can be divided into a strong coupling region ( $\sigma \air -
\infty$)
and a weak coupling asymptotic  region ($\sigma \air + \infty$). It is
sometimes
useful to think of the strength of the coupling as providing a coordinate
invariant notion of one's location in this one-dimensional world. The
vacuum Penrose diagram is illustrated in \fthree.
\subsec{Eternal Black Holes}
To introduce the black hole solution of this theory it is useful
to introduce light-cone coordinates (the relation of these coordinates
to the previous ones will be discussed momentarily)
\eqn\thtwo{
x^\pm = x^0 \pm x^1 ,}
and to choose conformal gauge $g_{\mu \nu} = e^{2\rho}\eta_{\mu \nu}$,  or
in light-cone coordinates
\eqn\ththree{
g_{+-} = -{1 \over 2} e^{2 \rho} , \qquad g_{++} = g_{--} = 0.}
We then have $R=8 e^{-2 \rho} \partial_+ \partial_- \rho$ and the
equations of motion become
\eqn\thfour{\eqalign{
\phi &: \qquad  e^{-2(\phi+\rho)} \left[ -4 \p+ \pmi \phi +
 4 \p+ \phi \pmi \phi + 2
        \p+ \pmi \rho + \lambda^2 e^{2 \rho} \right] = 0, \cr
\rho &: \qquad  e^{-2 \phi} \left[ 2 \p+ \pmi \phi - 4 \p+ \phi \pmi \phi
                     - \lambda^2 e^{2 \rho} \right] =0. \cr }}
Note that these two equations imply
\eqn\thfive{
\p+ \pmi (\rho -\phi) = 0, }
so that $(\rho - \phi)$ is a free field. Since we have gauge fixed
$g_{++}$ and $g_{--}$ to zero we must also impose their
equations of motion as constraints. This gives
\eqn\thsix{\eqalign{
e^{-2 \phi} ( 4 \p+ \rho \p+ \phi - 2 {\p+}^2 \phi ) &= 0, \cr
e^{-2 \phi}( 4 \pmi \rho \pmi \phi - 2 {\pmi}^2 \phi ) &=0 . \cr }}
Now \thfive\ implies that $\rho$ and $\phi$ are equal up to the sum of
a function purely of $x^+$, $f_+(x^+)$ and a function purely of $x^-$,
$f_-(x^-)$. But
a coordinate transformation $ x^\pm \air \tilde x^\pm (x^\pm)$
preserves the conformal gauge \ththree\ and can be used to set $f_\pm=0$.
Thus we can choose $\rho=\phi$ in analyzing the equations of motion.
With this choice the remaining equations and constraints reduce to
\eqn\thseven{\eqalign{
& \pmi \p+ (e^{-2 \rho}) = -\lambda^2, \cr
& {\p+}^2 (e^{-2 \rho}) = {\pmi}^2 (e^{-2 \rho}) = 0, }}
which has the general solution (up to constant shifts of $x^\pm$)
\eqn\theight{
e^{-2 \phi} = e^{-2 \rho} = {M \over \lambda} - \lambda^2 x^+ x^-, }
where $M$ is a free parameter which will turn out to be the
mass of the black hole.

Calculating the curvature we find
\eqn\thnine{
R = 8e^{-2 \rho} \p+ \pmi \rho = {4 M \lambda \over M/\lambda - \lambda^2
                                  x^+ x^-} ,}
which is divergent at $x^+ x^- = M/\lambda^3$. This solution  has the same
qualitative features
as the $(r,t)$ plane of the Schwarzschild black hole. The Penrose
diagram is in fact the same as that in \ffive\ with $(\bar u , \bar v)$
replaced by $(x^- , x^+)$.

Region I in fig. 5 should asymptotically approach the flat space
vacuum. To see this we can introduce coordinates
\eqn\forty{\eqalign{
\lambda x^+ &= e^{\lambda \sigma^+} ,\cr
\lambda x^- &= - e^{-\lambda \sigma^-}. \cr }}
Note that the range $- \infty < \sigma^+ , \sigma^- < +\infty$ covers
only region I of \ffive. It is also important ro remember that in these
coordinates $\rho$ will no longer equal $\phi$ since $\phi$ transforms as
a scalar under coordinate transformation while $\rho$ does not. In these
coordinates we find that
as $\sigma = (\sigma^+ - \sigma^-)/2 \air \infty$
\eqn\foone{\eqalign{
\phi & \air -\lambda \sigma - {M \over 2 \lambda} e^{-2 \lambda \sigma}, \cr
\rho & \air 0 - {M \over 2 \lambda} e^{-2 \lambda \sigma}, \cr }}
and the solution approaches the vacuum up to exponentially small corrections.
It is also important to note that $g_s = e^{\phi} \air 0$ as
$\sigma \air \infty$ and that at the horizon $x^- =0$, $g_s =
\sqrt{\lambda/M}$. Thus we are in weak coupling throughout region
I for sufficiently massive black holes ($ M >> \lambda$).
\subsec{Coupling to Conformal Matter}
So far all we have constructed is an ``eternal'' black hole
solution. To determine whether such solutions form from
non-singular initial conditions and to study Hawking radiation
we must couple in some dynamical matter degrees of freedom.
%
%
To study this process in our $1+1$ dimensional model we modify
\twseven\ by adding a matter term of the form
\eqn\fotwo{
S_M = -{1 \over 4 \pi} \sum_{i=1}^N \int d^2 x \sqrt{-g} (\nabla f_i)^2, }
where the $f_i$ are a set of $N$ massless matter fields
For the moment we take $N=1$ and will consider general $N$
when we discuss Hawking radiation and back reaction.
In conformal gauge the $f$ equation of motion is simply
\eqn\fothree{
\p+ \pmi f =0. }

\ifig\ften{Penrose diagram for formation of a dilaton black hole
by an $f$ shock-wave.}{\hskip .5in \epsfysize=3.5in \epsfbox{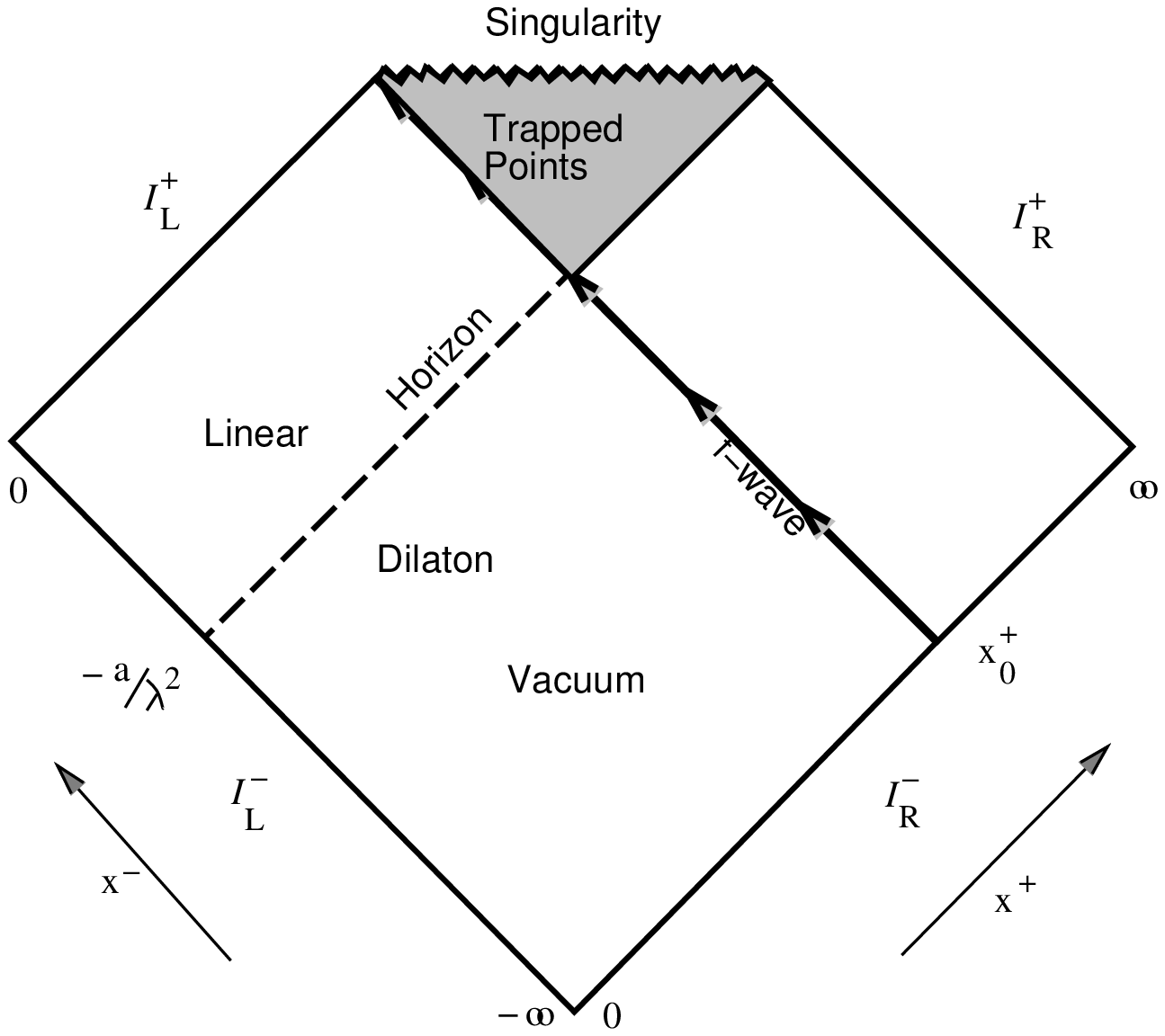}}

Let us consider sending in a pulse of energy from the
right. Although we could consider taking $f$ to be
some function of $x^+$ with finite width \refs{\CGHS}, to simplify
the calculation we take the $f$ pulse to be a
shock-wave traveling in the $x^-$ direction with magnitude $a$
described by the stress tensor
\eqn\fofour
{\half \partial_+ f\partial_+ f= a\delta (x^+ - x^+_0)\ .}
The only modification in the equations of motion and
constraints due to the matter fields in this case
is in the $g_{++}$ constraint which becomes
\eqn\fofive{
e^{-2 \phi} ( 4 \p+ \rho \p+ \phi - 2 {\p+}^2 \phi) = -{ 1 \over 2}
             \p+ f \p+ f.}
For $x^+ < x_0^+$ we assume we are in the vacuum, while for $x^+ > x_0^+$ we
know that the solution must be of the form
\theight. Matching the discontinuity across $x_0^+$ we obtain
the solution
\eqn\fosix{ e^{-2\rho} = e^{-2\phi} =-a(x^+ - x^+_0)
\Theta(x^+ - x^+_0) -\lambda^2 x^+ x^- . }
For $x^+ > x^+_0$ this is identical to a black hole of mass $ax^+_0 \lambda$
after
shifting $x^-$ by $a/\lambda^2$.
The Penrose diagram for this
spacetime closely resembles that of the Vaidya spacetime (\fstar )
and is shown in \ften.

\subsec{Hawking Radiation and the Trace Anomaly}

So far we have achieved a satisfying description of the classical
formation of a $1+1$-dimensional black hole from collapsing matter. However
the real motivation for
studying this model is to understand quantum effects. We will do this
in several parts. To begin with we will analyze the quantum effects
of matter fields in the fixed classical background of a black hole formed by
collapsing matter.

In two dimensions there is a beautiful relation between the trace
anomaly and Hawking radiation discovered in
\CrFu. For a massless scalar field
the trace of the energy-momentum tensor is zero classically,
$T \equiv T^\mu_\mu = 0$. Quantum mechanically there is a one-loop anomaly
which relates the expectation value of the
trace of the energy-momentum tensor to the
Ricci scalar
\eqn\foseven{\langle T \rangle= {c \over 24} R,}
where $c=1$ for a massless scalar and $c=1/2$ for a Majorana fermion.
In conformal gauge with $T= -4 e^{-2 \rho} T_{+-}$ this gives for
$N$ $c=1$ scalars
\eqn\foeight{ \langle T_{+-}^f \rangle = - {N \over 12} \p+ \pmi \rho .}
Given the expectation value of $T_{+-}$ as above we can use energy-momentum
conservation to determine $T_{++}$ and $T_{--}$. We have
\eqn\fonine{
\p+ T_{--}+\pmi T_{+-}-\Gamma^-_{--}T_{+-}	      =0,}
and similarly for $T_{++}$.
Using $\Gamma^+_{++}= 2 \p+ \rho$, $\Gamma^-_{--} = 2 \pmi \rho$
the solution is found as
%
\eqn\fifty
{\eqalign{\vev{T^f_{++}}& = -{N \over 12} \left(
 \partial_+\rho \partial_+\rho - \partial^2_+\rho +
t_+(\sigma^+)\right)\ ,\cr
          \vev{T^f_{--}}& =-{N \over 12} \left(
  \partial_-\rho\partial_-\rho - \partial^2_-\rho +
t_-(\sigma^-)\right)\ .\cr}}
The functions of integration $t_\pm$  are not determined
purely by energy-momentum conservation and must be fixed by
imposing physical boundary conditions.
(In the next subsection we will see that they are related to the Casimir
energy of the matter fields.)
For the collapsing
$f$-wave, $t_\pm$  are fixed by requiring that $T^f$  vanish
identically in the linear dilaton region, and that
there be no incoming radiation along ${\cal I}^-_R$ except for the
classical $f$-wave at $\sigma^+_0$.

We now
 turn to a calculation of Hawking radiation from a ``physical''
black hole formed by collapse of an infalling $f$ shock-wave as in
\fofour.
The calculation and its physical interpretation is clearest in coordinates
where the metric is asymptotically constant on ${\cal I}_R^{\pm}$.
We thus set
\eqn\fione
{\eqalign{e^{\lambda y^+}& =\lambda x^+, \cr
          e^{-\lambda y^-} &= -\lambda x^- - {a\over\lambda}.\cr}}
This preserves the conformal gauge \onetwo\ and gives for the new
metric
\eqn\fitwo{-2 g_{+-} = e^{2 \rho} = \cases{[1+{a \over \lambda} e^{\lambda
y^-}]^{-1},
             & if $y^+ < y_0^+$; \cr
            [1+ {a \over \lambda} e^{\lambda (y^- -y^+ +y_0^+)}
]^{-1}
             & if $y^+ > y_0^+$ \cr}}
with $\lambda x_0^+ = e^{\lambda y_0^+}$.

The formula for $\rho$, together with the boundary conditions
on $T^f$ at ${\cal I}^-_{L,R}$ then implies
\eqn\fifthree
{t_+ = 0, \qquad t_- = {-\lambda^2 \over 4} [1- (1+a e^{\lambda
y^-}/\lambda )^{-2} ]. }
The stress tensor is now completely determined, and one can read off
its values on ${\cal I}^+_R$ by taking the limit $y^+\to \infty$:
\eqn\fiffour
{\eqalign{ \vev{T^f_{++}} &\to 0, \qquad\vev{T^f_{+-} }\to 0,\cr
          \vev{T^f_{--}} &\to { N \lambda^2 \over48} \left[
1-{1\over\left(1+a e^{\lambda y^-}/\lambda\right)^2}\right]~~.}}
The limiting value of $T^f_{--}$ is the flux of $f$-particle energy across
${\cal I}^+_R$. In the far past of ${\cal I}^+_R$ $(y^- \to-\infty)$ this
flux vanishes exponentially while, as the horizon is approached, it
approaches the constant value $N \lambda^2/48$.
This is nothing but Hawking radiation. The result
that the Hawking radiation rate is asymptotically
independent of mass is peculiar to the model defined by
\twseven\ and does not hold for a generic model.

Although we have established that there is a net flux of energy
which starts at zero and builds up to a constant value (ignoring
backreaction) the skeptical reader might wonder whether this
is in fact thermal Hawking radiation. In order to show that this is
indeed the case, we must describe the full quantum state, which is the
subject of the next section.

\subsec{The Quantum State}

Quantum states of the matter field $f$ are constructed with right and
left-moving $f$ creation and annihilation operators: The right-moving
operators are
\eqn\cran{\eqalign{a_w & = -\frac{i}{2\pi} \int \frac{dz^-}{\sqrt{2w}}\
f(z^-) \overleftrightarrow\partial_- e^{iwz^-}~,\cr
a^\dagger_w & = \frac{i}{2\pi} \int \frac{dz^-}{\sqrt{2w}}\, f(z^-)
\overleftrightarrow\partial_- e^{-iwz^-}~,}}
where $w>0$,
and obey
\eqn\ccrs{\left[a_w, a^\dagger_{w^\prime}\right] = \delta (w-w^\prime)\ .}
The vacuum is then defined by the condition that
\eqn\vdef{a_w | 0_z\rangle = 0\ .}

The definition \cran\ of the creation and annihilation operators depends
on a choice of coordinates, here denoted $z$.  The state $|0_z\rangle$
is defined with respect to these operators and so
will also depend on the choice of coordinates.  What appears to be a
vacuum in one coordinate system, will be a many-particle state (obtained
by a Bogolubov transformation) in another.  This reflects the physical
fact that observers in the state $|0_z\rangle$
 which are not inertial with respect to $z$
coordinates will detect particles.

In describing Hawking radiation in the shock-wave geometry, the matter
state is taken to be the ``inertial'' vacuum state prior to the shock wave,
in which inertial observers detect no particles.  This will be the case
if the vacuum is defined with respect to coordinates \forty\ in which the
metric is simply
\eqn\iimt{ds^2 = -d\sigma^+ d\sigma^-}
below the shock wave.  Since $f$ is a free field, this defines the
right-moving part of the quantum state everywhere, including above the
shock wave.

We are now in a position to investigate thermal properties of the
quantum state on ${\cal I}^+$.\foot{The following argument is due to
L.~Thorlacius \refs{\laref}.} These follow from the two-point correlation
function which is simply
\eqn\fcor{\left\langle 0_\sigma|f(\sigma^-)\, f(\sigma^{\prime -}) |
0_\sigma\right\rangle =
\ln (\sigma^-- \sigma^{\prime -})}
in $\sigma$ coordinates.  To interpret this we should transform to the
inertial coordinate $y^-$ of \fione\ on ${\cal I}^+_R$, which is related to
$\sigma^-$ by
\eqn\zsig{\sigma^- = -\frac{1}{\lambda} \ln \left(e^{-\lambda y^-} +
\frac{a}{\lambda}\right)\ .}
Evaluating \zsig\ at late retarded times $(y^-\to\infty)$ and inserting in
\fcor\ one finds
\eqn\wllp{\left\langle 0_\sigma | f(y^-)\, f(y^{-\prime}) |
0_\sigma\right\rangle = \ln \left( {1 \over a} e^{-\lambda y^{-\prime}}
-{1 \over a}e^{-\lambda y^-}\right)\ .}
This correlation is periodic in imaginary time with period
$\beta=2\pi/\lambda$, indicating that $|0_\sigma\rangle$ indeed approaches a
thermal state with temperature $T=\lambda/2\pi$ at late times. This has
also been seen \GiNe\ in a direct computation of the quantum
state on ${\cal I}^+$.

The expression for the quantum state of the $f$-field also provides a
different way of understanding the non-zero expectation value for $\langle
T^f_{--}\rangle$ in \fiffour. Clearly,
\eqn\tzero{\left\langle 0_\sigma |: T^f_{--}:_\sigma\, |
0_\sigma\right\rangle=0,}
where $:T^f_{--}:_\sigma$ denotes the operator $T^f_{--}$ normal ordered with
respect to creation and annihilation operators in $\sigma^-$ coordinates.  It
is well known that for $N~~c=1$ matter fields
the normal ordering constant in different coordinate
systems is related by the Schwarzian derivative
\eqn\schw{:T^f_{--}:_y = \left(\frac{\partial \sigma^-}{\partial
y^-}\right)^2 :T^f_{--}:_\sigma -{N \over 12} \left(\frac{\partial
\sigma^-}{\partial y^-}\right)^{3/2}
\left(\frac{\partial}{\partial \sigma^-}\right)^2 \left(\frac{\partial
\sigma^-}{\partial y^-}\right)^{1/2}\ .}
This implies, using \zsig\ and \tzero,  that on ${\cal I}^+$
\eqn\fiveten{\left\langle 0_\sigma | :T^f_{--}:_y\, | 0_\sigma\right\rangle
=
\frac{N\lambda^2}{48}\left[1-\frac{1}{(1+ae^{\lambda y^-}/\lambda)^2}
\right],}
in agreement with \fiffour. Thus the quantities $t_\pm$ in the previous
section arise because the coordinates which define the vacuum and those
which are asymptotically inertial do not agree, resulting in an
expectation value for the stress tensor normal ordered in inertial
coordinates.

So far we have not discussed the left-moving part of the
quantum state\foot{In Section 3.10 models with a boundary
condition relating left and right movers will be considered. Such models
more closely resemble the four-dimensional situation.}, which
contains a collapsing matter wave.  This
does not directly enter into the preceding
description of the right-moving quanta
which appear on ${\cal I}^+_R$, except insofar as it supplies the stress energy
which distorts the metric and produces the mismatch between inertial
coordinates on ${\cal I}^+_R$ and ${\cal I}^-_L$. The left-moving part
of course has excited quanta even before the
inclusion of gravitational effects, which may be described by
a coherent state
\eqn\fiveeleven{|f^c\rangle = A: e^{\frac{i}{\pi} \int d\sigma^+ \partial_+
f^c(\sigma^+) f (\sigma^+)}\ :_\sigma |0_\sigma\rangle}
for a wave with profile given by the function $f^c(\sigma^+)$. $A$ here is a
normalization factor, and the normal ordering is in asymptotically
inertial $\sigma^+$ coordinates. A shock wave is obtained in a limit in which
$f^c(\sigma^+)$ is very sharply peaked.

\subsec{Including the Back-Reaction}

If expression  \fiffour\ is integrated along all of ${\cal I}^+_R$
to obtain the total energy emitted in Hawking radiation an infinite
answer is obtained. This is obviously nonsense:  the black hole can not
radiate more energy than it owns.

The reason for this nonsensical result is simple: the backreaction of
the Hawking radiation on the geometry has been neglected. While this
should be unimportant at early times when the Hawking radiation is weak,
ultimately it should be important enough to terminate the radiation
process when the mass reaches zero.

As a first stab at including the backreaction, let us simply include the
quantum stress tensor
\foeight, \fifty\ to act as a source for the classical metric equations. For
example the ${\rho}$ equation \thfour\ is modified to read

\eqn\mrho{{e^{-2\phi}}({2}{\partial_+}{\partial_-}{\phi} - {4}{\partial_+}
{\phi}{\partial_-}{\phi} {-} {\lambda^2} {e^{2\rho}}) = {{N}\over{12}}
{\partial_+}{\partial_-}{\rho},}

\noindent while the constraint equations are modified by the addition
of \fifty. These modified equations can be derived from the non-local
action \poly
\eqn\plact{S={S}_D-{{N}\over{96\pi}} {\int} {d^2}{x} {\sqrt{-g}} {R}
\ {\sq}^{-1}{R},}
\noindent where ${\sq^{-1}}$ is the scalar Greens function. Note that
in conformal gauge ${\sq^{-1}}{R} = -2{\rho}$, so that \plact~~becomes local:
\eqn\sfin{\eqalign{
{S}&= {{1}\over{\pi}} {\int} {d^2}{\sigma}
\biggl[{e^{-2\phi}}({2}{\partial_+}
{\partial_-}{\rho} - {4}{\partial_+}{\phi}{\partial_-}{\phi}+ {\lambda^2}
{e^{2\rho}})\cr
&{-} {{N}\over{12}}\,{\partial_+}{\rho}{\partial_-}{\rho} + {{1}\over{2}}
{\sum_{i=1}^N} {\partial_+}{f_i}{\partial_-}{f_i})\biggr],\cr}}

There is another, equivalent, method of deriving the extra term in \plact.
The quantum theory is defined by the functional integral in conformal
gauge
\eqn\zzz{{Z} = {\int}{\cal D}({b},{c},{\rho},{\phi}){\CD}{f_i}
{e^{{i}({S}_D{+}{S_{bc}}+{S_M})}},}
\noindent where ${b}$ and ${c}$ are Fadeev-Popov ghosts arising from gauge
fixing to conformal gauge, and ${S_{bc}}$ is their action. In order to define
the measures in ${Z}$ one must introduce a short distance regulator. This
should be done in a covariant manner, which implies that the measures
will depend on ${\rho}$ and so should be denoted e.g. ${\CD}_{\rho}{f_i}$.
This dependence of the measure on ${\rho}$ is given by
\eqn\ddd{{\CD}_{\rho}{f_i} = {\CD}_0{f_i}{e^{{-}{{iN}\over{12\pi}}
{\int}{\partial_+}{\rho}{\partial_-}{\rho}}},}
\noindent where ${\CD}_0$ is the measure with ${\rho} = {0}$. The
term in the exponent is precisely the
extra term in \sfin. Thus we see that this extra term arises from
the metric dependence of the functional measure on the matter fields.
Similar terms arise from the ghost-gravity measure, but in the following
section we will see that they can be suppressed.

\subsec
{\it The Large ${N}$ Approximation}

The quantum-modified equation \mrho\ does {\it not} provide a
consistent description of the quantum theory to leading order in an $\hbar$
expansion. The problem is that the left hand side is order $\hbar^0$
while the right hand side is order $\hbar^1$\foot{
In fact not even all order $\hbar^1$ terms are included in
\mrho: For example the corrections from the ghost-gravity measure are
omitted.}. Exact solutions to this
equation would involve all powers of $\hbar$, but higher powers of
$\hbar$ in such solutions would be affected by order $\hbar^2$ corrections to
the equation. To put it another way, the qualitative nature of a solution
cannot be affected by perturbative corrections if, as required by validity of
the
perturbation expansion, the corrections are indeed small.  Thus we cannot
expect to describe a black hole  which disappears through evaporation in a
perturbative expansion about a static, classical black hole.

The solution to this dilemma is to expand the theory in $1/N$ (rather than
$\hbar$) with ${N}{e^{2\phi}}$
held fixed \CGHS . Both sides of \mrho\ are then of the same order ${N}^1$,
and it is easily seen that all corrections\foot{Including those from the
ghost-gravity measure.} are order ${N^0}$ and therefore
negligible to leading order. Furthermore, since the entire action is
large the stationary
phase approximation is valid, and we need merely solve the semiclassical
equations. The semiclassical ${\rho,\phi}$ equations can be cast in
the form
\eqn\phin{2\left(1-\frac{N}{12} e^{2\phi}\right)
{\partial_+}{\partial_-}{\phi} =  ({4}
{\partial_+}{\phi}{\partial_-}{\phi}+{\lambda^2}{e^{2\rho}})
\left(1-\frac{N}{24}\, e^{2\phi}\right)\ ,}
\eqn\rhon{2\left(1-\frac{N}{12}\, e^{2\phi}\right)
{\partial_+}{\partial_-}{\rho} = ({4}{\partial_+}
{\phi} {\partial_-}{\phi} + {\lambda^2}{e^{2\rho}}),}
The ${++}$ constraint equation is
\eqn\consn{\eqalign{{T_{++}}&=
{e^{-2\phi}}({4}{\partial_+}{\phi}{\partial_+}{\rho}
- {2}{\partial_+^2}{\phi}) + {{1}\over{2}} {\sum_{i=1}^{N}}{\partial_+}
{f_i}{\partial_+}{f_i}\cr
&{-}{{N}\over{12}} ({\partial_+}{\rho}{\partial_+}{\rho} -
{\partial_+^2}{\rho}) + {t_+} = {0},\cr}}
and a similar equation holds for ${T_{--}}$.

An immediately obvious feature of \phin\ and \rhon\ is
\refs{\RST,\BDDO}\ that
$\left(1-\frac{N}{12}\, e^{2\phi}\right)$
on the left hand side vanishes at the critical value
of the dilaton field:
\eqn\pher{{\phi_{cr}} = {{1}\over{2}} \ln {{12}\over{N}}.}
Unless the right hand sides of \phin\ and \rhon\ vanish when ${\phi}$
reaches ${\phi_{cr}}$ the second derivatives of $\rho$ and $\phi$
will have to
diverge. While the RHS of \phin\ and \rhon\ do vanish for the vacuum \thone,
this will not be the case for perturbations of the vacuum, and singularities
will occur.
These singularities can be viewed as a quantum version of the classical
black hole singularities \RST . Classical singularities occur when the
area $e^{-2\phi}$ goes to zero along a spacelike line, quantum
singularities occur when the quantum corrected area, $(e^{-2\phi} -
\frac{N}{12})$, goes to zero.

It is important to stress that the large-$N$ approximation can not be
trusted in regions where the fields themselves grow to be of order $N$.
In particular the semiclassical equations must break down before the
singularity is reached, and one cannot reliably conclude that a
real singularity does indeed exist (though we shall
continue to refer to the regions where the large-$N$ approximation
breaks down as a singularity).  To probe the region near the
singularity requires a more complete treatment of the quantum
theory.

To see the singularity
explicitly, consider a matter shock wave at ${x_0^+}$
as given by equation \fofour. Beneath the shock wave $({x^+ < x_0^+})$,
the geometry is the vacuum. The equations imply that ${\rho}$ and ${\phi}$, but
not their first derivatives ${\partial_+}{\rho}$ and ${\partial_+}{\phi}$,
are continuous across the shock wave. The geometry above the shock wave can
then be perturbatively computed
in a Taylor expansion about the
shock wave.  One finds that just above the shock wave \refs{\RST,\BDDO}
\eqn\phsk{{\partial_+}{\phi}(x_0^+,x^-) =
{{1}\over{{2}{x_0^+}}}\,\left({{M/\lambda^2}\over
{\sqrt{\left(\lambda x^+_0 x^-\right)^2 + Nx^+_0 x^{-}/12}}}\, -1\right)
\ ,}
where by continuity ${\phi}({x_0^+}, {x^-})$ is given by its
vacuum value $-\half \ln (-\lambda^2 x^+_0 x^-)$ .

There are two notable features of this expression. The first is that
${\partial_+}{\phi}$ diverges when the shock wave crosses the timelike line
in the vacuum where ${\phi} = {\phi_{cr}}$. Before
diverging, however, it must cross zero at an earlier value ${x_H^-}$ of
${x^-}$. This point marks the beginning of
an apparent horizon, as defined in \tpnt. Behind this horizon and above the
shock wave there
is a region of trapped points, or an apparent black hole. The
singularity at ${\phi} = {\phi_{cr}}$ is thus inside an apparent black hole.

In a region of trapped points lines of constant ${\phi}$ are spacelike.
Therefore the singularity at ${\phi} = {\phi_{cr}}$ leaves the shock
wave on a spacelike trajectory. It can also be seen analytically
 \RST\ that the apparent
horizon leaves the shock wave on a timelike trajectory, corresponding
to the fact that the black hole is radiating and shrinking.

\ifig\fnuma{Numerical simulation of black hole
formation and evaporation from reference \nummer. Initial conditions
are specified
along the left and lower boundaries of the plot corresponding to
a null $M=.5$ shock wave along the left boundary. The coordinates are
$\tau_\pm=\sigma^\pm$ of equation \forty.
The contours depict lines of
constant $\phi$ (the rippled dashes are an artifact of the plotting
routine). The interior of the black hole is the region where
these lines slope downward to the
right, and the apparent horizon is the boundary of this region.}
{\epsfysize=3.5in \epsfbox{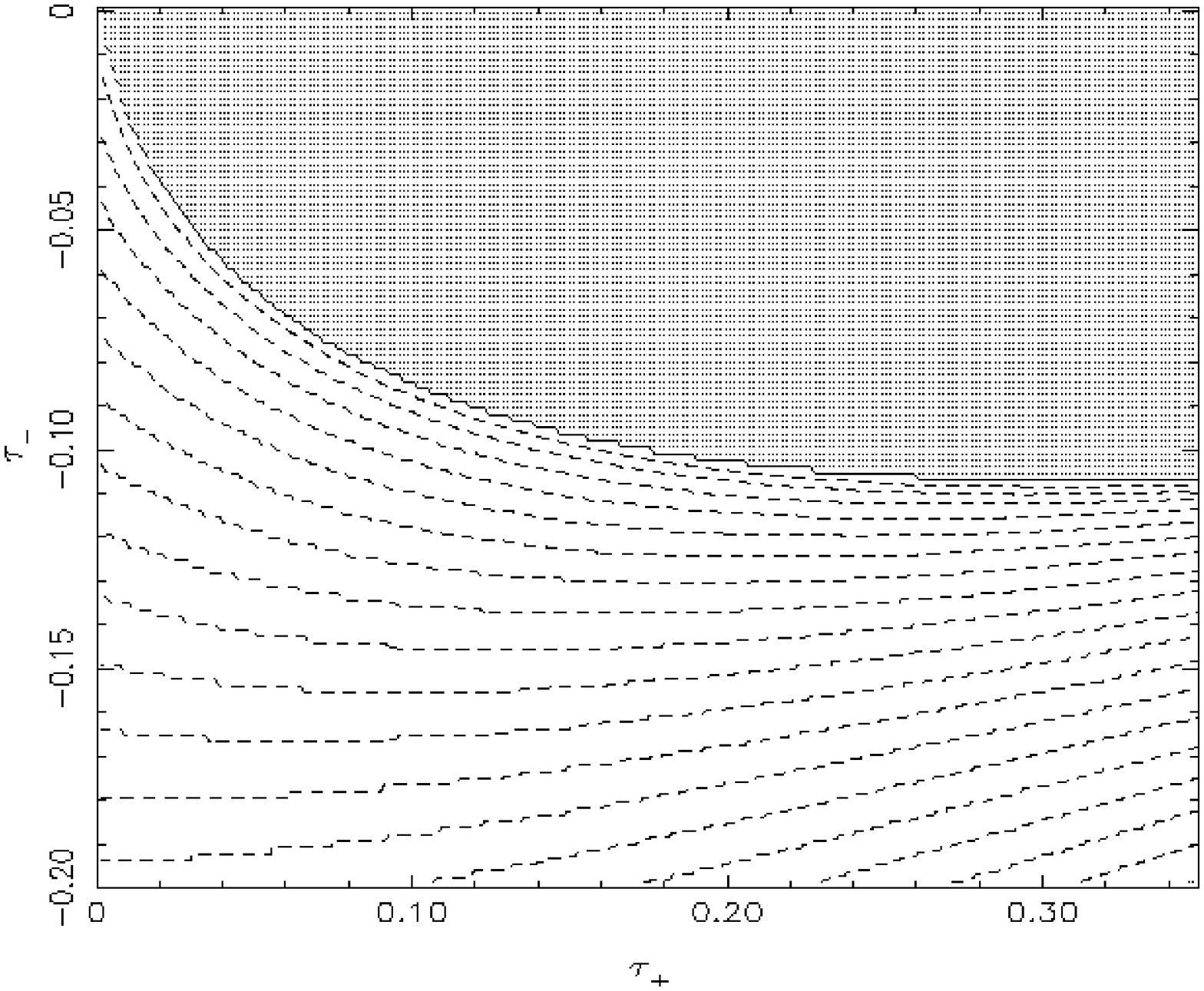}}
\ifig\fnumb{A plot from \nummer\ of the singularity line $\phi=\phi_{cr}$ and
the apparent
horizon line $\partial_+ \phi = 0$ for  step sizes $d\tau$
ranging between $4 \cdot 10^{-3}$ and $6.25 \cdot 10^{-5}$.
It is evident that the curves converge.}
{\epsfysize=3in \epsfbox{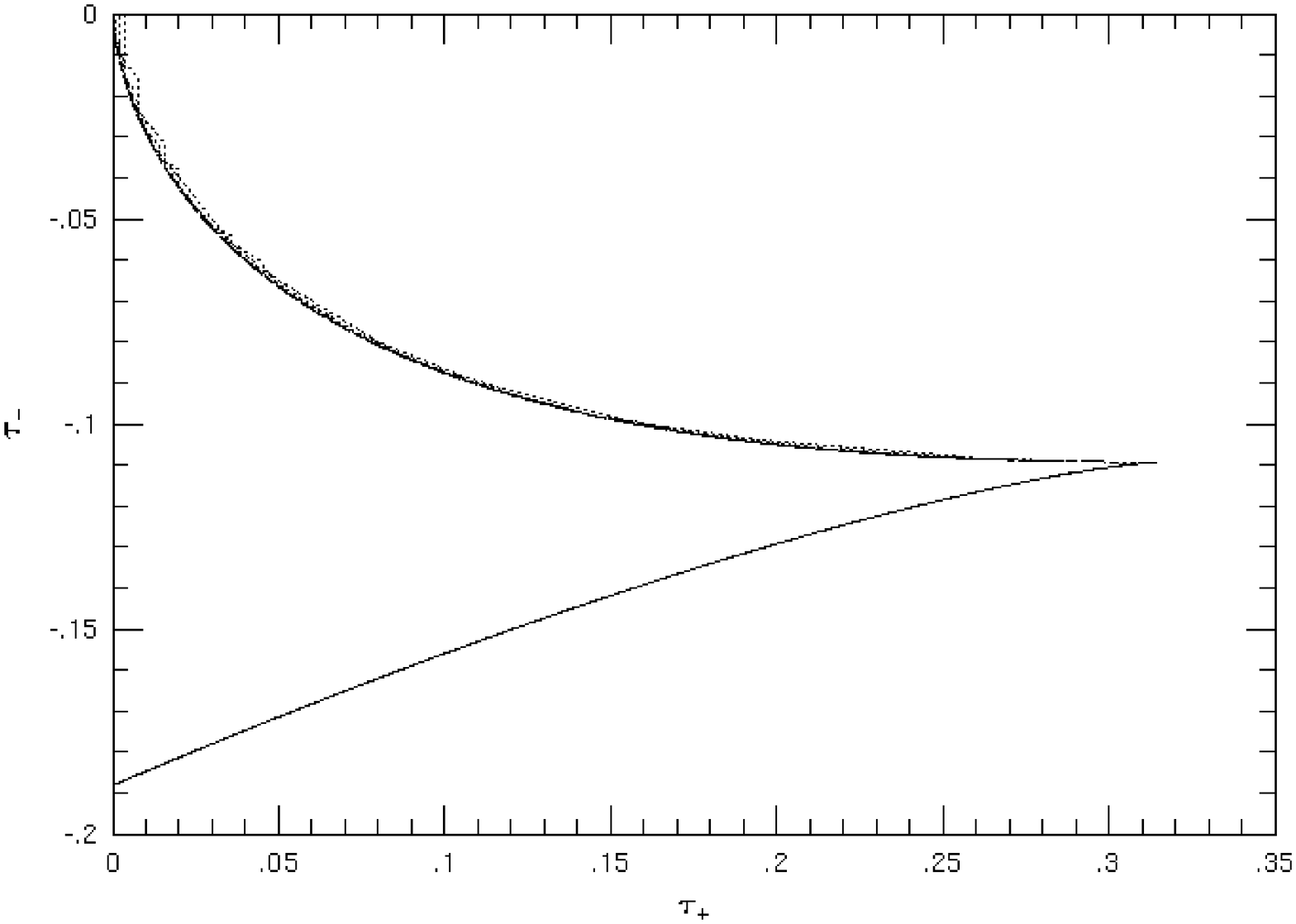}}
Numerical work is required to obtain the complete spacetime geometry
\refs{\hast,\lowe,\nummer,\tdu}, illustrated in \fnuma\ and \fnumb. The
apparent
horizon continues to
recede due to Hawking emission. After a finite proper time it meets the
singularity curve at the endpoint, where the black hole has shrunk to
zero size and the equations break down.

In order to continue to the causal future of the endpoint further
physical input, such as a boundary condition, is required.  This is best
discussed in the context of improved ``soluble'' models, as will be
discussed in the following two sections.  However we have already
learned
one important lesson.  When the black hole reaches zero size, its
interior is still large in the sense that much of the left-moving
incoming quantum state from ${\cal I}^-$ evolves directly into
 the black hole, and has
not been scattered up to ${\cal I}^+$. This feature is not specific to the
model discussed here \hhay, and will be important when we
discuss the information puzzle in Section 4.

\subsec{Conformal Invariance and Generalizations of Dilaton Gravity}

The quantization of dilaton gravity discussed in the previous sections,
is not unique.
If the quantum theory is defined as an expansion in ${e^{2\phi}}$,
there are new finite, renormalizable, counterterms at every order
in perturbation theory. For example at ${n}$th order there is the
term ${e^{{2}({n}-{1}){\phi}}}({\nabla\phi})^2$. While some important
constraints
on these terms will be discussed, they are far from being completely
fixed.

One elementary constraint is that the theory should have a stable ground
state. In fact it is quite easy to destabilize the ground state in the
process of adding
terms to the action. General criteria for the existence of a positive
energy theorem are discussed in \past .

Further properties of the quantum theory follow from the connection between
two-dimensional gravity and conformal field theory \refs{\jptb,
\deAli,\BiCa,\qtdg}. This connection
is best understood by quantizing the theory in conformal gauge:
\eqn\eee{\eqalign{{g_{+-}}&= {-} \half\,{e^{2\rho}},\cr
{g_{++}}&= {g_{--}} = {0}.\cr}}
\noindent This gauge leaves unfixed a group of residual diffeomorphisms
for which
\eqn\hhh{\eqalign{
{\delta}{g_{++}}&= {\nabla_+}{\zeta_+} = {g_{+-}}{\partial_+}
{\zeta^-} = {0},\cr
{\delta}{g_{--}}&= {\nabla_-}{\zeta_-} = {g_{+-}}{\partial_-}{\zeta^+}
= {0}.\cr}}
\noindent These equations imply
\eqn\YYY{{\zeta^{\pm}} = {\zeta^{\pm}} ({\sigma^{\pm}}),}
\noindent and that the residual diffeomorphisms generate the conformal
group. Correspondingly the moments of ${T_{++}}$ and ${T_{--}}$ generate
Virasoro algebras.

Invariance of the quantum theory under the residual
symmetry group can be insured, for example, by constructing a BRST charge
${Q}$ which obeys ${Q^2} = {0}$ and identifying physical states as
${Q} $ - cohomology classes.

At this point it should be clear that -- although a slightly different set of
words is being used -- what is being constructed here is a ${c} = {26}$
conformally invariant sigma model with ${\rho}, {\phi}$ and ${f_i}$
as fields living in an ${N} + {2}$ dimensional target space. If one
demands that the matter fields ${f_i}$ constitute a free ${c} = {N}$ conformal
field theory, then the ${\rho}, {\phi}$ sigma model must be conformally
invariant with ${c} = {26} - {N}$.

Letting ${X^\mu} = ({\rho}, {\phi})$, the ${\rho}, {\phi}$ sigma model
can be written in the form:
\eqn\FFF{
{S} = {-} {{{1}\over{2\pi}}} {\int} {d^2}{x} {\sqrt{{-}{\hat g}}} [{\cal G}
_{\mu\nu}{\nabla}{X^\mu}{\nabla}{X^\nu}
+ {{1}\over{2}}
{\Phi}{\hat R} + {T}],}
\noindent ${\hat g}$ here is a fiducial metric and ${\cal G}$, ${\Phi}$
and ${T}$ are functions of ${X^\mu}$. The couplings ${\cal G},
{\Phi}$ and ${T}$ are severely restricted by conformal
invariance. Namely, the beta functions must vanish:
\eqn\bfuns{\eqalign{0&=\beta^G_{\mu\nu}  =
2\nabla_\mu\nabla_\nu \Phi + {\cal R}_{\mu\nu} +\cdots , \cr
                      0&=     \beta^\Phi =
              {(\nabla\Phi)^2} - \half \nabla^2\Phi+{N-24 \over 3}+\cdots ,\cr
                            0&= \beta^T   =
-2\nabla \Phi\cdot \nabla T +8T+\nabla^2 T +\cdots,\cr}}
where ${\cal R}$ is the curvature of $\cal G$.
These equations are indeed obeyed, to leading order in $1/N$, by the
${\cal G}, {\Phi}$ and ${T}$ implicit in \sfin.
While conformal invariance severely constrains the quantum theory, there are
still an infinite number of solutions. This may be viewed as an initial
data problem in which initial data is specified as a function of
${\phi}$ at fixed ${\rho}$, and the beta function equations are
then used to solve for ${\cal G}, {\Phi}$ and ${T}$ at every
value of ${\rho}$.

In order to correspond to the theory of dilaton gravity that we are
interested in, the values of ${\cal G},~{\Phi}$ and ${T}$ at
weak coupling $({\phi}{\rightarrow}{-}{\infty})$ should agree with
those implicit in \twseven. One particularly interesting set of
values will be discussed in the next section.

\
\subsec{The Soluble RST Model}

In the preceding section it was argued that there are an infinite
number of inequivalent theories of dilaton gravity, all of which reduce
to \twseven\ at weak coupling. For large ranges of parameter values,
these inequivalent  theories have qualitatively similar physical
behavior: The existence of black holes does not depend in a sensitive
manner on details of the couplings.  However, it was pointed out by
d'Alwis \refs{\deAli} and Bilal and Callan \refs{\BiCa} (see also
\refs{\qtdg,\kzm}) that for very
special values of the couplings, the theory becomes exactly soluble. A
particularly elegant and simple model of this type was discovered by
Russo, Susskind, and Thorlacius \refs{\rst}, as follows.

The classical action for the RST model is, in conformal gauge,
\eqn\cact{\eqalign{S_{cl} ={1\over\pi} \int d^2x
\Bigg[&(2e^{-2\phi}-{N\over12}\phi)
\partial_+\partial_-\rho\cr
&+e^{-2\phi}(\la^2e^{2\rho}-4\partial_+\phi\partial_-\phi)
+{1\over2}\sum^N_{i=1}\partial_+f_i\partial_-f_i\Bigg],}}
where $\rho$ is the conformal factor, $\phi$ is the dilaton and $f_i$
are $N$
scalar  matter fields. This differs from the classical action \twseven\
by the second term, which is proportional to $N/12$.  It is convenient to
define\foot{Our conventions differ
slightly from
\refs{\rst}. They are chosen so that  $\chi$ and $\Om$ are
held fixed as $N$ is taken to infinity.}
\eqn\odef{\eqalign{
\Om &={12\over N}e^{-2\phi}+{\phi\over2}+{1\over4}\ln{N\over 48}\ ,\cr
\chi &={12\over N}e^{-2\phi}+\rho-{\phi\over2}-{1\over4}\ln{N\over 3}
\ .}}
In the large-$N$ limit, with $\chi$ and $\Om$ held fixed, the quantum
effective
action is then
\eqn\nact{S ={1\over\pi}\int d^2 x\Bigg[{N\over12}(-\partial_-\chi
\partial_+\chi+\partial_+\Om\partial_-\Om
+\la^2e^{2\chi-2\Om})+{1\over2}\sum^N_{i=1}\partial_+f_i\partial_-f_i\Bigg]
\ .}
When rewritten in terms of $\rho$ and $\phi$, \nact\  is seen to differ
from
the classical action \cact\  by the term
${N\over12}\partial_+\rho\partial_-\rho$ responsible for
Hawking radiation. (The effects of ghosts may be ignored in the
large-$N$
limit.) \nact\ describes a conformally invariant field theory.  In fact
the theory described by \nact\ can be exactly solved as a conformal
field theory without restriction to the large $N$ limit.  Unfortunately
we shall see below that certain boundary conditions must be imposed,
which prevent exact solubility outside of the large $N$
limit. Attempts to solve the full theory with boundary conditions
when $N=24$ can be found in
\refs{\verl}.

The residual conformal gauge invariance \hhh\ remains unfixed in \nact. We fix
this by the ``Kruskal gauge'' choice
\eqn\gchc{\chi=\Om\ ,}
which implies
\eqn\rphi{\rho=\phi+{1\over2}\ln{N\over12}\ .}
In Kruskal gauge the equations of motion are simply
\eqn\oeom{\partial_+\partial_-\Om=-\la^2\ ,}
and the constraints
reduce to
\eqn\cstr{\partial^2_\pm\Om=-\hat T_{\pm\pm}\ ,}
where
\eqn\tdef{\hat T_{\pm\pm}={6\over N}\sum^N_{i=1}\partial_\pm f_i\partial_\pm
f_i+\hat t_\pm\ .}
The functions $\hat t_\pm(x^\pm)$ are fixed by boundary conditions,
and the normalizations are chosen so
that $N$ scales out of the final equations.

The linear dilaton vacuum solution
\eqn\pldv{\phi=-{1\over2}\ln\Bigg[{-\la^2 Nx^+x^-\over12}\Bigg],}
\eqn\tpm{\hat t_\pm^0=-{1 \over4 (x_\pm)^2},}
corresponds to
\eqn\oldv{\Om=-\la^2x^+x^--{1\over4}\ln[-4\la^2x^+x^-]\ .}
The solution corresponding to general incoming matter from
$\cI^-$ is
\eqn\gsol{\eqalign{
\Om &=-{\la^2}x^+(x^-+{1\over\la^2}P_+(x^+))+{1\over\la}M(x^+)\cr
&-{1\over4}\ln[-4\la^2x^+x^-], }}
where
\eqn\mdef{\eqalign{
M(x^+) &= \la\int_0^{x^+} d \tilde x^+ \tilde x^+(\hat T_{++}-\hat t^0_+),
\cr
P_+(x^+) &=\int_0^{x^+} d \tilde x^+(\hat T_{++}-\hat t^0_+)\ .  }}
and $\hat t_-=\hat t_-^0$.
By transforming back to $\rho,\ \phi$ variables it can be seen for large
$M$
that this corresponds at early times to a black hole which forms and
evaporates.

However, the late-time behavior of \gsol\ is unphysical. Viewed as a
function
of $\phi$, $\Om$ has a minimum at
\eqn\sixtwentythree{\eqalign{
\phi_{cr} &=-{1\over2}\ln{N\over 48},\cr
\Om_{cr} &={1\over 4}. }}
There is no real value of $\phi$ corresponding to $\Om<\Om_{cr}$.
At late times the the solution \gsol\ evolves in to this region.
$\Om=\Om_{cr}$
should be regarded as the analog of the origin of radial coordinates
and the end
of the spacetime, rather
than continuing to negative radius. Reflecting boundary conditions,
consistent with energy conservation should
be imposed. RST accordingly  require
\eqn\rbc{\eqalign{
f_i&|_{\Om=\Om_{cr}} =0\ ,\cr
\partial_\pm\Om&|_{\Om=\Om_{cr}}=0\ .}}
The line $\Om=\Om_{cr}$ along which the boundary conditions are imposed
undergoes dynamical motion in
the $x^+,x^-$ plane. Of course this boundary  line
could be moved to a fixed timelike coordinate line {\it e.g.\/}
$x^+=x^-$ by a
conformal transformation. However, this would be incompatible with
Kruskal gauge
 and does not simplify the analysis.

Actually, subsequent to the work of RST, it was realized that the
boundary conditions \rbc\  are not conformally invariant even to leading
order in $1/N$ \refs{\bc,\asst}. Conformally invariant boundary
conditions do exist \refs{\bc}. They differ from \rbc\ by terms proportional
to $\partial_+^2 \hat
x^-(x^+)$, where $\hat
x^-(x^+)$ is the boundary curve, on the RHS of the $\Omega$
boundary condition. These corrected boundary conditions
lead to qualitatively similar
conclusions (in the present context) and are somewhat more complicated.  Thus
for our present
purposes it is simplest to stick with \rbc.

It follows from the equations of motion that the boundary curve $\hat
x^-(x^+)$
obeys
\eqn\ccc{\la^2\partial_+\hat x^-(x^+)=-\partial_+P_+(x^+)+{1\over4(x^+)^2}\ .}
If $\partial_+P_+$ is small enough, the right hand side is positive and the
boundary
curve is a timelike line. No black holes are formed: incoming matter is
benignly reflected up to future null infinity. A similar behavior occurs
in
four-dimensional general relativity in that sufficiently weak scalar
$S$-waves
can simply pass through  the origin without collapse.

On the other hand, if $\partial_+P_+$ exceeds the critical value
$1/4(x^+)^2$, the
boundary curve turns to the right (towards spatial infinity) and becomes
spacelike as in the shock wave geometry of \fstar. It can be seen
that the scalar curvature diverges along the spacelike
segments of the boundary curve. It is not possible to
implement the boundary condition
\rbc\  along these segments. Such spacelike boundary segments
necessarily bound regions of future trapped points
where $\partial_+\Om<$ and $\partial_-\Om<0$, which is
the interior of a black hole. Thus these spacelike singularities
resemble in every way the singularities inside four-dimensional black
holes.

The trajectory of a spacelike segment of the boundary curve is
determined, not
by boundary conditions, but by the initial conditions on $\cI^-$. If the
incoming energy is finite, the boundary curve will eventually revert to
a
timelike  trajectory. This is the ``endpoint" at which the future
apparent
horizon---the boundary dividing the regions $\partial_+\Om>0$ and
$\partial_+\Om<0$---meets the singularity, and the black hole has
evaporated to
zero size. After the endpoint the boundary  conditions \rbc\  are
immediately
imposed. The analytic solution is given in \refs{\rst} and
the Penrose diagram depicted in fig. 10.

In conclusion, the RST model embodies all the features of black hole
evaporation anticipated by Hawking. Black holes form and evaporate in a
finite
time, leaving nothing behind. Information is lost behind a global event
horizon.

For a time, many people (including the author) interpreted the RST
construction as strong evidence for the existence of fully consistent
theories of quantum gravity which destroy information.  However, rather
recently it was realized \refs{\lpst} that the RST model is in fact
inconsistent even at large $N$.\foot{This problem goes beyond the one
mentioned below \rbc, which is fixed in reference \bc.}
The problem is that there
is actually an infinite energy ``thunderbolt''
(denoted by the thin dashed line in fig. 10) which emanates from the endpoint
and is associated with the mismatch of the quantum state of the matter fields
above and below the null line $x^- = x^-_E$ emanating from the endpoint
toward ${\cal I}^+$. To see this consider the two point function,
\eqn\twpf{G(\epsilon)\equiv \left\langle f_-(x^-_E+\epsilon)f_- (x^-_E -
\epsilon)\right\rangle,}
of two right-moving matter fields just above and below the thunderbolt.
The reflecting boundary conditions \rbc\ can be used to relate this to a two
point function of incoming {\it left}-moving fields back on ${\cal I}^-$.
The image point of $x^-_E+\epsilon$ is obtained by reflection off the
post-black-hole boundary segment, while the image point of
$x^-_E-\epsilon$ is obtained by reflection off the pre-black-hole
boundary segment, leading to
\eqn\rtup{G(\epsilon) = \left\langle f_+\bigl({x^+_E \over
1-4\lambda^2x^+_E \epsilon}\bigr) f_+ \bigl({x^+_E\over
1+4x^+_E (P_++\lambda^2\epsilon) }\bigr)\right\rangle\ ,}
where $P_+\equiv P_+(\infty)$ is the total incoming Kruskal
momentum. These image points do not approach one another on
$\ci^-$ and
$G(\epsilon)$ is non-singular as $\epsilon\to 0$
\eqn\gstr{ G(\epsilon)\to \ln \bigl(x^+_E- {x^+_E\over
1+4x^+_E P_+}
\bigr).}
This is very strange
behavior for the two point function on ${\cal I}^+$: in any smooth
state, the two-point function should diverge logarithmically as the
points approach one another.  Any state for which this is not the case
must differ at arbitrarily high frequencies from the vacuum, and have
correspondingly infinite energy.\foot{This phenomenon was first noticed
by Anderson and DeWitt \refs{\ande}, and will be discussed in more
generality in Section 4.6.} Thus an infinite-energy
thunderbolt emanates from the
endpoint\foot{This is distinct from the finite-energy thunderpop
discussed in \refs{\rst}.}, and the RST model badly fails to conserve
energy.

In sections 4.7 and 4.8
we will discuss how this problem can be fixed.  We shall
argue that a proper, energy-conserving implementation of the endpoint
boundary condition leads to a radically different picture, in which
information is not lost after all, but is rescued from the
black hole interior and reradiated up to ${\cal I}^+$.
\ifig\frst{Collapsing radiation forms a large apparent
black hole (shaded
region) which evaporates, shrinks down to $r=0$ at $x_E$, and
subsequently disappears. This is Hawking's picture of four-dimensional
black hole evaporation. It is explicitly realized in the $RST$ model,
for which $r=0$ corresponds to $\phi=\phi_{cr}$, and there is
an energy non-conserving ``thunderbolt'' emanated from $x_E$
to $\ci^+$ along the thin
dashed line. The  spacelike surface
$\Sigma$ (thick dashed line) is placed so that it intersects
the apparent horizon
after the black hole has lost almost all of its initial mass, yet is still
well above the Planck mass so that the curvatures everywhere on and
in the past of $\Sigma$
are subplanckian.}
{\epsfysize=4.50in \epsfbox{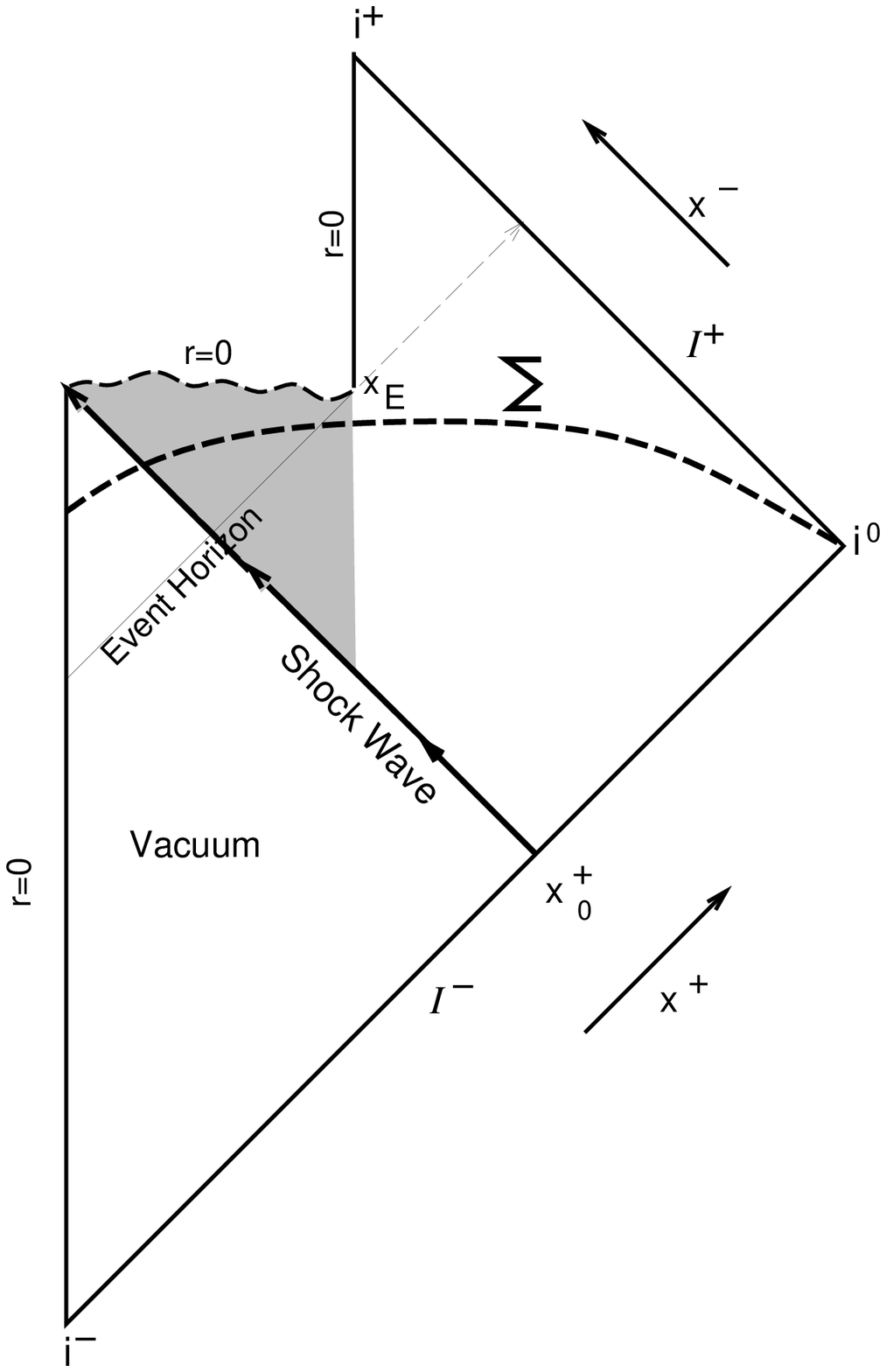}}

\newsec{The Information Puzzle in Four Dimensions}

In the previous sections we studied black hole formation and evaporation
in detail in a two-dimensional model using a semiclassical expansion.
We found that black holes form and evaporate, and eventually
approach a singular region which is the quantum cousin of
the classical black hole singularity. New physical input is required
to continue past the singularity. One proposal for such is the
endpoint boundary condition of the $RST$ model.
Corrections to the semiclassical expansion were suppressed by
powers of $1/N$. Armed with this sharpened insight, we now turn to
four dimensions and the information puzzle.

How similar is the four-dimensional problem to the two-dimensional
problem? A $1/N$ expansion of gravity coupled to matter fields is
also possible in four dimensions\tomb . At leading order one finds that
quantum fluctuations of the gravitational field are suppressed,
and that the quantum state of all the fields is a coherent state governed by
semiclassical equations. At subleading order some kind of finite
cutoff will be needed
because of the nonrenormalizability of quantum gravity. However
the cutoff-dependence
should be small as long as the local curvatures are small, as in any
process involving weak gravitational fields.
The real problem is that even the leading-$N$ semiclassical equations
are far too complicated to solve analytically (although some
numerical headway has recently been made in \tsvpa). The best one
can do is understand their qualitative behavior. The main features are clear:
Large black holes can be formed in an essentially classical manner.
They then slowly emit Hawking radiation
and - by energy conservation - simultaneously shrink.
Ultimately they become planckian and the approximations break down.

Of course in the real world $N$ takes some fixed value,
and it may not be correct to treat $N$ as large. Nevertheless
the semiclassical expansion can still be controlled in some regions
by an expansion in
$1/M$, where $M$ is the black hole mass. The expansion will then
break down when curvatures become large and $M$ shrinks down to the
Planck mass $M_p$ (at large $N$ one can continue on to
$M_p/N$). This takes us up to the surface $\Sigma$ in \frst.
Further input is required to go much beyond $\Sigma$. In the next
subsection we will discuss the information flow prior to $\Sigma$.
Following that we will discuss the possibilities for what may
happen beyond $\Sigma$.

\subsec{Can the Information Come Out Before the Endpoint?}

A central question in discussions of the information problem is
as follows. Consider an incoming state which collapses to form a
large, macroscopic black hole. Is
detailed information about the matter which  collapsed to form the
black hole available outside the apparent horizon
{\it before} the black hole becomes planckian
and the semiclassical expansion breaks down? To make this question
more precise, consider the spacelike slice $\Sigma$ depicted in \frst.
This slice begins at the origin, leaves the black hole at a time when
most of the initially large mass has evaporated but it is still well above the
Planck mass, and then continues out to spatial infinity. The region
outside of the black hole contains the Hawking radiation emitted
by the black hole over its long lifetime. The local
curvatures on and everywhere in the past of this slice are
subplanckian. One therefore expects that quantum gravity is
unimportant, quantum fluctuations of the metric are small, and that
semiclassical calculations are reliable for calculating the quantum
state $\psi_\Sigma$ on $\Sigma$.  Of course, $\psi_\Sigma$ is a pure
state obtained by unitary evolution from $\ci^-$ to $\Sigma$. However, not
all the information in $\psi_\Sigma$ is accessible to observers outside
of the black hole.  Let us formally divide the Hilbert space on $\Sigma$
into portions $\psi^{\rm ext}$ and $\psi^{\rm int}$ exterior and interior
to the black hole
\eqn\psdc{\psi_\Sigma = \sum\limits_{ij} a_{ij} \psi^{\rm ext}_i
\psi^{\rm int}_j\ .}
Observations outside the black hole are then determined by the exterior
density matrix obtained by tracing over the interior Hilbert space
\eqn\rext{\rho^{\rm ext} = \sum\limits_{ijk} a^*_{ik} a_{jk} \psi^{\rm
*
ext}_i \psi^{\rm ext}_j\ .}
In particular $\rho^{\rm ext}$ contains all information about the
quantum state of the
Hawking
radiation emitted prior to $\Sigma$.

The question now is, given the quantum state \rext\ outside the black
hole, can the incoming state from ${\cal I}^-$ be (almost completely)
reconstructed? If so then one would say that the information is outside the
black hole.

The impossibility of such a reconstruction follows from the
impossibility of {\it quantum xeroxing} or {\it quantum
bleaching}.\foot{The following argument is of course essentially due to
Hawking, but the version presented here recapitulates conversations held
at
the 1992 Aspen Conference on Quantum Aspects of
Black Holes, and follows a lucid and more
detailed presentation of Preskill \refs{\abdi}.}  A quantum xerox machine
takes any incoming state $|A\rangle$ into two copies of itself
\eqn\xrx{|A\rangle \to |A\rangle \otimes |A\rangle\ .}
One might hope that the evaporating black hole acted as a quantum
xerox machine, encoding the information
that falls in to the black hole in  the Hawking radiation
outside the black hole. The interior and exterior quantum state on
$\Sigma$ could then {\it both} be unitary transformations of the incoming
state, and
the initial state could be determined from measurements either
inside or outside the black hole.

This is impossible because quantum xeroxing violates the superposition
principle. If
\eqn\sixthirtyone{|A\rangle\to |A\rangle \otimes |A\rangle\ ,}
and
\eqn\sixthirtytwo{|B\rangle\to |B\rangle\otimes |B\rangle\ ,}
then the superposition principle implies
\eqn\sixthirtythree{\eqalign{|A\rangle + |B\rangle &\to |A\rangle \otimes
|A\rangle + |B\rangle \otimes |B\rangle\cr
&\not= (|A\rangle + |B\rangle) \otimes (|A\rangle + |B\rangle)\ .}}
so the information can not be {\it both} inside and outside the black
hole at a given time.

One may still  hope that the information is outside the black hole.
As just argued, if
it is outside, it is not inside, so the interior must be in a unique
quantum state which has been ``quantum bleached'' of all information
about the initial state. This is unreasonable.  In smooth
coordinates\foot{Of course there are coordinate systems (such as
Schwarzchild) in which the
horizon appears singular and in such coordinates it is not
obvious that quantum bleaching can not occur. However coordinate
invariance, together with the existence of coordinate
systems which are regular at the horizon, implies that the horizon is a
truly non-singular place
in both the classical and the quantum theories.}, the horizon is a smooth
place at which all curvatures are subplanckian.  There are no guards
stationed there which strip intruders of all information. Surely some
information can be carried across the horizon, and quantum bleaching
can not occur.

We accordingly reach the conclusion that information indeed
falls into the black hole, and does not get out before the black hole
becomes planckian.

A quantitative measure of the lost information is given by the entropy of
$\rho^{\rm ext}$
\eqn\sixthirtyfour{S_{\rm ext} = -tr \rho^{\rm ext} \ln\rho^{\rm ext}\ .}
$S_{\rm ext}$ depends only on the two sphere (on the apparent horizon)
at which $\Sigma$ is divided
into interior and exterior portions, and not on the shape of the rest of
$\Sigma$ (because deformations of $\Sigma$ which leave its intersection
with the horizon fixed correspond to unitary
transformations of $\rho^{\rm ext}$). $S_{\rm ext}$ is non-zero due to
correlations between the
interior and exterior portions of the quantum state $\psi_\Sigma$. As
argued by Hawking, the Hawking
radiation outside the black hole looks thermal when its correlations with
the internal quantum state are ignored. The value of $S_{\rm ext}$ can
then be estimated by integrating standard formulae for blackbody
radiation over the black hole lifetime.  This gives (in four dimensions)
\refs{\zur}
\eqn\zur{S_{\rm ext} \sim \frac{16\pi M^2}{3}\ .}

In two dimensions this can be made very precise \refs{\fpst}.
$S_{\rm ext}$ has been
computed exactly \fpst\
 at large $N$ in the RST model\foot{The troubles with the RST model discussed
in section 3.10 do not affect this computation, as $\Sigma$ is
prior to the endpoint.}, where backreaction
effects are incorporated.  It is given by $\frac{2\pi M}{\lambda}$ (plus
subleading in $\frac{1}{M}$ corrections which can be found in
\refs{\fpst}).  Taking into account the
difference between two- and four-dimensional thermodynamics, this exact
large $N$ calculation agrees with the estimate \zur\ based on adiabatic
reasoning. In particular, this calculation shows that
at least in two dimensions inclusion of back reaction does not
significantly alter the information content of the Hawking radiation, as
had been previously advocated by some authors.

Despite the plausibility of the preceding arguments that information
falls into a black hole and does not leave it before the black hole
becomes planckian, they have been
repeatedly questioned.
The most frequently raised objection to these arguments is as
follows\refs{\theo,\infloss,\blshift,\verl}. Consider a
typical quantum of Hawking radiation
on the portion of $\Sigma$ outside the black hole.  This quantum started
out life as a virtual mode of the vacuum on ${\cal I}^-$ which is
eventually scattered into a quantum of real radiation via interactions
with the gravitational field.  A typical such mode will be redshifted
over a long period during which it hovers near the horizon The energy of the
mode on ${\cal I}^-$ is accordingly related to the energy of the Hawking
quanta on $\Sigma$ by an enormous blueshift factor, of order $e^{16\pi M^2/3}$.
Thus we apparently need to understand the incoming state at
incredibly short, ultra-planckian distances in order just to find the
quantum state of ordinary Hawking modes on $\Sigma$.  Low-energy reasoning is
therefore inadequate for determining how much information is outside the
black hole.\foot{The energy per quanta is suppressed by a
factor of $1/N$ in a $1/N$ expansion,
so this objection does not apply to large $N$ theories.
Nevertheless
it would suggest that the $1/N$ expansion could
break down sooner than anticipated,
and would lead one to question the physical relevance of the
large $N$ approximation.}

This reasoning is incorrect in general\foot{The remainder of this
subsection is based on extended conversations with J. Polchinski, E. Verlinde,
and the Les Houches summer school
students.}.  To see why, consider a closed, flat universe
with matter fields in their vacuum state for $t < 0$.  Next let the
universe slowly expand at a rate $H$ for $0< t < t_0$, where $t_0$ is a
very long time, and then turn off the expansion.

Can the low-energy part of the
 quantum state of the matter field be found for $t>0$
without solving ultra-planckian dynamics? Field modes with energy
${\cal E}$ for $t> t_0$ started out life as modes with energies of order
$e^{Ht_0}{\cal E}$. For very long $t_0$, low-energy modes at $t>t_0$ will
have started out life as ultra-planckian modes for $t<0$, even if $H$ is
small.  Thus, according to the preceding argument, the low-energy
quantum state for $t>t_0$ cannot be found without analyzing Planck-scale
physics.

In fact -- as might be intuitively obvious --
the post-expansion state {\it can} be found, using the adiabatic
theorem, without solving Planck-scale physics\foot{The notion of an
adiabatic vacuum for a slowly varying spacetime
was introduced by Parker \prkr\ and the adiabatic
approximation  was developed in the 70's. A
review with references can be found in \brdv.}.  Matter energy is not conserved
during $0< t < t_0$ because the matter Hamiltonian is time
dependent due to the background expansion.  However, the scale of energy
violation is given by $H$. So only modes with energies of order $H$ can
be kicked out of their ground state and acquire life as real quanta.  An
ultra-planckian mode enters the region $t> 0$ in its vacuum state.
It remains there until it is redshifted down to the scale $H$, at which
point it may become excited by interactions with the background
geometry. The adiabatic theorem gives us all the information we
need about these ultra-planckian quanta at $t<0$: they remain in their
adiabatic ground state until they are redshifted down to the scale given
by the local rate of change of the background geometry. This example
(together with several others) is explicitly worked out in
Birrel and Davies \brdv.

An even simpler example, which does not involve gravity,
is as follows. Consider a box with reflecting walls of initial size
$L^3$ with interior fields in their ground state.
Now expand the box very slowly until it reaches the size
$(\gamma L)^3$.
For a fixed, slow expansion rate, $\gamma$ can be
made as large as one wishes by just
continuing the expansion for a long time. Post-expansion modes
of frequency $\omega$ started out life as (possibly ultra-planckian)
modes of frequency
$\gamma \omega$. One might jump to the false conclusion that
Planck scale physics is therefore
required to determine the final quantum state
of the box. The adiabatic theorem guarantees that this is not the case.
Indeed, if it {\it were} the case, there would be no need for
the LHC at CERN: Physics above the weak scale could be cheaply
explored with expanding boxes!

The black hole case is more involved
than these examples, but qualitatively similar.  It is
possible to find a set of smooth spacelike slices, labeled by a time
$T$, which begin just above ${\cal I}^-$ and culminate at $\Sigma$.
The slices can be arranged so that the intrinsic curvature is
everywhere subplanckian. The quantum state of the high-energy modes on
each of these slices is then the adiabatic ground state.  The energy
of these modes (as measured by $T$) is slowly redshifted as the black
hole evolves. Modes do not get excited until their wavelengths reach
the scale
set by the evolving black hole geometry. The full quantum state of
subplanckian modes on $\Sigma$ can thus be found without recourse to
planckian physics.

Having said this it is important to add that, as emphasized in \theo,
to date every $explicit$ calculation of Hawking evaporation involves
a reference to high frequencies at some stage in the calculation.
In practice it is awkward to adapt the calculation to
the adiabatic time  slicing. In four dimensions it is probably
impossible in practice. In two dimensions
such an explicit calculation may be feasible, but has not been
carried out. It would certainly be of great interest to do so.

Of course it is a logical possibility that, even though the low-energy
analysis is self-consistent and does not predict its own demise, that
there are nevertheless corrections from Planck scale physics which
become important for reasons which are peculiar to black holes.
That is, while the low-energy laws of physics are of course capable of
describing all low-energy phenomena observed so far, it is possible that black
hole
dynamics are strange enough that new corrections to those laws,
unobservable elsewhere, come in to play.  This
point of view is advocated in \refs{\blshift}, wherein it is argued that
string theory is required to understand the information flow, even
{\it before } the geometry becomes planckian.  Our view
is that new laws of physics should not be invoked to explain a
phenomenon unless it cannot be understood in the context of the old
ones. We will argue below that there is a self-consistent resolution
of the information puzzle which does not require the intervention of
planckian dynamics in low-energy processes.

In conclusion, the full quantum state, and in particular  the flow of
information, can be consistently analyzed with low-energy effective
theory up until the time that the black hole becomes very small and the
curvature becomes planckian.  It is seen that a large portion of the
information in the initial state remains within the black hole up until
this time.  If all the information is going to appear outside the black
hole, it must do so after this time. How or if this might happen will be
discussed in the following sections.

\subsec{Low-Energy Effective Descriptions of the Planckian Endpoint}

In the preceding it has been argued that the low-energy laws of physics
are sufficient for understanding the evolution of an evaporating
black hole as long as it is much larger than the Planck length. However
eventually it must shrink down to the Planck size, and quantum gravity must be
solved to continue the evolution in detail. We refer to this point as the
{\it endpoint}
(because it is the endpoint of the semiclassical evolution), even though
the system may still undergo further evolution. As quantum gravity is poorly
understood, it might seem that one should simply give up on the problem
past the endpoint.  However,
it still makes sense to ask what a low-energy
experimentalist who makes black holes and measures the outgoing
radiation could observe, and to try to describe this by some kind of effective
dynamics.
It should be possible to summarize
our ignorance about Planck scale physics in a phenomenological boundary
condition (or generalization thereof) which governs how low-energy
quanta enter or exit the planckian regions at and after the endpoint.

In principle this effective description should be derived by a
coarse-graining procedure from a complete theory of quantum gravity such
as string theory.  But this is not feasible in practice.  Instead we
shall consider all the different possible
descriptions, and find that they can be highly constrained by low-energy
considerations alone.

A classic example of this type of approach is the analysis of the
Callan-Rubakov effect
\refs{\caru,\jpia}, in which charged $S$-wave fermions are scattered off of a
$GUT$ magnetic monopole.  Even at energies well below the $GUT$ scale, the
scattering cannot be directly computed from a low-energy effective field
theory, because the fermions are inexorably compressed into a small region in
the monopole core in which $GUT$ interactions become important.
Initially the $GUT$ scale physics was analyzed in some detail.
The results were then coarse-grained and summarized in an effective boundary
condition for fermion scattering at the origin.  It was subsequently realized
that the detailed $GUT$ scale analysis was largely unnecessary for
understanding the low-energy scattering:
up to a few free parameters (a matrix in flavor space) the effective
description is determined by low-energy symmetries.

In the following sections we turn to the
black hole problem with this philosophy in mind, and consider all the
possible effective descriptions. As in the
Callan-Rubakov effect, we shall find that
the possibilities are extremely constrained just by
self-consistency of the low-energy theory.
\subsec{Remnants?}
One logically possible outcome of gravitational collapse is that planckian
physics shuts off the Hawking
radiation when the black hole reaches the Planck mass, and the
information about the initial state is eternally stored in a planckian
remnant.  As there are infinite numbers of ways of forming black holes
and letting them evaporate, this remnant must have an infinite number of
quantum states in order to encode the information in the initial state.
In an
effective field theory
these remnants would resemble an infinite number of species of
stable particles, and be governed by an effective lagrangian of the form
\eqn\remin{{\cal L}_{\rm eff}=-\sum_{i=0}^{\infty}\bigl( (\nabla \phi_i)^2
+M_p^2 \phi_i^2 +...\bigr).}
The operators $\phi_i$ create and annihilate a remnant in the $i$'th state.
The $+...$ represents interaction terms which we shall argue below must be
quite
important.

This raises the so-called ``pair-production problem''. Since the
remnants carry mass\foot{Massless remnants would create even worse
difficulties.}, it must be possible to pair-produce them in a
gravitational field.  Naively (ignoring the interactions in \remin)
the total pair-production rate
is proportional to the number of remnant species, and therefore infinite.
It is easy to hide a Planck-mass particle, but it is hard to hide an
infinite number of them. Thus it would seem that remnants can be
experimentally ruled out by the observed absence of copious pair-production.

However this formal argument is at odds with
an explicit semiclassical calculation \garst\ of the pair production rate.
The specific process considered in \garst\ was the production of
charged Reissner-Nordstrom black holes in an electromagnetic field,
so we first mention some pertinent facts about charged black holes.
The Hawking evaporation of a charged black hole, unlike that of a neutral black
hole, shuts off when it reaches a finite value of the mass $M$
equal to the charge $Q$. In \stas\ it was shown that
the charged black holes
have an infinite degeneracy of stable quantum states with
$M=Q$, {\it i.e.} there are remnants.
For large charge, these states can (unlike their neutral planckian cousins
discussed above)
be described with weakly-coupled, semiclassical perturbation theory.
These states can be created with the infinite number of ways
of throwing matter in to the black hole and then letting it
Hawking evaporate back to $M=Q$.
The precise description of the states depends on how the spacetime is sliced.
They
may be viewed
as
(greatly
redshifted) matter excitations which are either
hovering just outside the horizon (see {\it e.g.} \sbgl ),
and/or as actually inside the horizon (see {\it e.g.} \refs{\banrev,\stas}).
In any case the important point is that the
infinite degeneracy potentially leads to unacceptable
rate of pair-production, so the  charged remnants
provide an excellent laboratory for analyzing the pair-production
problem\foot{This was also stressed in \sbg.}.

In \garst\ an exact euclidean instanton was found describing the
pair creation process\foot{A different instanton was found in
\refs{\gibint,\remn}. However this instanton contains planckian regions and is
accordingly destabilized by locally divergent one-loop corrections. It
therefore cannot be used in a semiclassical evaluation of
pair-production\sbgl.}. The instanton is a complete, smooth geometry
when (and only when) the horizons of
the oppositely-charged pair-created black holes are identified.
It contains no high-curvature planckian regions (for weak external fields).
It also contains no region corresponding to the interior of
the black hole horizon.

To first approximation the pair creation rate is given by the
exponential of minus the instanton action. This is a finite
number which agrees
with the Schwinger result in the appropriate limit. At next order one
must compute the one-loop determinant. This has not been explicitly
computed, but it will also be finite after renormalization\foot{Except for
the usual divergence from the infinite volume of the background spacetime,
which should be subtracted off to get the production rate.} because the
geometry is everywhere smooth and there are no internal
infinite-volume regions. Thus this calculation predicts a finite rate
of pair production.

So what happened to the infinite number of remnant states which were
supposed to make the rate diverge?
Ordinarily the one-loop determinant counts the number of states,
so that is where a divergence might be anticipated. In fact
if the theory is defined with a cutoff, the one loop determinant will
indeed have a divergence as the cutoff is removed
corresponding to the infinite number of
high-frequency (but low-energy because of the redshift) states near the
horizon.
However this divergence does not appear in the production
rate after renormalization. It is
absorbed by renormalization of Newton's
constant: the state-counting divergence of the one-loop determinant is
precisely cancelled by the divergence arising in the classical instanton
action when it is
reexpressed in terms of the renormalized (rather than the bare) Newton's
constant\foot{These divergences are both proportional to the area of the
black hole horizon. The existence of a term proportional to
the bare Newton's constant times the horizon area was demonstrated in
\ggs\ with an exact computation of the classical instanton action.
The fact that the cancellation occurs in the manner
described here is essentially equivalent to
general arguments relating divergences
in the entropy to renormalization of Newton's
constant \refs{\suent,\fpst}.}. Hence this potential divergence
in the pair production
rate is eliminated in a standard fashion by renormalization.

One may also be concerned about
the infinite number of states behind the black hole horizon.
These simply do not appear in the calculation: As for
euclidean Schwarzchild, the instanton is
complete and smooth, but contains no region corresponding to the
interior of the
horizon. So, according to this calculation, such states simply
have no effect on the pair production rate \bos.
Of course, since they are causally separated from the exterior
spacetime,  they also have no effect on any Lorentzian scattering
process. Indeed, since these states lie in a region causally disparate
from the external spacetime, one expects that they can
be ignored and should not show up in the pair production rate.
It is satisfying that this
expectation is realized in the instanton calculation of \garst.

What could be wrong with the naive effective field theory argument?
It is hard to answer this question in detail because
so far no one has succeeded in deriving a useful effective
field theory description of the remnant states.
The naive effective field theory argument ignores the interactions --
the ``+...''  -- in \remin. However it appears that
these interactions must have important effects
and can not be ignored. To see why,  suppose \bos\ we had two remnants which
-- unbeknownst to us -- are in the same
quantum state. Then, it follows from
the effective field theory \remin\ without interactions
that we can discover that they are identical in a finite time
by quantum interference experiments. If this were indeed possible,
we would be
learning information about the quantum state behind the event horizon.
But this violates causality, and so cannot actually be possible.
We therefore conclude that the leading term in the effective field theory
in \remin\ is simply inadequate for a qualitative or quantitative
description of remnant dynamics \bos. The remnant states can not
-- at least in the charged Reissner-Nordstrom case --
effectively be thought of as an infinite collection of weakly interacting
particle species. Remnants are a new
kind of animal: Their behavior is quite different than that of
ordinary point particles.

Is there a good effective description of the type \remin ? At present certainly
not.
A proper
effective description may require treating the
the infinite number of remnant states as modes in an
internal remnant field theory
rather
than as an infinite number of distinct particle states.
This is natural because
the region near and inside a black hole can (unlike ordinary solitons)
contain a large volume
and many low-lying excitations.  Such a description
-- in which the discrete remnant species index labels momentum modes in an
internal dimension -- was partially developed for charged dilaton
black holes in \dxbh.
This example has interactions among
remnant states which are non-local in time along the
remnant worldline, corresponding to massless modes in the internal
remnant field theory. Effects
of these interactions could alter the
state counting estimate of the production rate.

Certainly
more remains to be understood on this
topic, and it remains controversial. However, it is clear
that the standard argument that infinite
pair-production is inevitable for all types of remnants is too naive, and
arguments/calculations have been given that
in some theories the pair production rate is
finite.  Further discussion can be found
in \refs{\sbg,\sbgl,\remn} and the reviews
\refs{\banrev,\sbgtr}.

A more inescapable objection to eternal remnants is the lack of any
plausible mechanism to stabilize them.  In quantum mechanics what is not
forbidden is compulsory. There cannot be a conservation law
forbidding remnant decay since that would also forbid remnant formation.
In the absence of a conservation
law, it is hard to understand why matrix elements connecting a massive
remnant to the vacuum plus outgoing radiation should be exactly zero.
Nature contains no example of such unexplained zeroes. Moreover, a
formal representation of quantum gravity as a
sum-over-geometries-and-topologies certainly includes such processes.
Eternal remnants are therefore highly unnatural.

An alternative to eternal
remnants is that the ``Planck soup'' which forms when
the black hole reaches the Planck mass continues to radiate in a manner
governed by planckian dynamics until all
the mass is dissipated. In principle, as we do not understand the
dynamics, the radiation emitted by the Planck soup could be correlated
with the earlier Hawking emissions and return all the information
back out to infinity. Energy conservation implies that the total energy
of the radiation emitted by the Planck soup is itself of the  order of the
Planck mass, and thus small relative to the initial mass of the black
hole.  It is very hard to encode all the information in the initial
state with this small available energy.  The only way to accomplish this
is to access very low-energy, long-wavelength states, which requires a
long decay time.  This leads to a lower bound of $\tau \sim M^4$ (in
Planck units) for the decay time of the
Planck soup \refs{\cash,\carwil,\abdi}. For a
macroscopic black hole this far exceeds the lifetime of the universe.
Hence, it is not possible for the information to be emitted in a
planckian burst at the end of the evaporation process.  In this scenario
one necessarily has a long-lived, but not eternal, remnant. Note that our
discussion
required no knowledge of planckian dynamics.  This is a prime example of
how low-energy considerations highly constrain the possible
outcome of gravitational collapse.

Of course, long-lived remnants are implausible without an explanation
for their long lifetime, or a mechanism for the Planck soup to reradiate
the information.  We shall encounter both below.

\subsec{Information Destruction?}
\ifig\fone{Collapsing radiation forms an apparent
black hole (shaded
region) which evaporates, shrinks down to $r=0$ at $x_E$, and
subsequently disappears.  The dashed wavy line is the region at which
Planck-scale physics becomes important, and is just prior to the
classical singularity. According to Hawking, information which crosses
the event horizon is irretrievably lost.}
{\epsfysize=4.50in \epsfbox{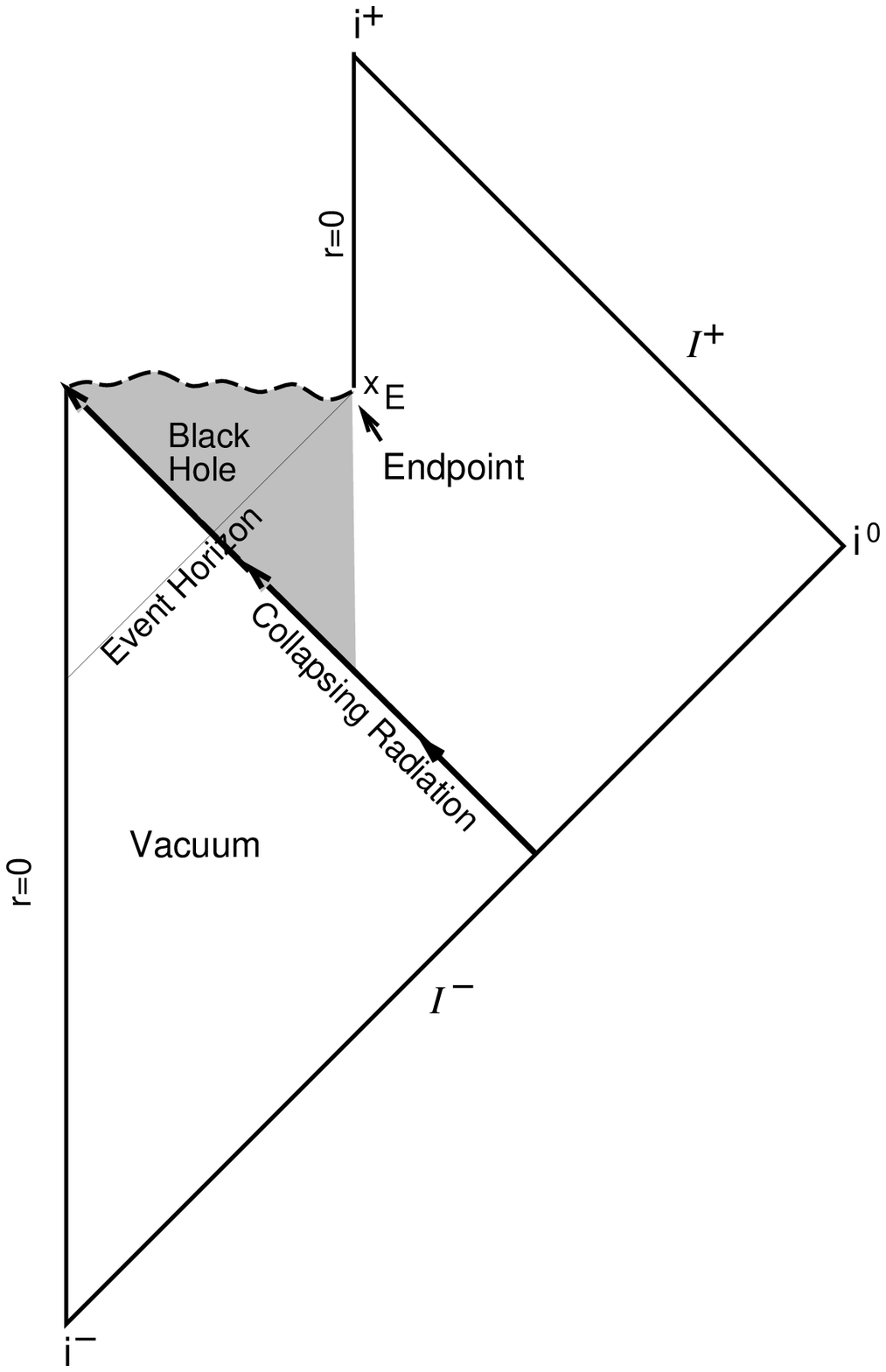}}

Faced with the apparent unpalatability of remnants, Hawking argued in
favor \hawktwo\ of a different possibility, depicted in \fone.  The black hole
disappears in a time of order the Planck time after shrinking to the
Planck mass, and the infalling information disappears with it.  After
all, in practice, information often escapes to inaccessible regions of
spacetime, even in the absence of gravity.  The inclusion of gravity,
Hawking argues, implies information is lost in principle as well as in
practice.

\ifig\ftwo{Hawking's rule for density matrix superscattering
for single
black hole formation.  The left (right) side of the diagram represents
the evolution of the ket(bra) of the density matrix.  The trace over
the
part of the Hilbert space which falls into the black hole is schematically
represented
by sewing together the left and right black hole interiors.}
{\epsfysize=2.50in \epsfbox{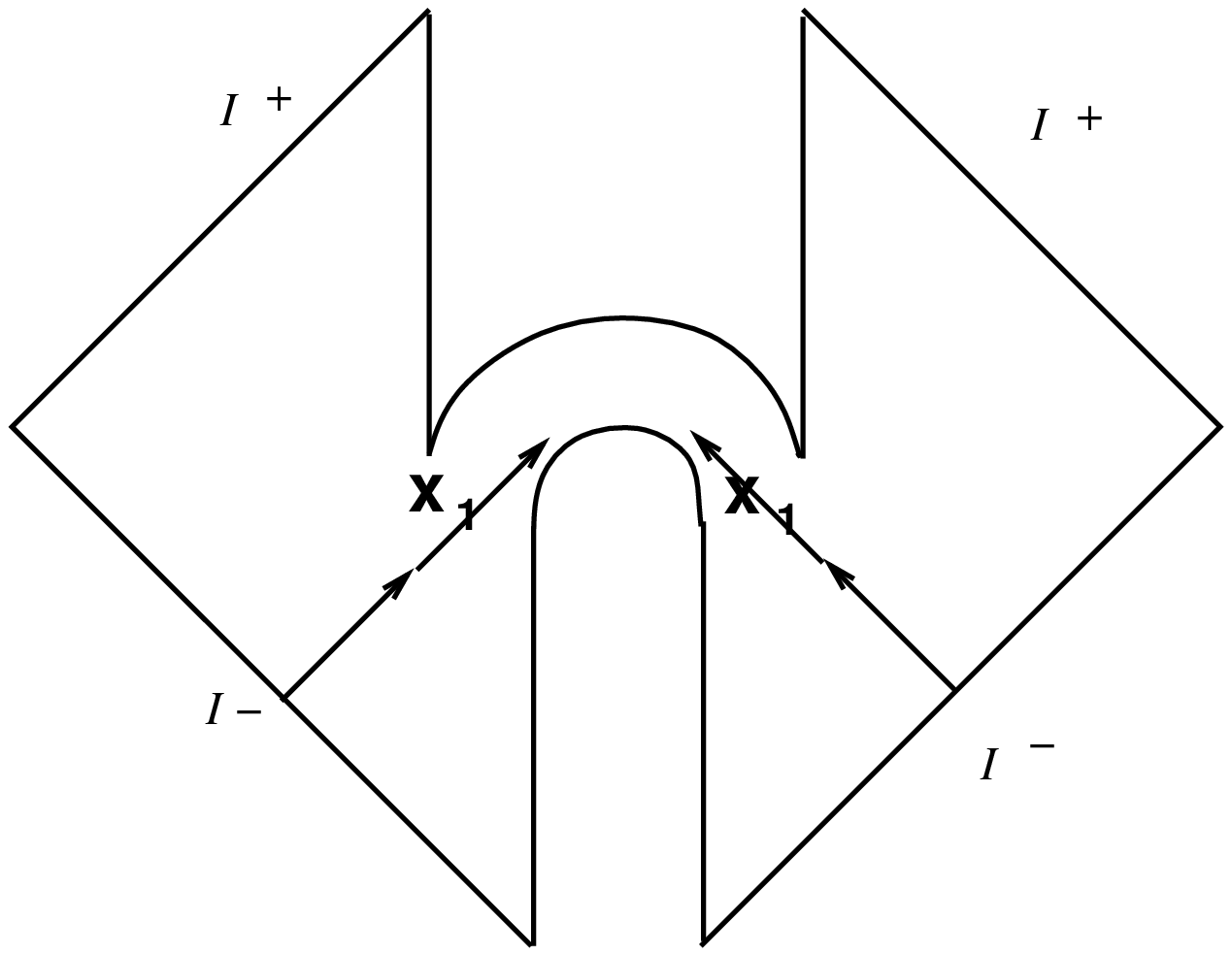}}

Since information is lost in this proposal, there can be no unitary
$S$-matrix mapping in-states to out-states.  Rather, Hawking suggests that a
``superscattering'' matrix, denoted ``$ {\not\kern-0.2em S} $'', which maps
in-density matrices (of the general form $ \rho=\sum \rho_{ij}|\psi_i\rangle
\langle \psi_j|$ )
to out-density matrices can be constructed as
\eqn\dss{\not\kern-0.2em S =tr_{BH} S\, S^\dagger\ .}
$ {\not\kern-0.2em S} $ will not in general preserve the entropy
$-{ tr}\rho \ln \rho$.
In components, $ {\not\kern-0.2em S} $ acts on an in-density matrix
as $\bigl(  {\not\kern-0.2em S}
\bigl[ \rho \bigr] \bigr)_{kl}=\bigl( {\not\kern-0.2em S}
 \bigr)_{kl}^{~~~ij}\rho_{ij}$.
$S$ here is a unitary operator which maps the in-Hilbert space to the
product of the out-Hilbert space with the Hilbert space of states which
falls into the black hole (defined, for example, as quantum states on
the event horizon in \fone ). $tr_{BH}$ is the instruction to trace
over these latter unobservable states.  Expressions of the form \dss
\ are familiar in physics, and arise, for example, in the computation of
$e^+e^-$ scattering in which the spins of the final state are not
measured. A diagrammatic representation of Hawking's prescription for
the case of one black hole appears in \ftwo .

\ifig\fthree{Hawking's rule for superscattering of two black
holes involves two
traces, one for each black hole.}
{\epsfysize=2.50in \epsfbox{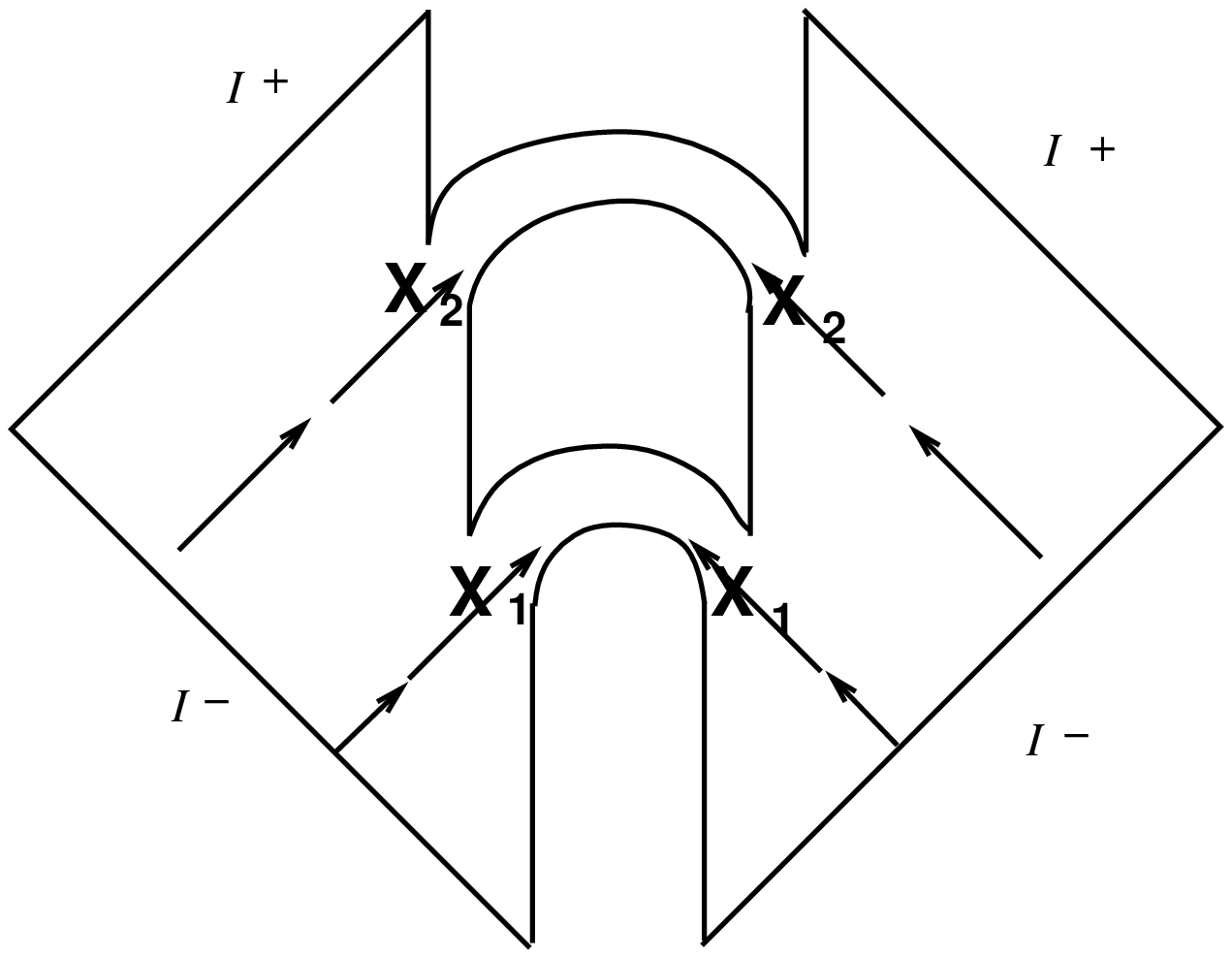}}

It is implicit in Hawking's proposal that the probabilistic outcome of
the formation/evaporation of an isolated black hole near the spacetime location
$x_1$ can in
this manner be computed from the portion of the quantum state which
collapses to form the black hole. In this case
the outcome of forming a second black
hole at a greatly spatially or temporally separated location $x_2$
is uncorrelated and the two-black hole $ {\not\kern-0.2em S} $-matrix can be
decomposed into a product of single black hole $\not\kern-0.2em S$-matrices
 (In other words, probabilities cluster.) The
corresponding diagrammatic representation of $\not\kern-0.2em S$ for the case
of two
black holes is given in \fthree .

\subsec{The Superposition Principle}
In fact as it stands Hawking's
proposal is not self-consistent\foot{The arguments of this and the following
section
may be related to those employed
in a somewhat different context in \bps\ and \sbg.}. The problem arises in its
sharpest form
when considering
superpositions of incoming states which form black holes at
different locations.
The superposition
principle of course implies that such states can be constructed.
To see the problem
note that there are inevitably
non-zero but possibly small quantum fluctuations in the location $x_1$
where the black hole is formed.  $tr_{BH}$ instructs one to trace by
equating the black hole interior states of the bra and the ket in
the density matrix, independently of the precise location where the
black hole is formed.  Now $x_1$ cannot be an
observable of the black hole interior Hilbert space, since  by
translation invariance the interior
state of the black hole does not depend on where it was formed. Hence
 the trace will
include contributions from black holes interiors which are in the same
quantum state, but which were formed at slightly different spacetime
locations.

\ifig\ffour{
Superscattering of an initial coherent superposition of
semiclassical states which form black holes near widely separated
locations $x_1$ and $x_2$.  The superposition principle and translation
invariance imply that all four diagrams contribute.}
{\epsfysize=3.50in \epsfbox{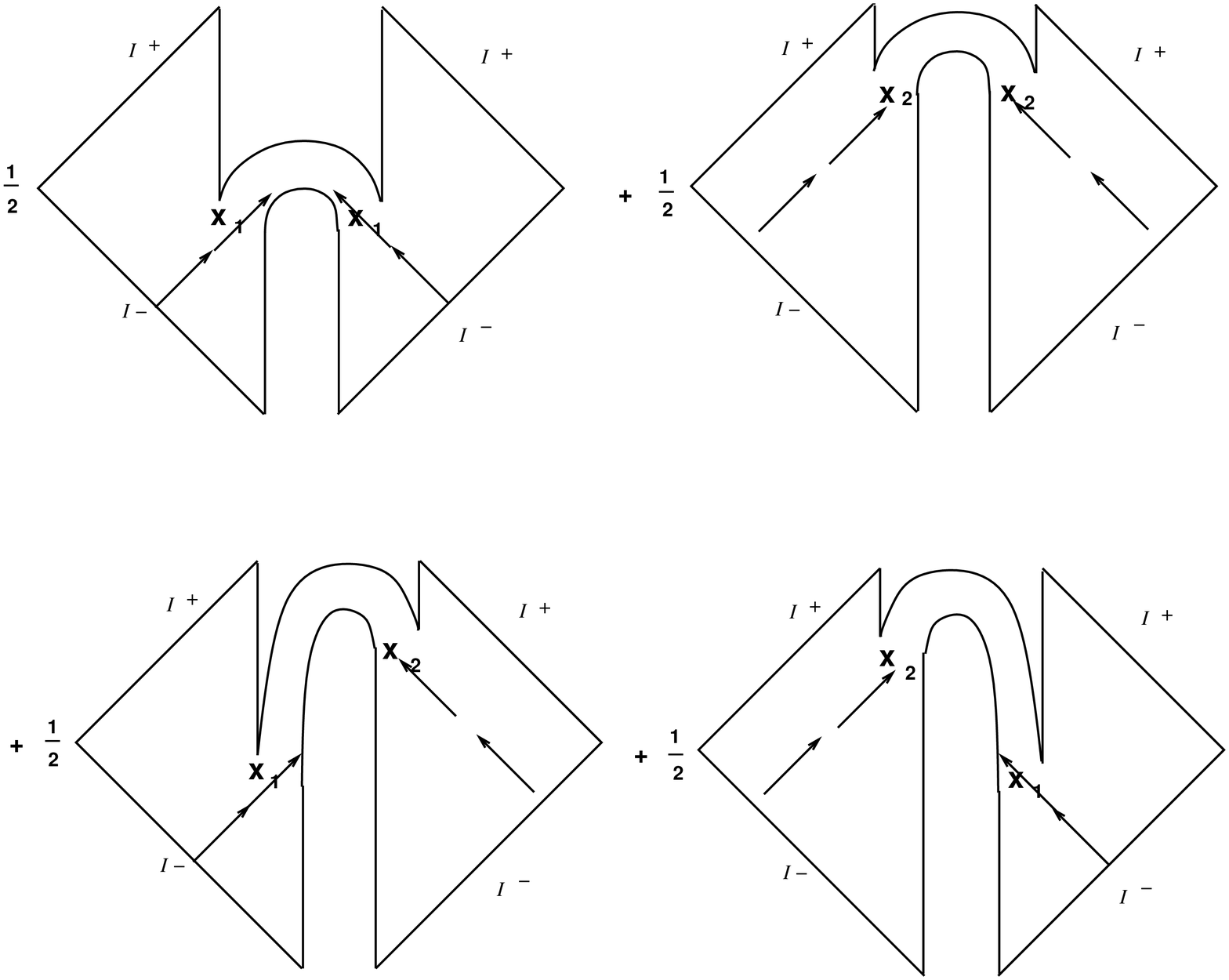}}

This phenomenon is more pronounced in initial states for which the
fluctuations in the location of the black hole are not small. Such states
can certainly be constructed.  For
example, let the in-state be the coherent superposition
\eqn\two{|\psi_{\rm in}\rangle= \frac{1}{\sqrt{2}} \left(|x_1\rangle +
|x_2\rangle\right),}
where $|x_i\rangle$ is a semiclassical initial state which collapses to
form a black hole near $x_i$, and $x_1$ and $x_2$ are very widely
separated spacetime locations.  By continuity
the construction of $\not\kern-0.2em S$
must include terms which equate the interior black hole
bra-state formed at $x_1$ with the ket-state formed at
$x_2$.  There are then four terms in $\not\kern-0.2em S$ as
illustrated in \ffour .

\ifig\ffive{The superposition principle implies that for two
black holes
this cross diagram must be added to that of \fthree , correlating
widely
separated experiments.}
{\epsfysize=2.50in \epsfbox{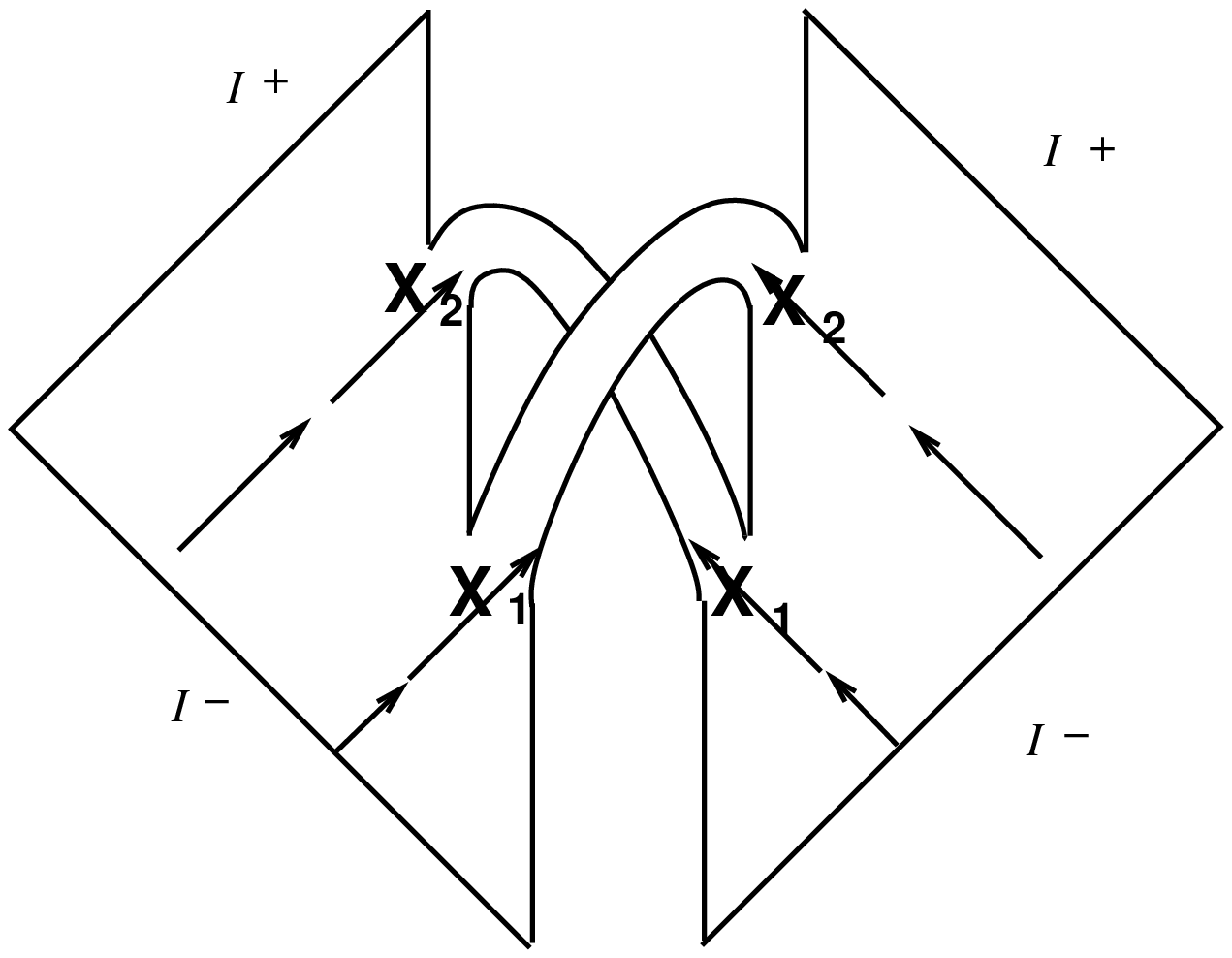}}

It may already seem rather strange
that $\not\kern-0.2em S $  should contain such
correlations between widely separated events, but matters become even
worse when one considers a semiclassical initial state $|x_1, x_2\rangle$
which collapses to form two black holes at the widely separated
locations $x_1$ and $x_2$.  The superposition principle then requires
that the cross diagram of \ffive\ be added to the diagram
of \fthree \foot{This extra  cross diagram will be
small if the parts of the incoming
states which form the two black holes are
very different and the black hole interiors have a correspondingly
small probability of being in the same state. On the other hand if they differ
only by a translation, \ffive\ will be similar in size to \fthree.}.
To see this, consider a smooth
one-parameter family of initial states $|x_1(s), x_2(s)\rangle$ in which
the locations $x_1$ and $x_2$ are interchanged as the parameter $s$ runs
from zero to one.  Let the in-state be
\eqn\three{|\psi_{\rm in}\rangle= \int^1_0 ds|x_1(s),
x_2(s)\rangle\ .}
Then the diagrams of \fthree\  and \ffive\ are interchanged as $s$ goes from
0 to 1 in the ket-state, so neither can be invariantly
excluded.

Thus the superposition principle implies that one cannot, in
the manner Hawking suggests, compute the probabilistic outcome of a
single experiment in which a black hole is formed.  Knowledge of all
past and future black hole formation events is apparently required
to compute the superscattering matrix
(although we shall see below that this is not as unphysical as it seems).
Again, it is striking that low-energy reasoning highly constrains
possible outcomes of black hole formation without requiring knowledge of
 planckian dynamics.

\ifig\fsix{When the evolution of spacelike slices (denoted by the dashed lines)
reaches the endpoint
$x_E$, the incoming slice, and the quantum state on the slice,
is split into exterior and interior portions.
This splitting process will be described using the operator $\Phi_J$ ($\Phi_K$)
which annihilates (creates) an incoming (outgoing)
asymptotically flat slice in the $J'th$ ($I'th$) quantum state and $\Phi_i$
which creates an interior slice in the $i'th$ quantum state.}
{\epsfysize=3.50in \epsfbox{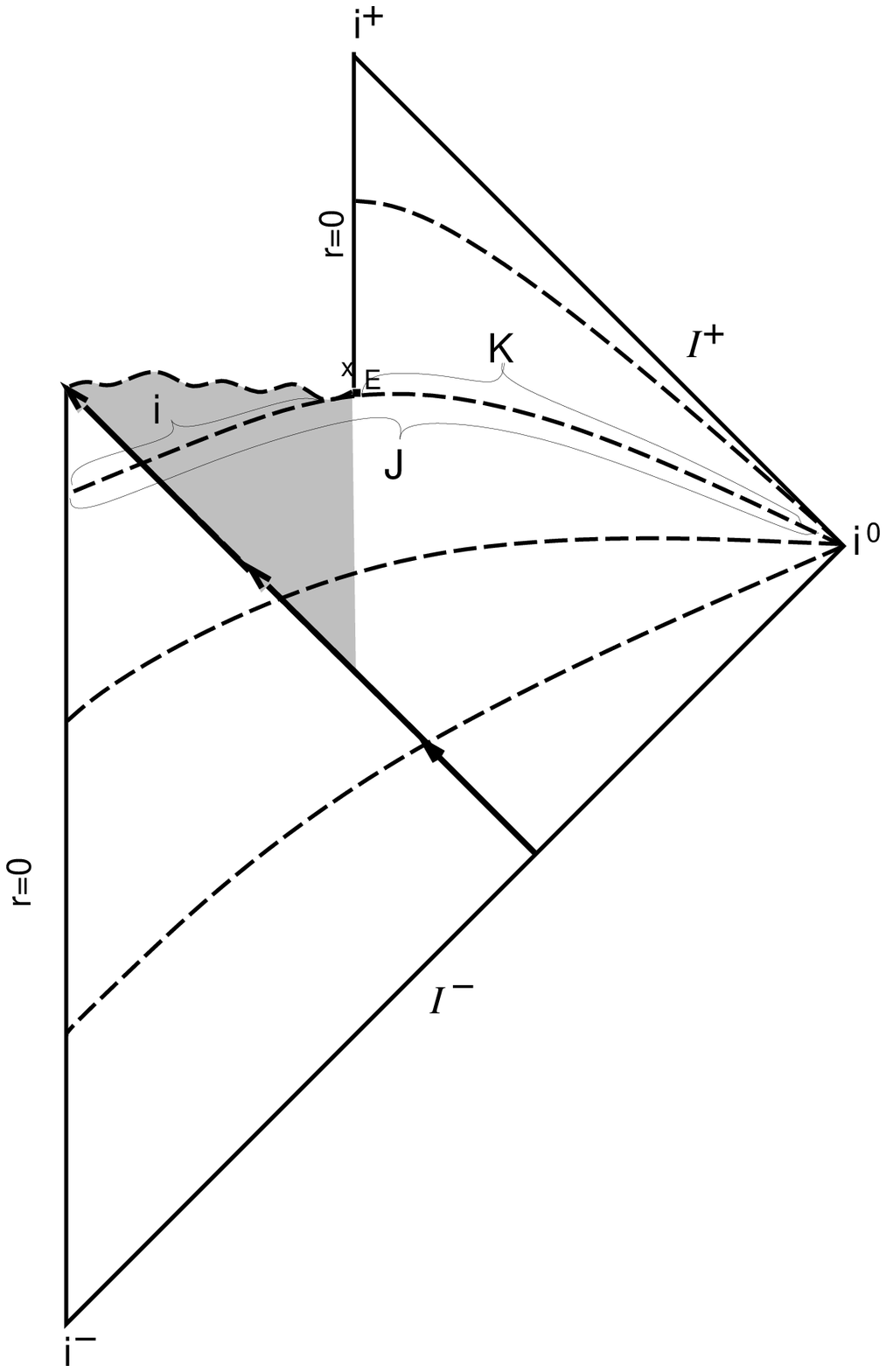}}

Note that our conclusions about difficulties with the usual
interpretation of Hawking's proposal
have derived from consideration of {\it superpositions} of semiclassical
states which form black holes.  These difficulties have not been so
evident in previous discussions simply because such superpositions are
not usually considered.

\subsec{Energy Conservation}
Although the superposition principle is restored with the extra cross diagram
of \ffive , correlations are introduced between arbitrarily widely separated
experiments, and clustering is violated \refs{\suss}.
Thus we seem to be faced with a choice:
abandon the superposition principle or abandon
clustering.  In fact we shall see below
that the breakdown of clustering
is a blessing in disguise, but first we need to
introduce a second refinement of Hawking's prescription required by
energy conservation\foot{I am grateful to S. Giddings for emphasizing
to me the
importance of understanding energy conservation in this context.}.
\ifig\fseven{Anderson and DeWitt studied a free field
propagating on a
geometry which is split into two at time $t=t_s$ by reflecting boundary
conditions at $x=0$. The sudden change in the Hamiltonian produces
infinite energy pulses which propagate along the dashed lines.}
{\epsfysize=3.00in \epsfbox{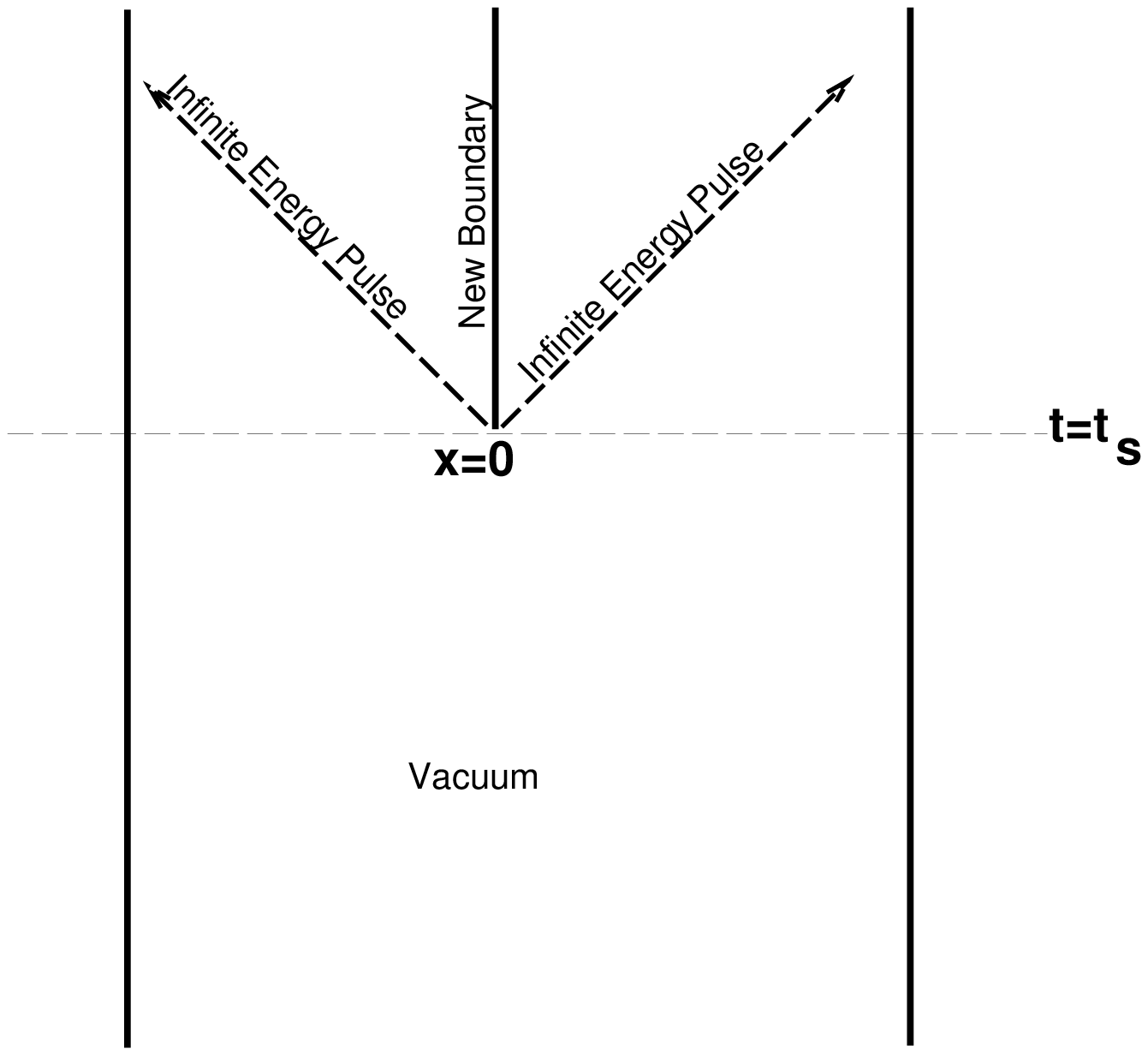}}
In computing the $\not\kern-0.2em S$-matrix, complete
spacelike slices are split into interior and exterior portions when they
encounter the evaporation endpoint at $x_E$, as illustrated in \fsix .
One imagines that the Hilbert
space on these slices is also split into the product of two
corresponding interior and exterior Hilbert spaces.  This requires some
new boundary conditions originating at $x_E$ (as in the RST model) : an
incoming light ray just prior
to $x_E$
falls into the black hole, while an incoming light ray just after $x_E$
reflects through the origin and back out to null infinity.  Implementing
this in practice immediately runs afoul of the Anderson-DeWitt
\refs{\ande} problem.  These authors considered the propagation of a free
conformal field in $1+1$ dimensions on the trousers spacetime of \fseven\
in which (as in the black hole case) spacelike slices are split into
two portions at some fixed time $t_s$, when reflecting boundary
conditions are turned on at $x=0$. They find that the vacuum state for $t<t_s$
evolves to a state with infinite energy for $t>t_s$.  This is not
surprising since the Hamiltonian changes at an infinite rate at $t=t_s$.

This phenomenon is not peculiar to two dimensions. A change in the Hamiltonian
in the form of new boundary conditions at a fixed spacetime location
violates general covariance and therefore energy conservation.
This problem should be expected
to affect the separation of Hilbert space into interior and exterior
portions at the evaporation endpoint $x_E$ for the black hole case.
Indeed the most concrete description given of this splitting
process --- that in
the $1+1$ dimensional RST model \rst --- suffers from exactly this problem
as discussed in 3.10.
Energy is not conserved in this model because the quantum state of the
matter field acquires infinite energy as it is propagated past $x_E$ \lpst.

\ifig\feight{A cosmic string decays into two pieces which end
at monopoles.  This process conserves energy, and the decay
Hamiltonian involves the fields $\phi_J$ which annihilates the incoming
string and $\phi_I, \phi_K$ which create the two outgoing strings.}
{\epsfysize=3.00in \epsfbox{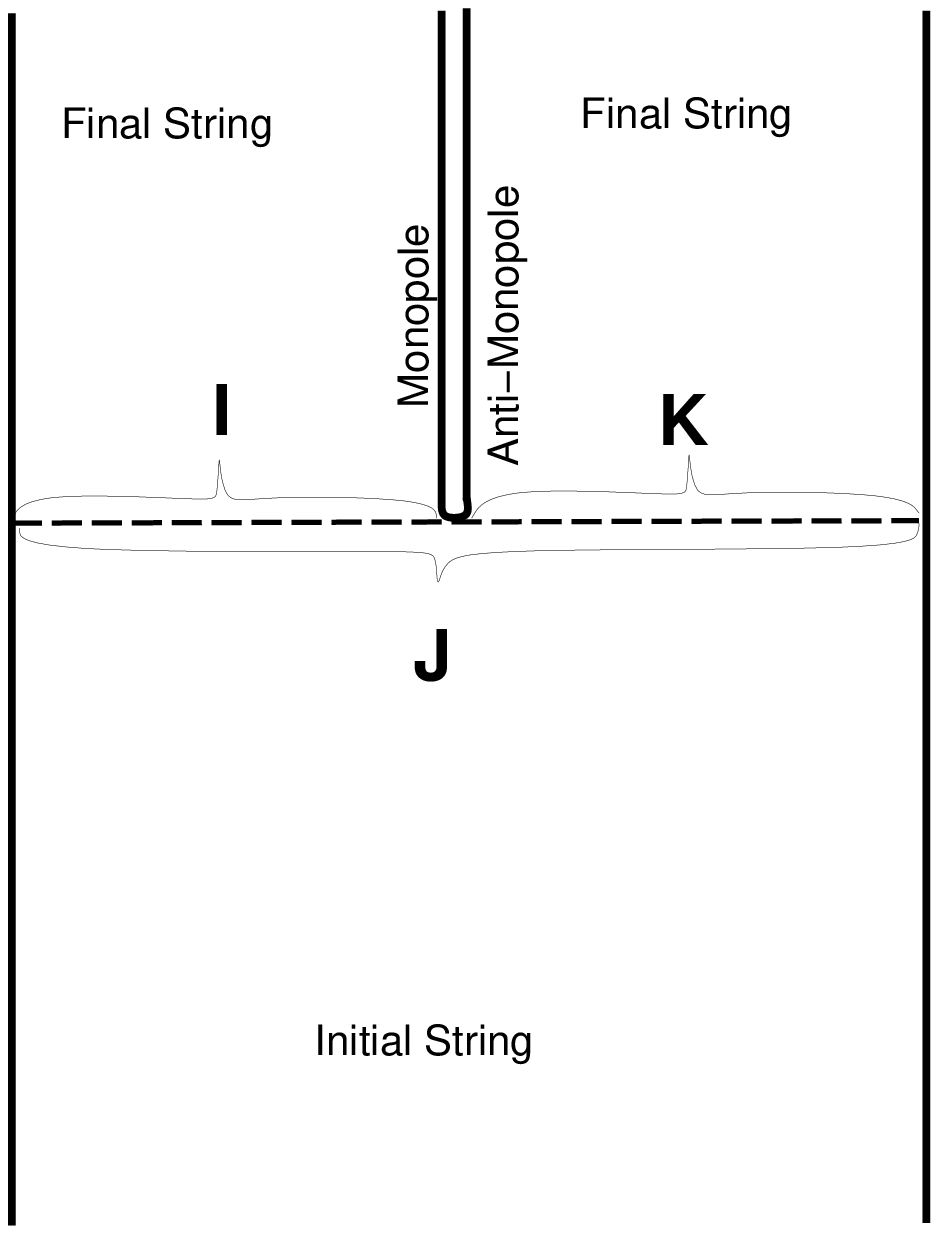}}

To remedy this, a smooth energy-conserving method
of splitting the incoming Hilbert space
into two portions is needed.  A physical example of a system which exhibits
such
a smooth splitting is given by cosmic string decay.  Consider, {\it
e.g.} a magnetic flux tube described by a
Nielsen-Olesen vortex.  At low energies it is
described by a $1+1$ dimensional quantum field theory whose massless
fields are the transverse excitations $X(\sigma)$ of the string.  Next suppose
that
the string can decay by the formation of a heavy monopole-anti-monopole
pair which divides the string into two parts.  Clearly such a process can
occur and will conserve energy.  It cannot, however, be simply described by
propagating the $1+1$ dimensional fields on the fixed geometry of \fseven\
(or superpositions thereof), as analyzed by Anderson and DeWitt.
Rather, the decay rate depends on the
final state after the split through initial and final wave function
overlaps appearing in decay matrix elements, and the decay time is thus
correlated with the quantum states on the two final strings.  This
decay process may be conveniently and approximately (at low energies)
described by the interaction Hamiltonian (see \feight )
\eqn\seven{{\cal H}_{\rm int} = \sum_{I,J,K} g\, \rho_{IJK}\phi_I
\phi_J\phi_K\ .}
In an appropriate basis, the mode of the field operator
\eqn\eight{\phi_I = a_I  + a^\dagger_I}
here creates or annihilates (from nothing) an entire string in the
$I$'th quantum state with wave function $u_I[X(\sigma)]$, and $[a_{I},
a^\dagger_J] = \delta_{I J}$. We emphasize that
$\phi_I$ is {\it not} an operator which acts on the single-string Hilbert
space. $\rho_{IJK}$
is the
overlap of the one initial and two final state wave functions $u_I, u_J, u_K$
for strings aligned as in
\feight.
$g$ is an effective low-energy coupling constant governing the decay
rate, in which our ignorance of the microscopic details of the splitting
interaction is hidden.

Despite many efforts, no other method of avoiding the Anderson-DeWitt
problem is known.  We accordingly {\it presume} that the disappearance of a
black hole is properly viewed as a  quantum decay process in which the
black hole interior and exterior are separated. We cannot {\it derive}
this presumption without solving quantum gravity. Nevertheless, it appears
to be forced on us by low-energy considerations. We know of no other
consistent effective description.

In this picture the decay does not then occur
instantaneously when the semiclassical evaporation endpoint $x_E$ is reached.
Rather the geometry itself decides when to split (some time after $x_E$)
in a quantum mechanical
fashion, controlled by the effective decay coupling constant as well as
phase space factors appearing in initial/final wave function overlaps.
The precise splitting time, like all other quantities, is then
subject to quantum fluctuations and correlated with the final state.
\subsec{The New Rules}
We have proposed two modifications of Hawking's prescription: the
inclusion of cross diagrams as in \ffive\
and the description of the final stages of
black hole evaporation as a quantum decay.
We shall see that these modifications have dramatic
consequences. In order to understand these consequences, it is useful
to note that the modified scattering rules are concisely summarized
by the tree diagrams\foot{The loop diagrams may be suppressed by adjusting
coupling constants, as in wormhole physics. A discussion of this and
the effects of
loops (if included) can be found in \strb.} of the theory defined by
\eqn\psiev{i\partial_T |\psi(T)\rangle = (H_0 + H_{\rm int}\rangle
|\psi(T)\rangle,}
\eqn\drule{\eqalign{\not\kern-0.2em S \bigl[|\psi_{\rm in}\rangle&
\langle \psi_{\rm in} |\bigr] = tr_{BH} |\psi_{\rm out}\rangle
\langle\psi_{\rm out}|,\cr
|\psi_{\rm in}\rangle&\equiv |\psi(-\infty)\rangle,\cr
|\psi_{\rm out}\rangle&\equiv |\psi(+\infty)\rangle,\cr}}
where  $H_0$ is the usual  gravitational Hamiltonian
which evolves the system along a set of spacelike slices labeled by time
coordinate $T$, but does not include the decay interaction.  The latter
is given, in precise analogy to the cosmic string case by
\eqn\eleven{H_{\rm int} = \sum_{i,J,K} g\, \rho_{iJK} \Phi_i \Phi_J \Phi_K\ .}
$\Phi_J$ here creates or annihilates an asymptotically flat spacetime in
the $J$'th quantum state. (It does {\it not} act on the flat space vacuum
to create the $J$'th excitation.) $\Phi_i$ creates or annihilates a
compact spacetime, {\it i.e.} a black hole interior, in the $i$'th quantum
state.
$\rho_{iJK}$ is the wave function overlap computed by aligning the
geometries as depicted in \fsix . $g$ is a decay coupling constant in
which our ignorance of Planck-scale physics is hidden.

The operators $\Phi_i=a_i  + a^\dagger_i$ generate a
multi-black-hole-interior Hilbert space $H_{\rm BH}$. If $[a_i,
a^\dagger_j] = \delta_{ij},\ |\psi_{\rm in}\rangle$ is taken to obey
$a_i|\psi_{\rm in}\rangle=0$ and $tr_{\rm BH}$ is the trace over $H_{\rm
BH}$, then the rule \drule\ for construction of ${\not\kern-0.2em S}$  contains
(with the
correct weighting) the cross diagrams required by the superposition
principle. To see how this works, suppose an initial state
$|\psi_{\rm in}\rangle$ collapses to
form two black holes (at different locations) which subsequently evaporate.
Then the out-state is of the general form
\eqn\psout{|\psi_{\rm out}\rangle=\sum_{i,j}a^\dagger_ia^\dagger_j|\psi_{\rm
out}^{ij}\rangle,}
where $a_k|\psi_{\rm out}^{ij}\rangle =0$. Using the commutation relations
$[a_i,
a^\dagger_j] = \delta_{ij}$, the out density matrix is
\eqn\dout{tr_{BH} |\psi_{\rm out}\rangle \langle\psi_{\rm out}|
=\sum_{i,j} \bigl(|\psi_{\rm out}^{ij}\rangle\langle\psi_{\rm out}^{ij}|
+|\psi_{\rm out}^{ij}\rangle\langle\psi_{\rm out}^{ji}|\bigr).}
The second term on the right hand side is precisely the cross diagram of
\ffive.

The $\Phi_i$'s may be simply viewed as a convenient mnemonic
for constructing the diagrammatic expansion of $\not\kern-0.2em S$.
 Alternately,
one may think of the black hole interiors as forming baby universes which
inhabit a ``third quantized'' Hilbert space \refs{\thrd,\strb}\
on which the $\Phi_i$'s act.
However, the detailed dynamics of these baby universes will not be
needed for our purposes because we view them as unobservable.
\subsec{Superselection Sectors, $\alpha$-parameters, and the Restoration of
Unitarity}
Next let us suppose that the initial state is in an ``$\alpha$-state''
obeying \worm
\eqn\twelve{\Phi_i | \{\alpha\}\rangle = \alpha_i | \{\alpha\}\rangle,}
where the $\alpha_i$'s are $c$-number eigenvalues, rather than $a_i
|\psi_{\rm in} \rangle=0$.
In such a state the operator $\Phi_i$ may be everywhere replaced by its
eigenvalue and
\eqn\thirteen{H_{\rm int} = \sum_{J,K} g_{JK} \Phi_J\Phi_K}
with
\eqn\fourteen{g_{\rm JK} = \sum_{i}\alpha_i\rho_{iJK} = c-{\rm numbers}\ .}
$H_{\rm int}$ {\it reduces to an operator on the Hilbert space of a single
asymptotically flat spacetime.}\foot{\thirteen\ may also
contain terms which create or destroy pairs of asymptotically flat universes.
But these can be ignored as they factor out of the
normalized evolution of a single connected universe.} It then follows
immediately from
\psiev\ that the out-state
\eqn\fifteen{|\psi_{\rm out}\rangle =
S_{\{\alpha\}}|\psi_{\rm in}\rangle}
is a unitary, $\alpha$-dependent transformation $S_{\{\alpha\}}$ of
the in-state. $S_{\{\alpha\}}$ here is obtained by solving \psiev,
which reduces to an ordinary Schroedinger-Wheeler-DeWitt equation in
an $\alpha$-state.

The reader may suppose that this result is of little interest
because the generic state is not an $\alpha$-state, rather it is a
coherent superposition of $\alpha$-states.  To understand the properties
of such superpositions, consider
\eqn\psdc{|\psi\rangle = \theta | \{\alpha\}\rangle + \theta^\prime |
\{\alpha^\prime\}
\rangle}
where
\eqn\aort{\langle\{\alpha\} | \{\alpha^\prime\}\rangle=0}
since $\alpha$-states are eigenstates of a hermitian operation with distinct
eigenvalues.

Observables ${\cal O}_i$ corresponding to measurements in the
asymptotically flat spacetime do not act on the multi-black-hole-interior
Hilbert space $H_{\rm BH}$.  Hence they commute with the $\Phi_i$'s and
leave the
$\alpha$-eigenvalues unchanged.  It then follows from \aort\ that
\eqn\ssct{\langle\{\alpha\}| {\cal O}_i | \{\alpha^\prime\} \rangle = 0}
and
\eqn\ssrt{\eqalign{&\langle\psi | {\cal O}_1 {\cal O}_2 \cdots {\cal O}_N |
\psi\rangle~~~~~~\cr
&~~~~~~ = |\theta |^2 \langle\{\alpha\} | {\cal O}_1 {\cal O}_2 \cdots {\cal
O}_N |\{\alpha\}\rangle\cr
&~~~~~~+|\theta^\prime|^2 \langle\{\alpha^\prime\}| {\cal O}_1 {\cal O}_2
\cdots {\cal O}_N |\{\alpha^\prime\}\rangle\ .\cr}}
A similar relation holds for more general superpositions of
$\alpha$-states, including the ``vacuum'' state obeying
$a_i|\psi\rangle=0$.

The content of \ssrt\ is that the $\alpha$'s label non-communicating
{\it superselection sectors}.  According to \ssrt, the amplitude for
repeating an experiment which measures an $\alpha$-value and obtaining a
different result the second time is
zero.\foot{In the Copenhagen interpretation, one would say that
measurement of an $\alpha$-value collapses the wave function to the
corresponding $\alpha$-eigenstate.} Once an experiment records a given
$\alpha$-value, all future experiments will agree.  There may be
parallel worlds with different $\alpha$-values, but we can never know
about them.  Hence {\it the $\alpha$'s are effectively  constants}
and {\it black hole formation/evaporation is an effectively unitary
process}.\foot{This argument parallels those in earlier work on baby universes.
In \refs{\wrloss}  it was argued, following \hawktwo, that virtual,
planckian baby universes destroy
information.  This conclusion was shown in \refs{\worm} to be false after
proper accounting of superselection sectors.  Following these
developments, many authors tried and failed to adapt the mechanism of
\worm\ to
avoid information destruction by black holes.  The missing ingredient in
these previous attempts to adapt the results of \refs{\worm}
 was the description of the Hilbert space split
as a quantum mechanical decay process.}

We find this result extremely satisfying.  Having modified Hawking's
superscattering rules so as to comply with the superposition principle and
energy conservation, we see that unitary is restored as a free bonus.
This attests to the robust nature of quantum mechanics, and the inherent
difficulty in finding self-consistent modifications.

The real significance of the very-long-range correlation produced by the
cross diagram of \ffive\ is now evident.  They simply conspire to
produce
infinite-range correlations between $\alpha$-values measured in widely
separated experiments.  They do {\it not} allow messages to be sent faster
than the speed of light, or money to be consistently won at the racetrack.

What are the $\alpha$'s in our universe?  Even an exact solution to
string theory could not answer this question: They can only be
determined by forming black holes and measuring the out-state\foot{Of
course in principle
the $\alpha$'s might be fixed by new considerations as in \sidcc,
but that is far beyond the scope of these lectures.}.  Until
they are known, the outcome of gravitational collapse is unpredictable.
The time reverse of this statement is that
information is lost in the sense that the in-state which formed a black
hole cannot be determined even from complete knowledge of the out-state.
This is certainly similar to, and could be regarded as a refinement of,
Hawking's original contention that information is lost in black hole
processes.
Indeed, if one performs a Gaussian average over $\alpha$'s one
recovers results similar to Hawking's (in that pure states
go into mixed ones)
for the case of a single black hole.  Thus the
difference between our proposal and Hawking's is in practice quite
subtle.

The following analogy may clarify the situation.  Consider scattering
photons off of a hydrogen atom.  Imagine that QED is perfectly
understood, except that the value of the fine structure constant is
unknown.  In this case it will not be possible to predict (retrodict)
the out-state (in-state) from the in-state (out-state) of a single
experiment, so that in a sense one could say that information is lost.
However, after performing many scattering experiments, the fine
structure constant is effectively measured, and no further information
loss occurs.

Information loss in black hole formation/evaporation is of exactly this
type.  It does not arise from a fundamental breakdown of unitarity,
rather it is associated with a lack of knowledge of coupling constants
(the $\alpha$'s or $g_{JK}$'s).  The only difference is that in the QED
case there was only one relevant coupling, while in the black hole case
many are needed (more than $e^{4\pi M^2}$ \polst\ ) even to predict the outcome
of
a single fixed in-state, and an enormous number of experiments would be
required to actually measure the parameters. Indeed, since there are an
infinite number of in-states which form black holes (of unrestricted
mass), it is never possible  to measure {\it all} the $\alpha$
parameters.

The alert reader may be concerned about the status of the
information/energy bounds discussed in 4.3, which constrain the rate at
which the information can be returned with the
small amount of energy available near and after the endpoint. The
arguments for these bounds are quite general and certainly apply to our
proposal.  Thus unitarity implies that our decay rate must be very slow.
One cannot simply
 explain this with a small $g$ as $g$ --- though hard to calculate ---
 is naturally order one
in Planck units. Rather it was shown explicitly in a two-dimensional model
in \refs{\polst} that
the decay is highly
suppressed by phase space factors: due to entanglement of the interior
and exterior states, the overlap between the initial and final state
wave function is small, providing for compatibility with the
information/energy bounds (see also \sbg ). Unitarity implies a similar phase
space suppression
in four dimensions: it is important to understand explicitly how this arises.
\newsec{Conclusions and Outlook}
In Section 3, two-dimensional models were analyzed with the aim
of gaining a more concrete understanding of black hole formation/evaporation
in a simplified context. Prior to the evaporation endpoint, these models
behave just as Hawking long ago argued that real four-dimensional black holes
would behave. Many criticisms of Hawkings calculation
(prior to the evaporation endpoint)
can be seen to be invalid in this simplified context. Thus the
results from the two-dimensional models strengthen our confidence in
Hawking's four-dimensional, pre-endpoint analysis. On the other hand, attempts
to
find a two-dimensional model which consistently implements
Hawking's post-endpoint prescription for throwing away the information which
falls into
the black hole have been notably unsuccessful.

Attempts to consistently realize Hawking' proposal in a concrete
fashion in two dimensions led to general insights which are
applicable in the four-dimensional context. In Section 4 we
reviewed arguments that
Hawking's proposal for information destruction by black
holes --- as usually interpreted ---  violates energy conservation
in addition to unitarity, and
does not provide a self-consistent rule for
evolving superpositions of states which form black holes at
different locations.  Refinements of (or reinterpretations of)
his proposal which restore the superposition principle and energy
conservation automatically restore unitarity, after the existence of
superselection sectors is properly accounted for.  This can be
accomplished without requiring that planckian dynamics become important
at low curvatures (as some have advocated). The resulting
description of quantum black hole dynamics
agrees exactly with Hawking's everywhere that
semiclassical
reasoning is valid, namely prior to the evaporation endpoint, but
differs thereafter.
It also does not invoke the existence of stable objects with no natural
right to eternal life: Rather it predicts the existence of long-lived
remnants whose long
lifetime may be naturally explained by phase-space suppression of the decay
rate. Thus a unitary, causal
description of black hole formation/evaporation appears to be
natural and compatible
with all known constraints of low-energy physics.

The arguments of Section 4 are general in nature.  Our understanding
would be greatly enhanced by the construction of an explicit
two-dimensional model
which realizes the picture of information flow described in Section 4.
Many of the tools required for such a construction
were developed in Section 3.
This is an interesting problem for future research.

In closing, we would like to raise an important issue which has
not been covered in these lectures, but which my be important
for future developments. In the nineteenth century,
Boltzmann derived the laws of thermodynamics from statistical
mechanics. In the early seventies, the laws of black hole mechanics
were derived from Einstein's equation and  differential geometry. It was
immediately noticed that the
laws of classical black hole mechanics are
identical to those of thermodynamics when the variables are renamed
({\it e.g.} the substitution of the entropy for the black hole area).
Shortly thereafter, with the discovery of Hawking evaporation, it was
realized that there is really only {\it one}
unified set of laws: in the presence
of quantum mechanical black holes, neither the laws of thermodynamics or
of classical black hole mechanics are separately valid.
For example, in the real world the horizon area $A$ may decrease
(because of Hawking evaporation) in violation of the area theorem
and the accessible entropy $S$ may decrease (by falling in to a black hole)
in violation of the second law.
However a combination of the two sets of laws appears to
remain intact. For example, there is good
theoretical evidence \bek\ that
the magical sum $S+A/4$ is always non-decreasing.

The derivations of the laws of thermodynamics and the
laws of classical black hole mechanics are both
extremely beautiful, but could hardly be more different.
The fact that they are united in the end crys out for a unified
treatment, in which the two sets of laws are not patched together,
but appear as different manifestations of the same underlying principle.
It is hard to imagine how this might be achieved. Some have advocated
that the laws of black hole mechanics are really statistical in nature,
and that the (exponential of) the horizon area literally counts black hole
microstates. Another possibility is that the entropy is a kind of
quantum area, and the second law of thermodynamics is a quantum
area theorem. Perhaps more likely is that a
totally new point of view is necessary. In any case the
resolution of this issue seems likely to lead to fundamental changes in our
view of quantum mechanics and gravity.
It will be fascinating to see how
or if this meshes with the picture of information flow developed in these
lectures.

In conclusion, quantum black hole physics is a fertile subject with
no shortage of fascinating and confusing questions.

\centerline{\bf Acknowledgments}

I am grateful to A. Anderson, T.~Banks, K. Becker, M. Becker,
C. Burgess, S. Coleman, J. Frolich, S.~Giddings, P. Ginsparg, J. Harvey,
S.~Hawking,
D.~Lowe, R. Myers,
J.~Polchinski,
J.~Preskill, M.~Srednicki, L.~Susskind, L.~Thorlacius, V. Rubakov,
E. Verlinde and the students at les Houches for stimulating
conversations and questions, and to the organizers for the invitation
to lecture. This work was supported in part by DOE grant DOE-91ER40618.

\listrefs

\end